\documentclass[10pt]{article}
\usepackage[utf8]{inputenc}
\usepackage{mathtools,amsmath,amssymb}
\usepackage{graphicx}
\usepackage{lmodern}
\usepackage[T1]{fontenc}
\usepackage[pagebackref=false]{hyperref}
\usepackage{bbm}
\usepackage{bm}
\usepackage{appendix}
\usepackage{comment}
\usepackage{mathtools}
\usepackage{array}
\usepackage{hyperref}
\numberwithin{equation}{section}
\numberwithin{equation}{subsection}

\usepackage[margin=2.5cm]{geometry}

\usepackage{xspace}

\usepackage{graphicx}

\usepackage{subcaption} 
\usepackage[numbers,sort]{natbib}

\usepackage{tikz}
\usetikzlibrary{decorations.pathreplacing}

\newtheorem{theorem}{Theorem}
\newtheorem{proposition}{Proposition}

\newtheorem{corollary}{Corollary}
\newtheorem{remark}{Remark}
\newcommand{\id}{I}
\newcommand{\Xt}{\widetilde{X}}
\newcommand{\co}{\;,}
\newcommand{\dt}{\;.}

\newcommand{\EE}{\mathrm{E}}

\newmuskip\pFqmuskip

\newcommand*\pFq[6][8]{
  \begingroup s
  \pFqmuskip=#1mu\relax
  \mathchardef\normalcomma=\mathcode`,

  \mathcode`\,=\string"8000

  \begingroup\lccode`\~=`\,
  \lowercase{\endgroup\let~}\pFqcomma

  {}_{#2}\phi_{#3}{\left[\genfrac..{0pt}{}{#4}{#5};#6\right]}%
  \endgroup
}
\newcommand{\pFqcomma}{{\normalcomma}\mskip\pFqmuskip}

\newcommand{\dd}{\mathcal{D}_{u(x)}}

\newcommand{\twoj}{\nu}
\allowdisplaybreaks

\begin{document} 
 
\begingroup

\begin{center}
 \begingroup\LARGE

\bf Duality for the multispecies stirring process 
 with open boundaries
\par\endgroup
 \vspace{3.5em}
 \begingroup\large \bf
 {Francesco Casini$^{(a,b)}$, Rouven Frassek$^{(a)}$, Cristian Giardin\`a$^{(a)}$}
 \par\endgroup
\vspace{2em}

\begingroup\sffamily

$(a)$ University of Modena and Reggio Emilia, \\
Department of Physics, Computer Science and Mathematics,\\
Via G. Campi 213/b, 41125 Modena, Italy\\ 
 \vspace{.5cm}
$(b)$ University of Parma, \\
Department of Mathematical, Physical and Computer Sciences,\\
Parco Area delle Scienze, 7/A, 43124 Parma, Italy\\
\par\endgroup
\vspace{2em}

\end{center}

\thispagestyle{empty}

\begin{abstract}
\noindent
We study the stirring process with $N-1$ species
on a generic graph $G=(V,\mathcal{E})$  with reservoirs.
The multispecies stirring process generalizes the symmetric exclusion process, 
which is recovered in the case $N=2$.
We prove the existence of a dual process defined on
an extended graph $\widetilde{G}=(\widetilde{V},\widetilde{\mathcal{E})}$ which 
includes additional extra-sites $\widetilde{V}\setminus V$ where
dual particles get absorbed in the long-time limit.
We thus obtain a characterization of the non-equilibrium
steady state of the boundary-driven system in terms of the absorption
probabilities of dual particles.  The process is integrable
for the case of the one-dimensional chain with two reservoirs
at the boundaries and with maximally one particle per site.
We compute the absorption probabilities by relying
on the underlying ${gl}(N)$ symmetry and the
matrix product ansatz. Thus one gets a closed-formula
for (long-ranged) correlations and for the 
non-equilibrium stationary measure.
Extensions beyond this integrable set-up
are also discussed.
\end{abstract}

\newpage
\tableofcontents

 \newpage 
\section{Introduction} 

\paragraph{Motivations.}
Boundary-driven particle systems are paradigmatic models for non-equilibrium statistical mechanics. In a boundary-driven system, the model is put in contact with reservoirs, 
and a current is generated through the system. In the long time limit, a non-equilibrium steady state sets in, with a stationary value of the current.
Our strategy to solve for the non-equilibrium steady state is of a probabilistic nature and relies on the use of {\em duality} \cite{schutzSandow}. Starting from the pioneering work of Kipnis, Marchiorio and Presutti \cite{KMP}, this approach
has been substantially developed in recent years to study stochastic processes in the boundary-driven set-up \cite{giardina2009duality,carinci2013duality}.
The boundary-driven process is mapped to a dual process with absorbing extra sites and the problem of describing the stationary state of the original system
is simplified to the problem of computing the absorption probabilities of the dual particles.  
 The aim of this paper is to develop the theory of stochastic duality and  study the stationary measure of certain symmetric boundary-driven {\em multispecies} particle systems with exclusion.

\paragraph{The model.}
The model we consider is the {\em  multispecies stirring process with open boundaries}, 
i.e. the multispecies analogue of the exclusion process as introduced in \cite{zhou2021orthogonal}, 
with additional boundary reservoirs as proposed in \cite{vanicat2017exact}. 
The hydrodynamic limit and fluctuations of the multispecies stirring process on $\mathbb{Z}^d$
has been recently studied in \cite{casini2023density}.
In our set-up, particles can be of $N-1$ types (or species) and they can move
on a generic connected and undirected graph $G=(V,\mathcal{E})$ with vertex set $V$ and
edge set $\mathcal{E}$. The maximal number of particles allowed at each vertex (also called \textit{maximal occupancy}) is denoted by $\nu\in \mathbb{N}$. 
Two particles do interact when they sit on two sites connected by a graph edge and this interaction consists in swapping them. 
Furthermore, reservoirs are attached at each site of the graph, creating and removing particles. For a precise definition of the rates of the process 
we refer to Section~\ref{sec1}. As we will see, the choice of the boundary reservoirs is motivated by the fact that they allow to determine an absorbing dual process, 
regardless of the graph $G$ on which the process is defined and of the maximal occupancy $\nu$ at each site.

A special case is obtained when the graph $G$ is a finite chain, 
with two reservoirs attached to the first and the last sites respectively, 
and the maximal number of particles allowed at each site is fixed to be $\nu=1$.
In this case the multispecies stirring process with open boundaries 
becomes integrable (see \cite{vanicat2017exact}). 
We recall that for the single species case,  the solution of the exclusion process with open boundaries \cite{derrida1993exact,1993JSP....72..277S} has been of crucial importance in our understanding of the structural properties of non-equilibrium state, such as long-range correlations \cite{spohn1983long} and non-local density large deviation functions \cite{derrida2007non,derrida1998exact} and current large deviations \cite{mallick2022exact,bodineau2005current}. One may argue that the results available for the exclusion process are rooted in the algebraic symmetries of the corresponding integrable spin $s=1/2$ Heisenberg chain, see e.g. \cite{schutzManyBody,SSEPReviewRagoucy} for two excellent reviews.
This integrability property can be combined with duality and leads to closed-form expressions for the correlations and the non-equilibrium steady state, as shown in \cite{frassek2020eigenstates,frassek2020duality}.
In this paper we focus on the multispecies stirring process with open boundaries.
We choose to restrict to the symmetric version of the model, as the formulation
of duality in the presence of both asymmetry and boundary driving is substantially
more involved (for recent results on the single species case see
\cite{schuetz2022reverse,barraquand2022markov}).

\paragraph{Organization of the paper and main results.}
We start in Section~\ref{sec1} by defineing the open multispecies stirring
process on a general graph and providing its Lie algebraic description in terms of the generators of the $gl(N)$ Lie algebra. In Section~\ref{sectionDuality} we prove an absorption duality (Theorem~\ref{thm-duality})
and show that the problem of characterizing the non-equilibrium
steady state can be reduced to the computation of the absorption probabilities similar to the single species case on the line  \cite{KMP} and for a general graph \cite{ frassek2020duality}.  
In Section~\ref{sectionIntegrabiliy}  we consider  the case of the one-dimensional chain with at most one 
particle per site ($\nu=1$), which is driven out-of-equilbrium by two reservoirs on the left and right boundaries. In this case we obtain a closed-form expression for the stationary correlations (Theorem~\ref{thm-correlations}). Here the computation
relies on the extension of the  Matrix Product Ansatz \cite{derrida1993exact} to multiple species \cite{vanicat2017exact} and the $gl(N)$ invariance of the process in the bulk. Finally in Section~\ref{sec5} we discuss a few extensions 
and generalizations of the model. 
The first generalization is the formula for the correlations for the boundary driven chain in the case when more than one particle is allowed at each site ($\nu > 1$). 
In this situation integrability is lost and we cannot derive explicit formulas for the absorption probabilities. However, in Appendix \ref{appendix-1pts-general}, we show how to compute the one-point correlation by using the properties of a random walk. The second generalization is the definition of a thermalized version of the boundary-driven multispecies stirring process and the formulation of the the absorbing duality with the same duality function of the multispecies stirring process (the details of the proof are reported in Appendix \ref{appendix-dualityThermalized}). Finally, the third generalization is the introduction of an ``ad-hoc'' interaction mechanism of reaction type, in order to have closed evolution equation for average particle densities (inspired by \cite{casini2022uphill,schutzReaction}). 
We report the details of the computations for the closure of the average density evolution equation in Appendix \ref{appendix-RD}. 
Again, an absorbing duality is proved with the same duality function of the multispecies stirring process.

\bigskip

\paragraph{Acknowledgments.} We acknowledge support from Istituto Nazionale di Alta Matematica (GNFM),  the INFN grant Gauge and String Theory (GAST),  the PRIN project CUP-E53D23002220006,  the FAR UNIMORE project CUP-E93C23002040005 and  the INdAM–GNFM project CUP-E53C22001930001. RF thanks  Rodrigo A. Pimenta for related discussions.

\section{The stirring process with open boundaries}
\label{sec1}
\subsection{Informal description of the process}
The process studied in this paper is the {\em multispecies stirring process}
with open boundaries. 
Each site
of a connected graph can host a maximal number of particles called $\nu\in \mathbb{N}$ .
The particles have a {\em type} (sometimes called {\em species} or {\em colour})
which can takes  values $\{1,2,\ldots,N-1\}$. {More precisely, at each site there are $\nu$ available places (called \textit{holes} or \textit{vacancies}) that can be either occupied by particles or not. As a consequence, each site can be totally empty (when no particles are present and thus $\nu$ holes are available), totally occupied (when $\nu$ particles, of any type, are present and thus no holes are available) or partially occupied (when some of the holes are occupied and some are available). All these configurations obey the \textit{exclusion constraint} that tells that, at each site and at any time, the sum of the number of particles of any type plus the number of holes is $\nu$.}

Then, the dynamics has two parts: 
\begin{itemize}
\item
on each edge of the graph, any two types of particles are swapped at rate $1$;
moreover a particle of any type and a hole are also swapped at rate $1$; 
\item
on each vertex $x$ of the graph, a particle of type $a \in \{1,\ldots,N-1\}$ is created  at rate $\alpha_a^x >0$;
additionally a particles of any type is replaced with a
particle of type $a$ at rate $\alpha_a^x >0$ and it is removed at rate $\alpha_N^x >0$.
\end{itemize}
The swap dynamics taking place on the edges is of Kawasaki-type 
with $N$ conservation laws
(the total number of particles of each type). 
The site-dynamics is instead of Glauber-type. 
In the long-time limit, a so-called non-equilibrium
steady state sets in.
In the case $N=2$, we retrieve the  boundary-driven version of the symmetric exclusion process \cite{schutzSandow,carinci2013duality}.

\subsection{The process generator}\label{subsectionGeneratorStr}
We now give the mathematical description of the multispecies stirring process.
We consider a connected graph $G=(V,\mathcal{E})$ with vertex set $V$ and edge set $\mathcal{E}$.
At each vertex site $x\in V$, we describe the occupation with an $N-$dimensional vector $n^{x}=(n_{1}^{x},\ldots,n_{N}^{x})$ in which the value of the $a$-th component $n_{a}^{x}$ {denotes the number of particles of species $a\in \{1,\ldots,N-1\}$, while the component $n_{N}^{x}$ counts the number of holes at site $x$.} On each vertex we {allow a total maximal number of particles $\nu\in \mathbb{N}$}. {In the following we use indices denoted by lowercase letters (for instance $a,b,c,d\in \{1,\ldots,N-1\}$) when only particles are taken into account. Moreover, we introduce indices denoted by uppercase letters (for instance $A,B,C,D\in\{1,\ldots,N\}$)  to also incorporate the holes, that correspond to the index $N$. The link between the two indices is given by 
\begin{equation}
	a\in \{1,\ldots,N-1\}\qquad \text{and}\qquad A=(a,N)\,.
	\end{equation} 
} 
The configuration space of the process on the graph $G$ is 
\begin{equation}\label{stateSpace}
    \Omega:=\bigotimes_{x\in V} \Omega_{x}
\end{equation}
where
\begin{equation}
\Omega_{x}:=\left\{n^x=(n_{1}^{x},\ldots,n_{N}^{x})\in\mathbb{N}_0^{N}\;:\; \sum_{A=1}^{N}n_{A}^{x}=\twoj\right\}\,.
\end{equation}
We denote a particle configuration on the graph as $\bm{n}\in \Omega$, where $\bm{n}=(n_{A}^{x})_{x\in V,\,A\in\{1,\ldots,N\}}$.
The infinitesimal generator of the process reads
\begin{equation}\label{Generator}
    \mathcal{L}=\sum_{(x,y)\in \mathcal{E}}\omega_{x,y}\mathcal{L}_{x,y}+\sum_{x\in V}\Gamma_{x}\mathcal{L}_{x}
\end{equation}
where  $ \omega_{x,y}\geq 0$ are so-called conductances and $\Gamma_{x}\geq 0$ are the couplings to reservoirs. 
The generator $\mathcal{L}_{x,y}$ is called the \textit{edge generator}, while $\mathcal{L}_{x}$ is called the \textit{site generator}. These linear operators act on functions $f:\Omega\to \mathbb{R}$ as follows
\begin{equation}\label{edgeGenerator}
\mathcal{L}_{x,y}f(\bm{n})=\sum_{A,B=1}^{N}n_{A}^{x}n_{B}^{y}\left[f(\bm{n}-\bm{\delta}^{x}_{A}+\bm{\delta}_{B}^{x}+\bm{\delta}_{A}^{y}-\bm{\delta}_{B}^{y})-f(\bm{n})\right]\co
\end{equation}
\begin{equation}\label{siteGenerator}
    \mathcal{L}_{x}f(\bm{n})=\sum_{A,B=1}^{N}\alpha_{A}^{x}n_{B}^{x}\left[f(\bm{n}+\bm{\delta}_{A}^{x}-\bm{\delta}_{B}^{x})-f(\bm{n})\right]\co
\end{equation}
where 
\begin{equation}
(\bm{\delta}_{A}^{x})^{y}_{B}=\begin{cases}
1\qquad &\text{if}\quad y=x,\;B=A\;,\\
0\qquad &\text{otherwise}\dt
\end{cases}
\end{equation}
Thus the dynamics consists in an exchange of particles between connected vertices at Poissonian times. 
This means that on the edge $(x,y)\in \mathcal{E}$  a particle {or a hole} indicated by $A$ at site $x$ is exchanged with a particle {or a hole} indicated by $B$ at site $y$ at rate $\omega_{x,y}n_{A}^{x}n_{B}^{y}$.
Moreover, each vertex  exchanges particles {or holes} with the external environment (reservoirs). 
Namely, on each site $x\in V$  a particle {or a hole} indicated by $B$ is replaced with a particle {or a hole} indicated $A$ at rate $\Gamma_{x}\alpha_{A}^{x}n_{B}^{x}$. {The transitions where the a particle is replaced by a hole/a hole is replaced by a particle is interpreted as a removal/injection of particles.}
\subsection{Reversible measures in the equilibrium set-up}
For a particular choice of the reservoir parameters one has an $N$-parameter family of reversible measures. More precisely
when the parameters are the same on each site, i.e.
\begin{equation}\label{reversibilityCondition}
\alpha_{A}^{x}=\alpha_{A}\qquad \forall x\in V\,,
\end{equation}
then the process described by the generator \eqref{Generator} is reversible with respect to the 
homogeneous product measure 
\begin{equation}
\label{reversibleMeasure}
\mu_{\text{rev}}=\bigotimes_{x\in V}\mu_{\text{rev}}^{x}
\end{equation}
with marginals $\mu_{\text{rev}}^{x}$ given by  the multinomial distribution
\begin{equation}
 \mu^{x}_{\text{rev}}\sim \text{Multinomial}\left(\twoj,\rho_{1},\ldots,\rho_{N}\right)\,.
\end{equation}
Here
 $$
\rho_{A}=\frac{\alpha_{A}}{|\alpha|}\co
$$
is the density of species $a$ and we used the notation $|\alpha|=\sum_{A=1}^{N}\alpha_{A}$. Explicitly, 
\begin{equation}
\mu_{\text{rev}}^{x}(n^{x})=\frac{\nu!}{\prod_{A=1}^{N}n_{A}^{x}!}\prod_{A=1}^{N}\rho_{A}^{n_{A}^{x}}\,.
\end{equation}
This can be proved  by checking that detailed balance is satisfied. 
If condition \eqref{reversibilityCondition} is not met, then in general reversibility is lost: indeed, in this situation, each reservoir at site $x$ has  its own set of densities vector $\rho^{x}=(\rho_{1}^{x},\ldots,\rho_{N}^{x})$ with components
\begin{equation}
	\label{rhox}
	\rho_{A}^x=\frac{\alpha_{A}^x}{|\alpha^x|}
\end{equation}
and  $|\alpha^x|=\sum_{A=1}^{N}\alpha_{A}^x$. {As a consequence, particles are injected and removed in the graph with different rates. When, for any $A\in \{1,\ldots,N\}$, there are at least two reservoirs at two different sites $x,y$ such that $\rho_{A}^{x}\neq  \rho_{A}^{y}$, a  non-zero current sets in, breaking the reversibility. Informally, one can say that the reservoirs try to impose their different densities at the boundaries of the the graph, putting the system out-of-equilibrium.} 
\begin{remark}
\label{rem-rev}
		{One can check that conditions \eqref{reversibilityCondition} are not the most general that implies reversibility. Indeed, by choosing the reservoir parameters as }
		\begin{equation}
		\label{cond-gen}
			\alpha_{A}^{x}\left(|\alpha^{y}|-\alpha_{A}^{y}\right)=\alpha_{A}^{y}\left(|\alpha^{x}|-\alpha_{A}^{x}\right)\qquad\qquad \forall x,y\in V,\quad \forall A\in \{1,\ldots,N\}
		\end{equation}
		where
		\begin{equation}
			|\alpha^{x}|=\sum_{A=1}^{N}\alpha_{A}^{x}\dt
		\end{equation}
		reversibility holds. 
		 In the case where the graph is a chain with nearest neighbors interactions and two reservoirs are attached to the two boundary sites
		then condition \eqref{cond-gen} is the multispecies counterpart of $\alpha\delta=\beta\gamma$ of the boundary driven partially excluded process introduced in \cite{schutzSandow} (see Section 3.2 of \cite{carinci2013duality}).
		\end{remark}
\subsection{Lie algebraic description of the process}

In this paper we will often use the fact that the Markov generator of the multispecies stirring process can be described in terms a Lie algebra ${gl}(N)$.
In this section we provide details about this.

Consider the Lie algebra ${gl}(N)$ with generators denoted by $\EE_{AB}$ with $A,B\in \{1,\ldots,N\}$ and commutation relations
\begin{equation}\label{eq:comgl}
\left[\EE_{AB},\EE_{CD}\right]=\EE_{AD}\delta_{CB}-\EE_{CB}\delta_{AD}\qquad \forall A,B\in \{1,\ldots,N\}\,.
\end{equation}
The finite-dimensional representations are labelled by partitions $\lambda=(\lambda_1,\lambda_2,\ldots,\lambda_N)$ of $\nu$ with  $\lambda_A\geq \lambda_{A+1}$,  $\lambda_A\in \mathbb{N}$ and $\sum_{A=1}^N \lambda_A = \nu\in\mathbb{N}$.  
We are interested in the {\em symmetric} finite-dimensional representations with 
\begin{equation}\label{eq:dynkin}
    \lambda=(\twoj,0,\ldots,0) \;.
\end{equation} 
The dimension $M_\twoj$ of this symmetric representations is given by the combination of $N$ objects in $\twoj$ positions with repetition, namely
\begin{equation}
	M_\twoj= \frac{(N+\twoj-1)!}{\twoj  !(N-1)!}\dt
\end{equation} 
The generators of the symmetric representations will be denoted by $E_{AB}$.
A basis of the vector space $\mathbb{C}^{M_\twoj}$ are the column vectors denoted by
\begin{equation}
  |n\rangle=  |n_{1},\ldots,n_{N}\rangle,\quad \text{with}\quad n_{A}\in\mathbb{N}_{0}\quad \text{such that}\quad \sum_{A=1}^{N}n_{A}=\nu\,.
\end{equation}
The basis vectors satisfy the orthogonality relation
\begin{equation}\label{ortho}
   \langle m|n \rangle =\langle m_{1},\ldots,m_{N}|n_{1},\ldots,n_{N}\rangle=\prod_{A=1}^{N}\delta_{m_{A},n_{A}}\co
\end{equation}
where  $ \langle m_{1},\ldots,m_{N}|$ is the row vector obtained by transposing $|m\rangle=|m_{1},\ldots,m_{N}\rangle$ and $\delta_{m_{A},n_{A}}$ is the Kronecker delta. 

The explicit action of the algebra generators on the basis  vectors $|n\rangle$ is the following:
\begin{equation}\label{actionE}
	\begin{cases}
		E_{AB}|n_{1},\ldots,n_{A},\ldots,n_{B},\ldots,n_{N}\rangle =n_{B}|n_{1},\ldots,n_{A}+1,\ldots,n_{B}-1,\ldots,n_{N}\rangle\qquad\qquad A\neq B\\[0.1cm]
		E_{AA}|n_{1},\ldots,n_{A},\ldots,n_{N}\rangle = n_{A} |n_{1},\ldots,n_{A},\ldots,n_{N}\rangle\dt
	\end{cases}
\end{equation}  
The matrices defined in this way satisfy the commutation relations \eqref{eq:comgl} and yield highest Dynkin weight \eqref{eq:dynkin}. \\
\\ \\
As mentioned above, the process with generator \eqref{Generator} can be described in terms of ${gl}(N)$ Lie algebra generators. The state space \eqref{stateSpace} is given by the $|V|$-fold tensor product of the vector space with basis elements $|n^x\rangle$ at a given site. Namely,
a vector $|{\bm{n}}\rangle \in \Omega$ can be written as

\begin{equation}
|{\bm{n}}\rangle=\left(\,\bigotimes_{x\in V}	|n_{1}^{x},\ldots,n_{N}^{x}\rangle\right)
\end{equation}
with $\sum_{A=1}^{N}n_{A}^{x}=\nu$ for any $x\in V$. For a fixed $x\in V$ we write $|n^{x}\rangle=|n_{1}^{x},\ldots,n_{N}^{x}\rangle$. The following orthogonality relation is a consequence of the single site relation \eqref{ortho}
\begin{equation}
    \langle {\bm{n}}|{\bm{m}}\rangle =\prod_{x\in V}\prod_{A=1}^N\delta_{n^x_{A},m^{x}_{A}}\,.
\end{equation}
We introduce the Hamiltonian operator
\begin{equation}\label{OriginalHamiltonian}
	\begin{split}
		H=\sum_{(x,y)\in \mathcal{E}}\omega_{x,y}\mathcal{H}_{x,y}+\sum_{x\in V}\Gamma_{x}H_{x}
	\end{split}
\end{equation}
where the edge Hamiltonian $\mathcal{H}_{x,y}$ that describes the interaction between two connected sites is
\begin{equation}\label{edgeHamiltonian}
\mathcal{H}_{x,y}=\sum_{A,B=1}^{N}\Big(E_{AB}^{x} E_{BA}^{y}-E_{BB}^{x} E_{AA}^{y}\Big)
 \end{equation}
  and where the site Hamiltonian $H_{x}$  is
 \begin{equation}\label{siteHamiltonian}
H_{x}=\sum_{A,B=1}^{N}\alpha_{A}^{x}\left(E_{AB}^{x}-E_{BB}^{x}\right)
\end{equation}
Here $E_{AB}^{x}$ denotes the generator $E_{AB}$ in \eqref{actionE} acting non-trivially on site $x$ (and as the identity on the other sites). 
The Hamiltonian in \eqref{OriginalHamiltonian} is stochastic and is linked to the Markov generator  by
\begin{equation}\label{Hamiltonian-Generator}
H=\mathcal{L}^{T} 
\end{equation}
where $T$ denotes transposition, see e.g.  \cite{belitsky2015self}.
The action of the generator on functions $f$ can then  be expressed as 
\begin{equation}
    \mathcal{L}f( {\bm{n}})=\langle f|H| {\bm{n}}\rangle
\end{equation}
where 
\begin{equation}
    \langle f|=\sum_{ {{\bm{m}}\in \Omega}}f( {\bm{m}})\langle  {\bm{m}}|\dt
\end{equation}
We can write the edge Hamiltonian \eqref{edgeHamiltonian} as a function of the coproduct of the quadratic Casimir of ${gl}(N)$
\begin{equation}\label{secondCasimir}
    C=\sum_{A,B=1}^{N}E_{AB}E_{BA}\co
\end{equation}
that acts diagonally as $C|{n}\rangle=\twoj(\twoj+N)|{n}\rangle$ on any state $|n\rangle$ and belogs to the center of ${gl}(N)$ (i.e. it commutes with all the algebra elements).  
More precisely,  considering the standard coproduct 
\begin{equation}
\begin{split}
\Delta:{gl}(N)\to {gl}(N)\otimes {gl}(N)
\end{split}
\end{equation}
with
\begin{equation}
E_{AB}\mapsto E_{AB}\otimes \mathbbm{1}+\mathbbm{1}\otimes E_{AB}\co
\end{equation} 
we have 
\begin{equation}
\Delta(C)=\sum_{A,B=1}^{N}\Delta(E_{AB})\Delta(E_{BA})=2\sum_{A,B=1}^{N}E_{AB}\otimes E_{BA}+C\otimes \mathbbm{1}+\mathbbm{1}\otimes C\,.
\end{equation}
Then, one can check that 
\begin{equation}\label{hamiltonianCasimir}
	\mathcal{H}_{x,y}=\frac{1}{2}\Delta_{x,y}(C)-\twoj(2\twoj+N)
\end{equation}
where $\Delta_{x,y}(C)$ denotes  $\Delta(C)$ acting on  the sites of edge $(x,y)\in \mathcal{E}$ and acting trivially on the other sites of the graph.

\subsection{Integrable process on a line segment}\label{subsection-description-process-LINE}
In this section we specialize the multispecies stirring process to the geometry of the one-dimensional chain with sites $\{1,\ldots,L\}$ where two reservoirs are attached to the boundary sites $1$ and $L$.  The reservoirs, exchanging particles with the external environment, put the chain out of equilibrium.  The  Hamiltonian that we consider here is obtained from \eqref{OriginalHamiltonian} assuming that the conductances are
\begin{equation}
	\omega_{x,y}=\begin{cases}
		1 \quad \text{if}\quad |x-y|=1\\
		0\quad \text{otherwise}
	\end{cases}
\end{equation}
and the coupling to reservoirs are
\begin{equation}
	\Gamma_{x}=\begin{cases}
		1\quad \text{if} \quad x\in \{1,L\}\\
		0\quad \text{otherwise}
	\end{cases}\dt
\end{equation}
The case $\nu=1$, i.e. one particle at most for each site, is integrable and has been considered previously in \cite{vanicat2017exact}. 
For  $\nu=1$ we denote the ${gl}(N)$  generators   as  $(e_{AB})_{A,B\in\{1,\ldots,N\}}$ obeying $(e_{AB})_{CD}=\delta_{AC}\delta_{BD}$.
The Hamiltonian can then be written as
\begin{equation}\label{hamiltonian}
	H=H_{\text{left}}+H_{\text{bulk}}+H_{\text{right}}
\end{equation}
where
\begin{equation}
	H_{\text{bulk}}=\sum_{x=1}^{L-1}\mathcal{H}_{x,x+1}\dt
\end{equation}
Here $\mathcal{H}_{x,x+1}$  denotes  the two-site Hamiltonian 
\begin{equation}\label{H-corsivo}
	\begin{split}
		\mathcal{H}=P-\id
	\end{split}
\end{equation}
with the permutation matrix
\begin{equation}
	P=\sum_{A,B=1}^Ne_{AB}\otimes e_{BA}\,,
\end{equation} 
acting non-trivially on the vertices of the edge $x,x+1$.
 In this context it will be useful to introduce the following notation for the occupation variables of the process. Each configuration is denoted by
\begin{equation}\label{Tau-Notation}
	|\bm{\tau}\rangle=|\tau_{1},\ldots,\tau_{L}\rangle
\end{equation} 
with $\tau_{x}\in \{1,\ldots,N\}$, $\forall x\in \{1,\ldots,L\}$. Since the {maximal occupancy at each site is $\nu=1$},
the configuration $\bm{n}$ introduced in Section~\ref{subsectionGeneratorStr} and $\bm{\tau}$ are related by 
\begin{equation}\label{notation-change-relation}
	n_{A}^{x}=\delta_{\tau_{x},A}\dt
\end{equation}
The state space of the process is now 
\begin{equation}
	\Omega^{'}=\left\{(\tau_{1},\ldots,\tau_{L})\;:\; \tau_{x}\in\{1,\ldots,N\}\right\}\dt
\end{equation}
The action of the Hamiltonian density $\mathcal{H}$ on the tensor product of the configuration of two sites follows immediately from \eqref{H-corsivo} and reads
\begin{equation}
	\begin{split}
		\mathcal{H}| \tau\rangle\otimes   |\tau'\rangle&=|\tau'\rangle \otimes |\tau\rangle-|\tau\rangle \otimes|\tau'\rangle\dt
	\end{split}
\end{equation}
The boundary terms of the Hamiltonian \eqref{hamiltonian}  are given by 
\begin{equation}
	H_{\text{left}}=\begin{pmatrix}
		\alpha_{1}-1&\alpha_{1}&\alpha_{1}&\ldots&\ldots&\alpha_{1}\\
		\alpha_{2}&\alpha_{2}-1&\alpha_{2}&\ldots&\ldots&\alpha_{2}\\
		\vdots&\vdots& &\ddots& &\vdots\\
		\alpha_{N-1}&\alpha_{N-1}&\ldots&\ldots&\alpha_{N-1}-1&\alpha_{N-1}\\
		\alpha_{N}&\alpha_{N}&\ldots&\ldots&\alpha_{N}&\alpha_{N}-1
	\end{pmatrix}
\end{equation}
and 
\begin{equation}
	H_{\text{right}}=\begin{pmatrix}
		\beta_{1}-1&\beta_{1}&\beta_{1}&\ldots&\ldots&\beta_{1}\\
		\beta_{2}&\beta_{2}-1&\beta_{2}&\ldots&\ldots&\beta_{2}\\
		\vdots&\vdots& &\ddots& &\vdots\\
		\beta_{N-1}&\beta_{N-1}&\ldots&\ldots&\beta_{N-1}-1&\beta_{N-1}\\
		\beta_{N}&\beta_{N}&\ldots&\ldots&\beta_{N}&\beta_{N}-1
	\end{pmatrix}
\end{equation}
where, without loss of generality,  we assume that the parameters satisfy
\begin{equation}\label{ratesConditions}
	\sum_{A=1}^{N}\alpha_{A}=1,\qquad\sum_{A=1}^{N}\beta_{A}=1\;.
\end{equation} 
Under these assumption the Hamiltonian~\ref{hamiltonian} is integrable. This has been schown in \cite{vanicat2017exact} within  the Quantum Inverse Scattering Method.  We will consider this process more extensively in Section~\ref{sectionIntegrabiliy}.

\section{Duality}\label{sectionDuality}
In this section we prove the first main result of this paper. We show that the multispecies stirring process with open boundaries
can be studied by means of a dual process. After recalling the definition of duality between two
Markov processes in Section~\ref{def-duality}, we formulate our duality result  in Theorem~\ref{thm-duality}. A proof of the theorem is given in Section~\ref{proof-th-duality}.
The duality result yields {to a dual process that} 
 has absorbing sites. Therefore, as shown in Corollary~\ref{Corolollary-ABS_Corr-abstract}, the study of correlations in the non-equilibrium steady state of the multispecies stirring process with open boundaries
is reduced to the study of the absorption probabilities of dual particles.

\subsection{Definition}
\label{def-duality}
Consider two Markov processes, $(\eta_{t})_{t\geq 0}$ defined on a state space $\Omega$ and $(\xi_{t})_{t\geq 0}$ defined on a state space $\widetilde{\Omega}$. We say that they are dual, with respect to a duality function $D:\Omega\times \widetilde{\Omega}\to \mathbb{R}$, if $\forall \eta\in\Omega$, $\forall \xi\in\widetilde{\Omega}$ and $\forall t> 0$ we have 
\begin{equation}\label{duality-expectation}
    \mathbb{E}_{\eta}\left[D(\eta_{t},\xi)\right]=\mathbb{E}_{\xi}\left[D(\eta,\xi_{t})\right]
\end{equation}
where $\mathbb{E}_{\eta}$ denotes the expectation with respect to the law of the Markov process $(\eta_{t})_{t\geq 0}$ initialized with the particle configuration $\eta$, whereas $\mathbb{E}_{\xi}$ denotes the expectation with respect to the law of the Markov process $(\xi_{t})_{t\geq 0}$ initialized with the particle configuration $\xi$.
The duality definition can also be formulated as a relation between the generators, provided some minimal technical conditions are satisfied \cite{jansen2014notion}. Call $\mathcal{L}$ the generator of $(\eta_{t})_{t\geq0}$ and $\widetilde{\mathcal{L}}$ the generator of $(\xi_{t})_{t\geq 0}$, then we say that these two processes are dual with respect to the duality function $D:\Omega\times \widetilde{\Omega}\to \mathbb{R}$ if $\forall \eta\in\Omega$ and $\forall \xi\in\widetilde{\Omega}$
\begin{equation}\label{dualityRelationGenerator}
    \left(\mathcal{L}D(\cdot,\xi)\right)(\eta)=\left(\widetilde{\mathcal{L}}D(\eta,\cdot)\right)(\xi)\dt
\end{equation}
In the specific case where $\mathcal{L}=\widetilde{\mathcal{L}}$ we say that the process is self-dual.
\begin{remark}
When the state spaces of the dual processes is countable, the generators of the processes and the duality function can be represented as matrices with elements $\mathcal{L}(\eta,\eta')$, $\widetilde{\mathcal{L}}(\xi,\xi')$ and $D(\eta,\xi)$ for arbitrary $\eta,\eta'\in\Omega$ and $\xi,\xi'\in \widetilde{\Omega}$. Therefore, we can write the duality relation \eqref{dualityRelationGenerator} as 
\begin{equation}
    \sum_{\eta'\in\,\Omega}\mathcal{L}(\eta,\eta')D(\eta',\xi)=\sum_{\xi'\in\, \widetilde{\Omega}}\widetilde{\mathcal{L}}(\xi,\xi')D(\eta,\xi')\co
\end{equation}
that can be read as
\begin{equation}\label{dualityIntertwines}
    \mathcal{L}D=D\widetilde{\mathcal{L}}^{\,T}
\end{equation}
where the superscript $T$ denotes the matrix transposition. Therefore, the duality relation \eqref{dualityIntertwines}  intertwines between two linear operators $\mathcal{L}$ and $\widetilde{\mathcal{L}}^{\,T}$. Working in terms of the Hamiltonian operators the duality relation reads 
\begin{equation}\label{DualityRelation}
    H^{T}D=D\widetilde{H}\dt
\end{equation}
\end{remark}
\subsection{Duality for the open multispecies stirring process}\label{statementDualitySubsection}
In this section we formulate duality for the multispecies stirring process $(\bm{n}(t))_{t\geq 0}$ with open boundaries, defined by the generator \eqref{Generator}.
The dual  process $(\bm{\xi}(t))_{t\geq 0}$ is defined on the \textit{enlarged graph} $\widetilde{G}=(\widetilde{V},\widetilde{\mathcal{E}})$ where 
\begin{equation}
	\widetilde{V}:=V\cup \left\{u(x)\,:\, x\in V\right\}\qquad \widetilde{\mathcal{E}}:=\mathcal{E}\cup \left\{(x,u(x))\,:\, x\in V\right\}\dt
\end{equation}
This means that to each site $x\in V$ we associate an ``extra-site'' via a bijection $u:V\to V$. We denote as $u(x)$ the extra-site associated to $x\in V$ . The configuration space of the dual process is the enlarged state space
\begin{equation}\label{dualStateSpace}
    \widetilde{\Omega}= \bigotimes_{x\in V} \widetilde{\Omega}_{x}\ = \bigotimes_{x\in V} (\Omega_{x}\times \mathbb{N}_{0}^{N-1})\dt
\end{equation}
Note that on the extra-site we allow an unbounded number of particles. Thus
dual particles will accumulate in these extra-sites in the course of time. 
We write the configurations $\bm{\xi} \in \widetilde\Omega$  as
\begin{equation}\label{row-dual-vectors}
    \bm{\xi}=\bigotimes_{x\in V} \left(\left(\xi_{1}^{x},\ldots,\xi_{N}^{x}\right)\otimes (\xi_{1}^{u(x)},\ldots,\xi_{N-1}^{u(x)} )\right) 
\end{equation} 
where the component $\xi_{a}^{x}$ is interpreted as the number of dual particles of type $a\in \{1,\ldots,N-1\}$ at site $x$, {while $\xi_{N}^{x}$ is interpreted as the number of holes at site $x$ of the dual process.}
The component $\xi_{a}^{u(x)}$  gives the number of dual particles of type $a\in \{1,\ldots,N-1\}$ at 
the extra-site $u(x)$ connected to $x\in V$. {We observe that the configuration variable at the extra site $u(x)$ does not have to satisfy any exclusion constraint, i.e. an unbounded number of particles can be hosted. Therefore, an infinite number of holes is available at each extra-site $u(x)$. As a consequence, the hole occupation variable $\xi_{N}^{u(x)}$ is not considered}. We state the following duality result.
\begin{theorem}[Absorbing duality]\label{thm-duality}
	The multispecies stirring process $(\bm{n}(t))_{t\geq 0}$ defined on the state space $\Omega$ with generator $\mathcal{L}$ defined in \eqref{Generator} is dual to the process $(\bm{\xi}(t))_{t\geq 0}$ defined on the enlarged state space $\widetilde{\Omega}$ with generator
	 \begin{equation}\label{DualGenerator}
		\widetilde{\mathcal{L}}=\sum_{(x,y)\in \mathcal{E}}\omega_{x,y}\mathcal{L}_{x,y}+\sum_{x\in V}\Gamma_{x}\widetilde{\mathcal{L}}_{x}
	\end{equation}
where 
$\mathcal{L}_{x,y}$ is defined in \eqref{edgeGenerator} and, for any function $f:\widetilde{\Omega}\to \mathbb{R}$ 
\begin{equation}\label{siteDualGenerator}
	\widetilde{\mathcal{L}}_{x}f(\bm{\xi})=|\alpha^{x}|\sum_{a=1}^{N-1}\xi_{a}^{x}\left(f(\bm{\xi}-\bm{\delta}_{a}^{x}+\bm{\delta}_{N}^{x}+\bm{\delta}_{a}^{u(x)})-f(\bm{\xi})\right)\dt
\end{equation}
The duality function is given by 
\begin{equation}\label{dualityElements}
	D(\bm{n},\bm{\xi})=\prod_{x\in V}\left(\frac{(\nu -\sum_{a=1}^{N-1}\xi_{a}^{x})!}{\nu!}\prod_{a=1}^{N-1}\frac{n_{a}^{x}!}{(n_{a}^{x}-\xi_{a}^{x})!}\left(\rho_{a}^{x}\right)^{\xi_{a}^{u(x)}}\,\right)
\end{equation}
where we recall the definition of the reservoir densities (cf. \eqref{rhox})
\begin{equation}
	\rho_{a}^{x}=\frac{\alpha_{a}^{x}}{|\alpha^{x}|}\dt
\end{equation}
\end{theorem}
The dual dynamics is described as follows. On one hand, the edge part $\mathcal{L}_{x,y}$ of the dual Markov generator gives rise to  the multispecies stirring dynamics on the graph. On the other hand, the site
part $\widetilde{\mathcal{L}}_{x}$ of the dual generator replaces a particle of any type $a\in\{1,\ldots,N-1\}$ at site $x$ with a particle of type $N$ and creates a particle of the same type $a$ at the extra-site $u(x)$. This last transition is performed with rate $|\alpha^{x}|\xi_{a}^{x}$. This means that eventually the dual process voids the graph, putting all the dual particles of species $\{1,\ldots,N-1\}$ in the extra-sites. In other words the extra-sites play the role of absorbing boundaries. 
\begin{remark} In the reversible situation, i.e. when $\forall x\in V$ we have $\rho_{a}^{x}=\rho_{a}$, the expectation  of the duality function  $D(\bm{n},\bm{\xi})$ with   $\bm{n}$ distributed as  $\mu_{\text{rev}} = \bigotimes_{x\in V}\text{Multinomial}\left(\twoj, \rho_{1},\ldots,\rho_{N}\right)$ is
\begin{equation}
\mathbb{E}_{\mu_{\text{rev}}}\left[D(\bm{n},\bm{\xi})\right]=\prod_{a=1}^{N-1}\left(\rho_{a}\right)^{\sum_{x\in V}\xi_{a}^{x}+\sum_{x\in V}\xi_{a}^{u(x)}}\qquad \forall \bm{n}\in \Omega,\quad\forall \bm{\xi}\in \widetilde{\Omega}\dt
\end{equation}
\end{remark}
\subsection{Proof of Theorem~\ref{thm-duality}}
\label{proof-th-duality}
To prove duality between the process $(\bm{n}(t))_{t\geq 0}$ and the process  $(\bm{\xi}(t))_{t\geq 0}$ we  show that \eqref{DualityRelation} is fulfilled.  To show this, we will use the Hamiltonians and their Lie algebraic description. Indeed, in this formalism the proof
reduces to finding symmetries of the generator (for the bulk duality) and group like transformations (for the ``boundary duality'').
To a configuration $\bm{\xi}$ of the configuration space  \eqref{dualStateSpace} of a dual process we associate the vector
\begin{equation}
    |\bm{\xi}\rangle=\bigotimes_{x\in V}\left(|\xi_{1}^{x},\ldots,\xi_{N}^{x}\rangle\otimes |\xi_{1}^{u(x)},\ldots,\xi_{N-1}^{u(x)}\rangle\right)\dt
\end{equation}
{Here $|\xi_{1}^{u(x)},\ldots,\xi_{N-1}^{u(x)}\rangle$ is the vector associated to the configuration at the extra site $u(x)$.  For $q_{1},\ldots,q_{N-1}\in \mathbb{N}_{0}$, we assume that it satisfies an orthogonality relation
\begin{equation}\label{ortho-extraSite}
	\langle q_{1}^{u(x)},\ldots,q^{u(x)}_{N-1}|\xi_{1}^{u(x)},\ldots,\xi_{N-1}^{u(x)}\rangle=\prod_{a=1}^{N-1}\delta_{\xi_{a}^{u(x)},q_{a}}\,.
\end{equation} }
{\begin{remark}\label{reamrk-extraSite}
	In the following, we denote the configuration vectors on the extra-site $u(x)$ by $|\xi_{1}^{u(x)},\ldots,\xi_{N-1}^{u(x)}\rangle$. This allows to stress that these vectors belong to the extra-space $N_{0}^{N-1}$ "attached" to site $x$. Moreover, this notation allows to directly connect the ket-vector $|\xi_{1}^{u(x)},\ldots,\xi_{N-1}^{u(x)}\rangle$ with the vector $(\xi_{1}^{u(x)},\ldots,\xi_{N-1}^{u(x)})$ defined in \eqref{row-dual-vectors}, in which we recall that the components $\xi_{a}^{u(x)}$ denote the number of dual particles of type $a$ at site $u(x)$.
\end{remark}} 
The Hamiltonian of the dual process reads
\begin{equation}\label{DualHamiltonian}
    \widetilde{H}=\sum_{x,y\in \mathcal{E}}\omega_{x,y}\mathcal{H}_{x,y}+\sum_{x\in V}\Gamma_{x}\widetilde{H}_{x}
\end{equation}
where $\mathcal{H}_{x,y}$ is the one defined in \eqref{edgeHamiltonian}, while 
\begin{equation}\label{siteDualHamiltonian}
    \widetilde{H}_{x}=|\alpha^{x}|\sum_{a=1}^{N-1}\left((\mathbf{a}^{+})_{a}^{u(x)}\,E_{Na}^{x}-E_{aa}^{x}\right)\dt
\end{equation}
Here we introduced the pair of bosonic operators $\mathbf{a},\,\mathbf{a}^{+}$ satisfying $[\mathbf{a},\mathbf{a}^{+}]=1$ and acting as
\begin{equation}
	\mathbf{a}^{+}|q\rangle=|q+1\rangle\qquad \mathbf{a}|q\rangle=q|q-1\rangle
\end{equation}
and 
\begin{equation}\label{bosonic_2}
	\langle q|\mathbf{a}^{+}=\langle q-1|\qquad \langle q|\mathbf{a}=(q+1) \langle q+1|
\end{equation}
on a generic vector $\langle q|$ with $q\in \mathbb{N}_{0}$, so that in \eqref{siteDualHamiltonian} 
$(\mathbf{a}^{+})_{a}^{u(x)}$ denotes  $\mathbf{a}^{+}$ acting on the extra-site $u(x)$ and on the species $a\in\{1,\ldots,N-1\}$. \\

We will show below that the Hamiltonians \eqref{OriginalHamiltonian} and \eqref{DualHamiltonian} are dual in the sense of \eqref{DualityRelation}. From an algebraic point of view, the duality matrix $D$ \eqref{dualityElements} is described as 
\begin{equation}\label{dualityMatrix}
    D=\prod_{x\in V}d_{x}\otimes \dd \;.
\end{equation}
Here
\begin{equation}\label{bulkElementDualityMatrix}
d_{x}=R_{x}\exp{(E^{x})}
\end{equation}
with  the diagonal part
\begin{equation}\label{Revmatrix}
    R_{x} =\sum_{n^{x}\in\Omega_{x}}\frac{\prod_{A=1}^{N}n_{A}^{x}!}{\nu!}|n_{1}^{x},\ldots,n_{N}^{x}\rangle\langle n_{1}^{x},\ldots,n_{N}^{x}|
\end{equation}
and
\begin{equation}\label{EquationEx}
E^{x}=\sum_{a=1}^{N-1}E_{aN}^{x}\,.
\end{equation}
Furthermore
\begin{equation}\label{dualityMatrix2}
\dd=\sum_{\xi_{1}^{u(x)},\ldots,\xi_{N-1}^{u(x)}=0}^{\infty}\prod_{a=1}^{N-1}\left(\rho_{a}^{x}\right)^{\xi_{a}^{u(x)}}\langle \xi_{1}^{u(x)},\ldots,\xi_{N-1}^{u(x)}|\dt
\end{equation}
We observe that the matrix $R_{x}$ is diagonal. Its elements are related to the inverse of the weights of the reversible measure \eqref{reversibleMeasure}. In particular, to obtain these elements, we have considered the weights of \eqref{reversibleMeasure} when all the parameters $\rho_{a}=\frac{1}{N}$. Then, the constant $\left(\frac{1}{N}\right)^{\nu}$ has been neglected, since it does not change the duality relation. This $R_{x}$   is associated to reversibility, see \cite{giardina2009duality}. \\

Since \eqref{dualityMatrix} is product over sites, proving \eqref{DualityRelation} is equivalent to showing that 
\begin{equation}\label{edgeDualRealtion}
    \mathcal{H}_{x,y}^{T}D=D\mathcal{H}_{x,y}\qquad \forall (x,y)\in \mathcal{E}
\end{equation}
and 
\begin{equation}\label{siteDualRelation}
    H_{x}^{T}D=D\widetilde{H}_{x}\qquad \forall x\in V.
\end{equation}
We perform the proof of duality in three steps: first we will show that matrix \eqref{dualityMatrix} has elements \eqref{dualityElements}; second we will prove the bulk duality \eqref{edgeDualRealtion}; finally we will show the boundary duality \eqref{siteDualRelation}. 
As a preliminary result we note that 
\begin{equation}\label{transpositionPropertyR}
(E_{BA}^{x})^{T}=R_{x}E_{AB}^{x}R_{x}^{-1}\qquad \forall x\in V\co
\end{equation} 
that follows immediately from the definition of $E_{AB}$. 
\paragraph{Elements of the duality matrix.}
Consider the matrix $D$ defined in \eqref{dualityMatrix}.
We aim to show that its matrix elements coincide with duality function \eqref{dualityElements}, i.e.
\begin{equation}\label{proofDualityElements}
\langle \bm{n}|D|\bm{\xi}\rangle=D(\bm{n},\bm{\xi})\qquad   \forall \bm{n}\in \Omega,\quad \bm{\xi}\in \widetilde{\Omega}\,.
\end{equation}
Fix an arbitrary site $x\in V$, then we have that 
\begin{align}
	 &\langle n^{x}_{1},\ldots,n_{N}^{x}|\left(d_{x}\otimes \dd\right)\left(|\xi_{1}^{x},\ldots,\xi_{N}^{x}\rangle\otimes|\xi_{1}^{u(x)},\ldots,\xi_{N-1}^{u(x)}\rangle\right)\nonumber\\=&\langle n_{1}^{x},\ldots,n_{N}^{x}| (\exp{(E_{N1}^{x}+\ldots+E_{N\,N-1}^{x}}))^{T}R_{x}\otimes\sum_{q_{1}^{u(x)},\ldots,q_{N-1}^{u(x)}=0}^{\infty}\prod_{a=1}^{N-1}\left(\rho_{a}^{x}\right)^{q_{a}^{u(x)}}\langle q_{1}^{u(x)},\ldots,q_{N}^{u(x)}|\nonumber
	 \\&|\xi_{1}^{x},\ldots,\xi_{N}^{x}\rangle \otimes |\xi_{1}^{u(x)},\ldots,\xi_{N-1}^{u(x)}\rangle
\end{align} 
where we used \eqref{transpositionPropertyR}. \\
On one hand, on the extra-site $u(x)$ we have 
\begin{align}
\sum_{q_{1}^{u(x)},\ldots,q_{N-1}^{u(x)}=0}^{\infty}\left(\prod_{a=1}^{N-1}\left(\rho_{a}^{x}\right)^{q_{a}^{u(x)}}\right)\langle q_{1}^{u(x)},\ldots,q_{N-1}^{u(x)}|\xi_{1}^{u(x)},\ldots,\xi_{N-1}^{u(x)}\rangle=\prod_{a=1}^{N-1}\left(\rho_{a}^{x}\right)^{\xi_{a}^{u(x)}}
\end{align}
{ 
where we used the orthogonality relation \eqref{ortho-extraSite}}. 
On the other hand, on the site $x$, we have 
\begin{align}
&\langle n_{1}^{x},\ldots,n_{N}^{x}|(\exp{(E_{N1}^{x}+\ldots+E_{N\,N-1}^{x})})^{T}R_{x}|\xi_{1}^{x},\ldots,\xi_{N}^{x}\rangle\nonumber\\&= \langle  n_{1}^{x},\ldots,n_{N}^{x}|\left(\sum_{k_{1}=0}^{\infty}\frac{\left\{\left(E_{N1}^{x}\right)^{T}\right\}^{k_{1}}}{k_{1}!}\ldots\sum_{k_{N-1}=0}^{\infty}\frac{\left\{\left(E_{N\,N-1}^{x}\right)^{T}\right\}^{k_{N-1}}}{k_{N-1}!}\sum_{s\in\Omega_{x}}\frac{s_{1}^{x}!\cdots s_{N}^{x}!}{\nu!}|s_{1}^{x},\ldots,s_{N}^{x}\rangle\langle s_{1}^{x},\ldots,s_{N}^{x}|\right)\nonumber
\\&
\hspace{2.cm}\qquad\times |\xi_{1}^{x},\ldots,\xi_{N}^{x}\rangle\nonumber
\\&=
\sum_{k_{1}=0}^{n_{1}^{x}}\ldots\sum_{k_{N-1}=0}^{n_{N-1}^{x}}\langle n_{1}^{x}-k_{1},\ldots,n_{N-1}^{x}-k_{N-1},\ldots,n_{N}^{x}+k_{1}+\ldots+k_{N-1}|\frac{n_{1}^{x}!\cdots n_{N-1}^{x}!}{(n_{1}^{x}-k_{1})!\cdots(n_{N-1}^{x}-k_{N-1})!}\nonumber
\\& 
\hspace{2.cm}\qquad\times \frac{1}{k_{1}!\cdots k_{N-1}!}\frac{\xi_{1}^{x}!\cdots \xi_{N}^{x}!}{\nu!}|\xi_{1}^{x},\ldots,\xi_{N}^{x}\rangle\nonumber
\\&=
\frac{(\nu-\sum_{a=1}^{N-1}\xi_{a}^{x})!}{\nu!}\prod_{a=1}^{N-1}\frac{n_{a}^{x}!}{(n_{a}^{x}-\xi_{a}^{x})!}
\end{align}
where we used the definitions of the action of $E_{AB}$  and the orthogonality relations \eqref{ortho}, together with the fact that $\xi_{N}^{x}=\nu-\sum_{a=1}^{N-1}\xi_{a}^{x}$. 
Finally, by taking the product over $x\in V$ \eqref{proofDualityElements} is proved.
\begin{flushright}
    $\square$
\end{flushright}

\paragraph{Proof of bulk duality \eqref{edgeDualRealtion}.}To show this relation we need two 'ingredients'. First the existence of a similarity transformation between the Hamiltonian $\mathcal{H}_{x,y}$  and its transposed. As we will show, this similarity transformation is $R_{x}R_{y}$.  Second, the possibility of finding a symmetry for the edge Hamiltonian. Exploiting \eqref{hamiltonianCasimir}, we can take any symmetry of the Casimir and apply the co-product. As we will see, it will be convenient to choose $\sum_{a=1}^{N-1}E_{aN}$. 

We first look for the similarity between $\mathcal{H}_{x,y}$ and its transposed. Using \eqref{transpositionPropertyR} we obtain  
\begin{equation}\label{transpositionPropertyH}
    \mathcal{H}_{x,y}^{T}=\left(R_{x}R_{y}\right)\mathcal{H}_{x,y}\left(R_{x}R_{y}\right)^{-1}
\end{equation}
therefore we have found the similarity transformation between $\mathcal{H}_{x,y}$ and its transposed. \\
We now look for a symmetry of $\mathcal{H}_{x,y}$, i.e. a matrix $S_{x,y}$ of the same dimension such that it satisfies 
\begin{equation}
	\mathcal{H}_{x,y}S_{x,y}=S_{x,y}\mathcal{H}_{x,y}
	\dt
\end{equation}
Using \eqref{hamiltonianCasimir}, we observe that $\mathcal{H}_{x,y}$ is proportional to the coproduct of the second Casimir, up to a diagonal term. Therefore, it is enough to look for a symmetry of $\Delta (C)$. Using the bilinearity of the coproduct operator and the fact that $C$ belongs to the centre of the Lie algebra, it is easy to show that for any linear combination of generators $E\in {gl}(N)$ the following holds:
\begin{equation}
 \text{if } \quad	[C,E]=0\qquad  \text{ then }\qquad \left[\Delta (C),\Delta(E)) \right]=0\dt
\end{equation}
Let
\begin{equation}
	E=\sum_{a=1}^{N-1}E_{aN}
\end{equation}
and, for a fixed $(x,y)\in \mathcal{E}$, we define
\begin{equation}
	S_{x,y}=\exp{(\Delta_{x,y}(E))}=\exp{(E^{x})}\exp{(E^{y})}
\end{equation}
where $\Delta_{x,y}(E)$ denotes the coproduct acting on sites $x$ and $y$. As a consequence, this operator $S_{x,y}$ satisfies
\begin{equation}\label{symmetryH}
	\left[S_{x,y},\mathcal{H}_{x,y}\right]=0
\end{equation}
i.e. it is a symmetry of $\mathcal{H}_{x,y}$. Exploiting these considerations we may write

\begin{equation}
    \begin{split}
        \mathcal{H}_{x,y}^{T}D&=(R_{x}R_{y})\mathcal{H}_{x,y}(R_{x}R_{y})^{-1}\left(d_{x}\otimes\dd\right)\left(d_{y}\otimes\mathcal{D}_{u(y)}\right)\prod_{z\in V\,:\, z\neq x,y}\left(d_{z}\otimes \mathcal{D}_{u(z)}\right)
        \\&=\left(R_{x}\exp{(E^{x})}\otimes \dd\right)\left(R_{y}\exp{(E^{y})}\otimes \mathcal{D}_{u(y)}\right)\mathcal{H}_{x,y}\prod_{z\in V\,:\, z\neq x,y}\left(d_{z}\otimes \mathcal{D}_{u(z)}\right)
        \\&=
        D\mathcal{H}_{x,y}
    \end{split}
\end{equation}
 where we used \eqref{transpositionPropertyH} and \eqref{symmetryH} in the second equality. Thus, \eqref{edgeDualRealtion} is proved. 
 \begin{flushright}
     $\square$
 \end{flushright}
 \paragraph{Proof of boundary duality \eqref{siteDualRelation}.} To prove \eqref{siteDualRelation} we transform via the Hadamard formula the transposed of the site Hamiltonian \eqref{siteDualHamiltonian} and then to  introduce properly a creation operator acting on an extra-site $u(x)$.
 Considering {$\mathcal{A},\mathcal{B}\in {gl}(N)$}, the Hadamard formula reads 
 \begin{equation}\label{HadamardFormula}
     \exp{(-\mathcal{B})}\mathcal{A}\exp{(\mathcal{B})}=\mathcal{A}-\left[\mathcal{B},\mathcal{A}\right]+\frac{1}{2!}\left[\mathcal{B},\left[\mathcal{B},\mathcal{A}\right]\right]-\frac{1}{3!}\left[\mathcal{B},\left[\mathcal{B},\left[\mathcal{B},\mathcal{A}\right]\right]\right]+\ldots
 \end{equation}
In the following we evaluate this formula for $\mathcal{B}=\sum_{a=1}^{N-1}E_{aN}$ defined in \eqref{EquationEx} and $\mathcal{A}=E_{CD}$  with $C,D\in\{1,\ldots,N\}$. We find  
\begin{align}
	\left[\mathcal{B},\mathcal{A}\right]=\sum_{a=1}^{N-1}\left[E_{aN},E_{CD}\right]=\delta_{CN}\sum_{a=1}^{N-1}E_{aD}-(1-\delta_{DN})E_{CN}
\end{align}
and 
\begin{align}
	\left[\mathcal{B}\left[\mathcal{B},\mathcal{A}\right]\right]=&\left[\sum_{a=1}^{N-1}E_{aN},\delta_{CN}\sum_{b=1}^{N-1}E_{bD}-(1-\delta_{DN})E_{CN}\right]\nonumber\\=&
	\sum_{a=1}^{N-1}\left(\delta_{CN}\sum_{b=1}^{N-1}\left[E_{aN},E_{bD}\right]-(1-\delta_{DN})\left[E_{aN},E_{CN}\right]\right)\nonumber\\=&
	\sum_{a=1}^{N-1}\left(-\delta_{CN}\delta_{aD}\sum_{b=1}^{N-1}E_{bN}-(1-\delta_{DN})\delta_{CN}E_{aN}\right)\nonumber\\=&
	-\left(\delta_{CN}(1-\delta_{DN})+(1-\delta_{DN})\delta_{CN}\right)\sum_{b=1}^{N-1}E_{bN}\nonumber
	\\=&
	-2\delta_{CN}(1-\delta_{DN})\sum_{b=1}^{N-1}E_{bN}\dt
\end{align}
From the third commutator on we always obtain zero, i.e. $[\mathcal{B},[\mathcal{B},[\mathcal{B},\mathcal{A}]]]=0$. All in all we have that 
 \begin{align}\label{HT-BA}
 	\exp{\left(-\sum_{a=1}^{N-1}E_{aN}\right)}E_{CD}\exp{\left(\sum_{a=1}^{N-1}E_{aN}\right)}=E_{CD}-\delta_{CN}\sum_{a=1}^{N-1}E_{aD}+(1-\delta_{DN})E_{CN}-\delta_{CN}(1-\delta_{DN})\sum_{b=1}^{N-1}E_{bN}\dt
 \end{align}

Using \eqref{transpositionPropertyR} we write the transpose of site Hamiltonian \eqref{siteHamiltonian} 
\begin{equation}
    \begin{split}
H_{x}^{T}=\sum_{A,B=1}^{N}\alpha_{A}^{x}\left(E_{AB}^{x}-E_{BB}^{x}\right)^{T}=R_{x}\sum_{A,B=1}^{N}\alpha_{A}^{x}\left(E_{BA}^{x}-E_{BB}^{x}\right)R_{x}^{-1}\dt
    \end{split}
\end{equation}
We multiply both sides by $R_{x}\exp{(E^{x})}$
\begin{equation}\label{intermediateTransposeSite}
    H_{x}^{T}R_{x}\exp{(E^{x})}=R_{x}\sum_{A,B =1}^{N}\alpha_{A}^{x}\left(E_{B A}^{x}-E_{BB}^{x}\right)\exp{(E^{x})}\dt
\end{equation}
By using \eqref{HT-BA} and the fact that 
$\sum_{a,b=1}^{N}E_{AB}$ is central for the algebra we have 

\begin{equation}\label{HadTransfBoundary}
\exp{(-E^{x})}\sum_{A=1}^{N}\sum_{B=1}^{N}\alpha_{A}^{x}\left(E_{B A}^{x}-E_{BB}^{x}\right)\exp{(E^{x})}=	|\alpha^{x}|\sum_{a=1}^{N-1}\left(\rho_{a}^{x}E_{Na}^{x}-E_{aa}^{x}\right)\dt
\end{equation}
Thus, we rewrite \eqref{intermediateTransposeSite} as
\begin{equation}\label{siteHadamardI}
H_{x}^{T}R_{x}\exp{(E^{x})}=R_{x}\exp{(E^{x})}|\alpha^{x}|\sum_{a=1}^{N-1}\left(\rho_{a}^{x}E_{Na}^{x}-E_{aa}^{x}\right)\dt
\end{equation}
Taking the tensor product of both sides of  \eqref{siteHadamardI} 
we obtain 
\begin{equation}\label{siteHadamardII}
    \begin{split}
    	&H_{x}^{T}d_{x}\otimes\left(\sum_{\xi_{1}^{u(x)},\ldots,\xi_{N-1}^{u(x)}=0}^{\infty}\left(\prod_{a=1}^{N-1}\left(\rho_{a}^{x}\right)^{\xi_{a}^{u(x)}}\right)\langle \xi_{1}^{u(x)},\ldots,\xi_{N-1}^{u(x)}|\right)\\=&
 d_{x}\,
|\alpha^{x}|\sum_{a=1}^{N-1}\left(\rho_{a}^{x}E_{Na}^{x}-E_{aa}^{x}\right)\otimes \left(\sum_{\xi_{1}^{u(x)},\ldots,\xi_{N-1}^{u(x)}=0}^{\infty}\left(\prod_{a=1}^{N-1}\left(\rho_{a}^{x}\right)^{\xi_{a}^{u(x)}}\right)\langle \xi_{1}^{u(x)},\ldots,\xi_{N-1}^{u(x)}|\right)
    \end{split}
\end{equation}
{With a slight abuse of notation and in the spirit of Remark \ref{reamrk-extraSite}, we here denote the occupation vector at extra-site $u(x)$ by $\langle \xi_{1}^{u(x)},\ldots,\xi_{N-1}^{u(x)}|$ 

.
} 
Recalling the action of the bosonic creation operator acting at site $u(x)$ and on the species $a\in \{1,\ldots,N-1\}$ we have that 
\begin{equation}\label{bosonicKX}
    \langle \xi_{1}^{u(x)},\ldots,\xi_{a}^{u(x)}+1,\ldots,\xi_{N-1}^{u(x)}|(\mathbf{a}^{+})^{u(x)}_{a}=  \langle \xi_{1}^{u(x)},\ldots,\xi_{a}^{u(x)},\ldots,\xi_{N-1}^{u(x)}|\dt
\end{equation}
Using the above equation we rewrite the right hand side term of \eqref{siteHadamardII} with Lie generator $E_{Na}^{x}$ as 
\begin{equation}\label{extra-site-trick}
    \begin{split}
&\sum_{\xi_{1}^{u(x)},\ldots,\xi_{N-1}^{u(x)}=0}^{\infty}\left(\prod_{b=1}^{N-1}\left(\rho_{b}^{x}\right)^{\xi_{b}^{u(x)}}\right)\rho_{a}^{x}\langle \xi_{1}^{u(x)},\ldots,\xi_{a}^{u(x)},\ldots,\xi_{N-1}^{u(x)}|
\\=&
\sum_{\xi_{1}^{u(x)},\ldots,\xi_{N-1}^{u(x)}=0}^{\infty}\left(\prod_{b=1\,:\,b\neq a}^{N-1}\left(\rho_{b}^{x}\right)^{\xi_{b}^{u(x)}}\right)\left(\rho_{a}^{x}\right)^{\xi_{a}^{u(x)}+1}\langle \xi_{1}^{u(x)},\ldots,\xi_{a}^{u(x)}+1,\ldots,\xi_{N-1}^{u(x)}|(\mathbf{a}^{+})_{a}^{u(x)}
\\=&
\sum_{\xi_{1}^{u(x)},\ldots,\xi_{N-1}^{u(x)}=0}^{\infty}\left(\prod_{b=1}^{N-1}\left(\rho_{b}^{x}\right)^{\xi_{b}^{u(x)}}\right)\langle \xi_{1}^{u(x)},\ldots,\xi_{a}^{u(x)},\ldots,\xi_{N-1}^{u(x)}|(\mathbf{a}^{+})_{a}^{u(x)}
    \end{split}
\end{equation}
where, in the up to last equality, we performed a change of summation variable and we used the fact that  
\begin{equation}
	\begin{split}
&\sum_{\xi_{a}^{u(x)}=0}^{\infty}(\rho_{a}^{x})^{\xi_{a}^{u(x)}}\langle \xi_{1}^{u(x)},\ldots,\xi_{a}^{u(x)},\ldots,\xi_{N-1}^{u(x)}|(\mathbf{a}^{+})_{a}^{u(x)}
=\sum_{\xi_{a}^{u(x)}=1}^{\infty}(\rho_{a}^{x})^{\xi_{a}^{u(x)}}\langle \xi_{1}^{u(x)},\ldots,\xi_{a}^{u(x)},\ldots,\xi_{N-1}^{u(x)}|(\mathbf{a}^{+})_{a}^{u(x)}\dt
\end{split}
\end{equation}
Therefore, inserting \eqref{extra-site-trick} in \eqref{siteHadamardII} we obtain 
\begin{equation}
    \begin{split}
H_{x}^{T}d_{x}\otimes \dd&=
     H_{x}^{T}d_{x}\otimes \sum_{\xi_{1}^{u(x)},\ldots,\xi_{N-1}^{u(x)}=0}^{\infty}\left(\prod_{a=1}^{N-1}\left(\rho_{a}^{x}\right)^{\xi_{a}^{u(x)}}\right)\langle \xi_{1}^{u(x)},\ldots,\xi_{N-1}^{u(x)}|
\\&=
\left\{d_{x}\otimes \sum_{\xi_{1}^{u(x)},\ldots,\xi_{N-1}^{u(x)}=0}^{\infty}\left(\prod_{a=1}^{N-1}\left(\rho_{a}^{x}\right)^{\xi_{a}^{u(x)}}\right)\langle \xi_{1}^{u(x)},\ldots,\xi_{N-1}^{u(x)}|\right\}\,|\alpha^{x}|\sum_{a=1}^{N-1}\left((\mathbf{a}^{+})_{a}^{u(x)}E_{Na}^{x}-E_{aa}^{x}\right)   
\\&=
\left(d_{x}\otimes \dd\right)\widetilde{H}_{x}\dt
    \end{split}
\end{equation}
Since the duality matrix \eqref{dualityMatrix} is product over sites, the above equality implies \eqref{siteDualRelation}. 
\begin{flushright}
$\square$
\end{flushright}

\subsection{Duality for the integrable process on a line segment}\label{integrableChain-duality}
Here we consider the chain defined in Section~\ref{subsection-description-process-LINE} with the hard-core constrain {(at most one particle per site)}, i.e. $\nu=1$. 
We recall that the geometry is a one dimensional chain with sites $\{1,\ldots,L\}$ where reservoirs are attached to $1$ and $L$. 
We also recall the notation for the occupation variable which is called $\bm{\tau}=(\tau_1,\ldots,\tau_L)$ with $\tau_x\in\{1,\ldots,N\}$.
The dynamics is described by an Hamiltonian that is written in terms of the first fundamental representation of ${gl}(N)$. The duality result stated in Section~\ref{statementDualitySubsection} is adapted to the present situation as follows. 
The extra-sites are denoted by $0$ and $L+1$, and they are connected with sites $1$ and $L$ of the chain respectively. The matrix $R_{x}$ defined in \eqref{Revmatrix} 
reduces to the identity, therefore the relation between the generators of the algebra and its transposed reduces to
\begin{equation}\label{transpostionPropertyFund}
	e_{AB}^T= e_{BA}\dt
\end{equation}
Therefore, duality matrix reads
\begin{equation}
	D=\mathcal{D}_{0}\otimes\prod_{x=1}^{L}\exp{\left(\sum_{a=1}^{N-1}e_{aN}^{x}\right)}\otimes \mathcal{D}_{L+1}
\end{equation}
where $\mathcal{D}_{0}$ and $\mathcal{D}_{L+1}$ are given in \eqref{dualityMatrix2}. 
The elements of this duality matrix are given by
\begin{equation}\label{ElementsDualityMatrixChain_1}
	D(\bm{\tau},\bm{\xi})=\left(\prod_{a=1}^{N-1}\alpha_{a}^{\xi_{a}^{0}}\right)\left(\prod_{x=1}^{L}\prod_{a=1}^{N-1}\mathbbm{1}_{\{\delta_{\tau_{x},a}\geq \xi_{a}^{x}\}}\right)\left(\prod_{a=1}^{N-1}\beta_{a}^{\xi_{a}^{L+1}}\right)\dt
\end{equation}
The dual Hamiltonian is
\begin{equation}
	\widetilde{H}=\widetilde{H}_{\text{left}}+\sum_{x=1}^{L-1}\mathcal{H}_{x,x+1}+\widetilde{H}_{\text{right}}
\end{equation}
where $\mathcal{H}$ is defined in \eqref{H-corsivo} and 
\begin{equation}\label{boundary-H-dual-chain}
	\widetilde{H}_{\text{left}}=\sum_{a=1}^{N-1}\left((\mathbf{a}^{+})_{a}^{0}\,e_{Na}^{1}-e_{aa}^{1}\right)\qquad 	\widetilde{H}_{\text{right}}=\sum_{a=1}^{N-1}\left((\mathbf{a}^{+})_{a}^{L+1}\,e_{Na}^{L}-e_{aa}^{L}\right)\dt
\end{equation}
In the long-time limit the dual process voids the chain, i.e. all particles of types $\{1,\ldots,N-1\}$ are eventually absorbed at the extra-sites $0$ and $L+1$ and replaced by types $N$. We also notice that, up the the bosonic creation operators $(\mathbf{a}^{+})_{a}$, the dual boundary Hamiltonians \eqref{boundary-H-dual-chain} are triangular. This simplification will be crucial in Section~\ref{sectionIntegrabiliy}.\\
Using duality, one can compute the  $m$-point correlations between non-empty particles in terms of the absorption probabilities of $m$ dual particles. Therefore, to determine the non-equilibrium steady state correlations of the integrable multispecies stirring process, it is enough to compute these absorption probabilities. In the following we denote by $\mu$ the non-equilibrium steady state distribution and we call $\bm{Y} = (Y_1,\ldots,Y_L)$ the random vector with law $\mu$. 
Furthermore we write $\mathbb{E}[\cdot]$ for the expectation with respect to $\mu$.
\begin{corollary}[Correlations via duality]\label{Corolollary-ABS_Corr-abstract}
Let $m\in \{1,\ldots,L\}$ and consider $m$ sites $1 \le x_{1} < x_2 < \ldots < x_m \le L $ and
{$m$ colours denoted by $a_{k}\in\{1,\ldots,N-1\}$  with $k=1,2,\ldots, m$, chosen among the $N-1$ available species.} Then, the $m$-point correlations in the non-equilibrium steady state read 
\begin{equation}\label{NESS-expectation-ABS}
\mathbb{E}\left[\prod_{k=1}^{m}\mathbbm{1}_{\{Y_{x_{k}}=a_{k}\}}\right]=\sum_{t_{1}=0}^{1}\ldots \sum_{t_{m}=0}^{1}\left(\prod_{k=1}^{m}\alpha_{a_{k}}^{t_{k}}\beta_{a_{k}}^{1-t_{k}}\right)\mathcal{P}_{x_1,\ldots,x_m}(t_{1},\ldots,t_{m})
\end{equation}
where
\begin{equation}\label{absProbabilitiesIntegrable}
	\mathcal{P}_{x_1,\ldots,x_m}(t_{1},\ldots,t_{m}):=\mathbb{P}\left(\bm{\xi}(\infty)=\sum_{k=1}^{m}\left(t_{k}\delta^{0}_{a_k}+(1-t_{k})\delta^{L+1}_{a_k}\right)\Big| \bm{\xi}(0)=\bm{\xi}\right)
\end{equation}
where the initial dual configuration is $\bm{\xi}=\sum_{k=1}^{m}\delta_{a_{k}}^{x_{k}}$, meaning that 
\begin{equation}
\label{xsi-init}
	\begin{cases}
		\xi_{A}^{x}=1\qquad \text{if}\;x=x_{k}\;\text{and}\; A=a_{k}\\
		\xi_{A}^{x}=0\qquad \text{otherwise}
	\end{cases}\dt
\end{equation}
In \eqref{absProbabilitiesIntegrable} the variable $t_k$ is $1$ (resp. $0$) when the dual particle of species $a_k$ initially positioned at site $x_{k}$ is absorbed at at the extra site $0$ (resp. $L+1$).
\end{corollary}
\textbf{Proof of Corollary~\ref{Corolollary-ABS_Corr-abstract}}:
For any $\bm{\tau}\in \Omega^{'}$ the duality matrix defined in \eqref{ElementsDualityMatrixChain_1} evaluated on the dual configuration $\bm{\xi}\in\widetilde{\Omega}$ given in \eqref{xsi-init} read
\begin{equation}
	D(\bm{\tau},\bm{\xi})
	=\prod_{k=1}^{m}\mathbbm{1}_{\{\tau_{x_{k}}=a_{k}\}}\dt
\end{equation}
Therefore, by ergodicity and duality  we have 
\begin{equation}\label{CorrelationByABS-prob}
	\begin{split}
\mathbb{E}\left[\prod_{k=1}^{m}\mathbbm{1}_{\{Y_{x_{k}}=a_{k}\}}\right]&=\lim_{t\to \infty}\mathbb{E}_{\bm{\tau}}\left[D(\bm{\tau}(t),\bm{{\xi}})\right]=\lim_{t\to \infty}\mathbb{E}_{\bm{\xi}}\left[D(\bm{\tau},\bm{{\xi}}(t))\right]
	\\&=\sum_{t_{1}=0}^{1}\ldots \sum_{t_{m}=0}^{1}\left(\prod_{k=1}^{m}\alpha_{a_{k}}^{t_{k}}\beta_{a_{k}}^{1-t_{k}}\right)\mathcal{P}_{x_{1},\ldots,x_{m}}(t_{1},\ldots,t_{m})\dt
\end{split}
\end{equation} 
\begin{flushright}
	$\square$
\end{flushright} 
{By using the correlations in the non-equilibrium steady state \eqref{exactCorrelations}, it is possible to determine the non-equilibrium steady probability distribution. }
    \begin{corollary}[Non-equilibrium steady state]\label{Corollary-NESS-ABS}
For any configuration $\bm{\tau}\in \Omega^{'}$  the probability mass function of the non-equilibrium steady states is fully determined by the correlations of equation \eqref{exactCorrelations} through the relation 
{
\begin{equation}\label{implicit-mass-function}
	\mu(\bm{\tau})=
	\sum_{m=0}^{L}\;\sum_{1\leq x_{1}<x_{2}<\ldots<x_{m}\leq L}\left(\prod_{x\notin\{x_{1},\ldots,x_{m}\}}\delta_{\tau_{x},N}\right)\sum_{b_{1},\ldots,b_{m}=1}^{N-1}\left(\prod_{k=1}^{m}(\delta_{\tau_{x_{k}},b_{k}}-\delta_{\tau_{x_{k}},N})\right)\mathbb{E}\left[\prod_{k=1}^{m}\mathbbm{1}_{\{Y_{x_{k}}=b_{k}\}}\right]
\end{equation}
where $\delta_{\tau_{x_{k}},b_{k}}-\delta_{\tau_{x_{k}},N}=(-1)^{\delta_{\tau_{x_{k}},N}}$.
}
\end{corollary}
\paragraph{Example for an $L=2$ chain.} To clarify equation \eqref{implicit-mass-function} we report the example of a chain with $L=2$, where we compute the probability mass functions of three configurations. Without loss of generality we write only the formulae involving particles of species $1$. We have: 
 \begin{itemize}
 	\item probability of two occupied site:
 	\begin{equation}
 		\mu(1,1)=\mathbb{E}\left[\mathbbm{1}_{\{Y_{1}=1\}}\mathbbm{1}_{\{Y_{2}=1\}}\right]\dt
 	\end{equation}
 	Here we have only one term corresponds to $m=2$ in formula  \eqref{implicit-mass-function}.

 	\item probability of the first site occupied and the second empty:
 	\begin{align}
 		\mu(1,N)=&\mathbb{E}\left[\mathbbm{1}_{\{Y_{1}=1\}}\mathbbm{1}_{\{Y_{2}=N\}}\right]=\mathbb{E}\left[\mathbbm{1}_{\{Y_{1}=1\}}\left(1-\sum_{a=1}^{N-1}\mathbbm{1}_{\{Y_{2}=a\}}\right)\right]\nonumber
 		\\=&
 		\mathbb{E}\left[\mathbbm{1}_{\{Y_{1}=1\}}\right]-\sum_{a=1}^{N-1}\mathbb{E}\left[\mathbbm{1}_{\{Y_{1}=1\}}\mathbbm{1}_{\{Y_{2}=a\}}\right]\dt
 	\end{align}
	{In the last equality we have two terms: the first corresponds to $m=1$ (with $x_{1}=1$) and the second corresponds to $m=2$ in formula \eqref{implicit-mass-function}.}
 	\item probability of both sites empty:
 	\begin{align}
 		\mu(N,N)=&\mathbb{E}\left[\mathbbm{1}_{\{Y_{1}=N\}}\mathbbm{1}_{\{Y_{2}=N\}}\right]=\mathbb{E}\left[\left(1-\sum_{a_{1}=1}^{N-1}\mathbbm{1}_{\{Y_{1}=a_{1}\}}\right)\left(1-\sum_{a_{2}=1}^{N-1}\mathbbm{1}_{\{Y_{2}=a_{2}\}}\right)\right]\nonumber
 		\\=&
 		1-\sum_{a_{1}=1}^{N-1}\mathbb{E}\left[\mathbbm{1}_{\{Y_{1}=a_{1}\}}\right]-\sum_{a_{2}=1}^{N-1}\mathbb{E}\left[\mathbbm{1}_{\{Y_{2}=a_{2}\}}\right]+\sum_{a_{1},a_{2}=1}^{N-1}\mathbb{E}\left[\mathbbm{1}_{\{Y_{1}=a_{1}\}}\mathbbm{1}_{\{Y_{2}=a_{2}\}}\right]\dt
 	\end{align}
 	{In the last equality we have four terms: the first term corresponds to $m=0$, the second and the third term correspond to $m=1$ (with $x_{1}=1$ and $x_{1}=2$ respectively) and the fourth term corresponds to $m=2$ in formula \eqref{implicit-mass-function}. } 
 \end{itemize}
\textbf{Proof of Corollary~\ref{Corollary-NESS-ABS}:}
Given $\bm{\tau}\in \Omega^{'}$,  we have 
\begin{align}
	\mu\left(\bm{\tau}\right)=&\mathbb{E}\left[\prod_{x=1}^{L}\mathbbm{1}_{\{Y_{x}=\tau_{x}\}}\right]
	=\mathbb{E}\left[\left(\prod_{x=1\,:\,\tau_{x}\neq N}^{L}\mathbbm{1}_{\{Y_{x}=\tau_{x}\}}\right)\left(\prod_{x=1\,:\,\tau_{x}=N}^{L}\mathbbm{1}_{\{Y_{x}=\tau_x\}}\right)\right]\nonumber\\
	=&\mathbb{E}\left[\left(\prod_{x=1\,:\,\tau_{x}\neq N}^{L}\mathbbm{1}_{\{Y_{x}=\tau_{x}\}}\right)\left(\prod_{x=1\,:\,\tau_{x}=N}^{L}\left(1-\sum_{a=1}^{N-1}\mathbbm{1}_{\{Y_{x}=a\}}\right)\right)\right]
\end{align}
where in the last equality we exploited the fact that 
\begin{equation}\label{prova}			\mathbbm{1}_{\{Y_{x}=N\}}=1-\sum_{a=1}^{N-1}\mathbbm{1}_{\{Y_{x}=a\}}\dt
\end{equation}
The configuration $\bm{\tau}$ has $\ell=\sum_{x=1}^{L}(1-\delta_{\tau_{x},N})$ occupied sites and then the remaining $L-\ell$ sites are empty. 
Therefore, we may write
\begin{align}\label{sum}
&\left(\prod_{x=1\,:\,\tau_{x}\neq N}^{L}\mathbbm{1}_{\{Y_x=\tau_{x}\}}\right)\left(\prod_{x=1\,:\,\tau_{x}=N}^{L}\left(1-\sum_{a=1}^{N-1}\mathbbm{1}_{\{Y_x=a\}}\right)\right)\nonumber\\
=&
\prod_{x=1\,:\,\tau_{x}\neq N}^{L}\mathbbm{1}_{\{Y_x=\tau_{x}\}}+\left(\prod_{x=1\,:\,\tau_{x}\neq N}^{L}\mathbbm{1}_{\{Y_x=\tau_{x}\}}\right)(-1)\left\{\sum_{x_{1}=1}^{L}\delta_{\tau_{x_{1}},N}\sum_{a_{1}=1}^{N-1}\mathbbm{1}_{\{Y_{x_{1}}
=
a_{1}\}}\right\}+\nonumber\\
+&\left(\prod_{x=1\,:\,\tau_{x}\neq N}^{L}\mathbbm{1}_{\{Y_x=\tau_{x}\}}\right)(-1)^{2}\left\{\sum_{1\leq x_{1}<x_{2}\leq L}\delta_{\tau_{x_{1}},N}\delta_{\tau_{x_{2}},N}\sum_{a_{1},a_{2}=1}^{N-1}\mathbbm{1}_{\{Y_{x_{1}}=a_{1}\}}\mathbbm{1}_{\{Y_{x_{2}}=a_{2}\}}\right\}+\ldots+
\nonumber\\
+&\left(\prod_{x=1\,:\,\tau_{x}\neq N}^{L}\mathbbm{1}_{\{Y_x=\tau_{x}\}}\right)(-1)^{L-\ell}\sum_{1\leq x_{1}<\ldots<x_{L-\ell}\leq L}\left(\prod_{k=1}^{L-\ell}\delta_{\tau_{x_{k}},N}\right)\sum_{a_{1},\ldots,a_{L-\ell}=1}^{N-1}\prod_{k=1}^{L-\ell}\mathbbm{1}_{\{Y_{x_{k}}=a_{k}\}}\dt
\end{align} 
{In the right-hand-side of the equation above the first addend is the product of the indicator of $\ell$ sites, since none of the empty sites of the configuration $\bm{\tau}$ has been considered. Similarly, the second addend is the product of the indicator of $\ell+1$ occupied sites. Indeed, in addition to the $\ell$ occupied sites of the configuration $\bm{\tau}$, one empty site is in turn chosen and its hole is filled with all possible species particles. The idea is repeated in the next addends with $2,3,...L-\ell$ empty sites that are, in turn, filled with all possible species of particles. We notice that the exponent of the factors (-1) is given by the number of holes that have been filled with all possible species of particles. We introduce an index $m\in \{\ell,\ell+1,\ldots,L\}$ that counts the number of correlated site and we define the coordinates $1\leq x_{1}<x_{2}<\ldots<x_{m}\leq L$. Then, we associate to each of these $m$ one of the addend in \eqref{sum}. In particular, for $m=\ell$ we associate the first addend and we rewrite it as
	\begin{align}
		\prod_{x=1\,:\,\tau_{x}\neq N}^{L}\mathbbm{1}_{\{Y_{x}=\tau_{x}\}}=&
		\sum_{1\leq x_{1}<\ldots<x_{\ell}\leq L}\;\left(\prod_{x\notin \{x_{1},\ldots,x_{\ell}\}}\delta_{\tau_{x},N}\right)\sum_{b_{1},\ldots,b_{\ell}=1}^{N-1}\left(\prod_{k=1}^{\ell}(\delta_{\tau_{x_{k}},b_{k}}-\delta_{\tau_{x_{k},N}})\right)\left(\prod_{k=1}^{\ell}\mathbbm{1}_{\{Y_{x_{k}}=b_{k}\}}\right)
	\end{align}
	Similarly, the second addend in \eqref{sum} is associated to $m=\ell +1$ and it is rewritten as 
	\begin{align}
&\left(\prod_{x=1\,:\,\tau_{x}\neq N}^{L}\mathbbm{1}_{\{Y_x=\tau_{x}\}}\right)(-1)\left\{\sum_{x_{1}=1}^{L}\delta_{\tau_{x_{1}},N}\sum_{a_{1}=1}^{N-1}\mathbbm{1}_{\{Y_{x_{1}}
	=
	a_{1}\}}\right\}
			\\=&
		\sum_{1\leq x_{1}<\ldots<x_{\ell+1}\leq L}\;\left(\prod_{x\notin \{x_{1},\ldots,x_{\ell+1}\}}\delta_{\tau_{x},N}\right)\sum_{b_{1},\ldots,b_{\ell+1}=1}^{N-1}\left(\prod_{k=1}^{\ell+1}(\delta_{\tau_{x_{k}},b_{k}}-\delta_{\tau_{x_{k},N}})\right)\left(\prod_{k=1}^{\ell+1}\mathbbm{1}_{\{Y_{x_{k}}=b_{k}\}}\right)
	\end{align}
	The idea goes on for $m=\ell+2,\ldots L$ and we obtain that 
\begin{align}
&\left(\prod_{x=1\,:\,\tau_{x}\neq N}^{L}\mathbbm{1}_{\{Y_x=\tau_{x}\}}\right)\left(\prod_{x=1\,:\,\tau_{x}=N}^{L}\left(1-\sum_{a=1}^{N-1}\mathbbm{1}_{\{Y_x=a\}}\right)\right)\nonumber\\
=&
\sum_{m=\ell}^{L}\;\sum_{1\leq x_{1}<x_{2}<\ldots<x_{m}\leq L}\left(\prod_{x\notin\{x_{1},\ldots,x_{m}\}}\delta_{\tau_{x},N}\right)\sum_{b_{1},\ldots,b_{m}=1}^{N-1}\left(\prod_{k=1}^{m}(\delta_{\tau_{x_{k}},b_{k}}-\delta_{\tau_{x_{k}},N})\right)\left(\prod_{k=1}^{m}\mathbbm{1}_{\{Y_{x_{k}}=b_{k}\}}\right)
\end{align}
The above sum can be extended until $m=0$ since all the terms $m=0,\ldots,\ell-1$ are vanishing. 
Finally, by taking the expectation with respect to the steady state distribution we obtain \eqref{implicit-mass-function}.} 
\begin{flushright}
	$\square$
\end{flushright}
\begin{remark}\label{Remark-extension-graph-nu}
	The results in Corollary~\ref{Corolollary-ABS_Corr-abstract} and in Corollary~\ref{Corollary-NESS-ABS}  can be generalized to arbitrary graphs $G=(V,\mathcal{E})$ and to {higher maximal occupancy}, i.e. $\nu >1$.
	However the computation of absorption probabilities of the dual process is possible only in the set-up of the integrable chain with $\nu =1$ and two reservoirs at the boundaries.
	Thus we chose to present them in such a context. For the chain with {more than one particle allowed at each site} $\nu >1$ see Section~\ref{Subsection-ss-nonI}.
\end{remark}

\section{Correlations and steady state  for the integrable chain}\label{sectionIntegrabiliy}
In this section we combine the matrix product ansatz (MPA) \cite{vanicat2017exact} and  the $gl(N)$ invariance of the process in the bulk, cf.~\eqref{symmetryH}, to exactly compute the correlations in the non-equilibrium steady state
 for the integrable version of the multispecies stirring process on a defined on a line segment, see Section~\ref{integrableChain-duality}. 
This allows further to determine the absorption probabilities and the probability mass-function of the non-equilibrium steady state in closed form.
We start by recalling the formulation of the MPA  and then we explain our strategy.

\subsection{Matrix product ansatz for the non-equilibrium steady state}
The matrix product ansatz for the multispecies stirring process has been formulated in \cite{vanicat2017exact}, here we briefly recall the main steps.
 Denoting by $|\Psi(t)\rangle$ the column vector that encodes the probability distribution of the chain with Hamiltonian \eqref{hamiltonian} at time $t\geq 0$, its evolution equation is given by 
 the master equation
\begin{equation}
    \frac{d|{\Psi}(t)\rangle}{dt}=H|{\Psi}(t)\rangle\dt
\end{equation}
This Markov chain is irreducible and positive recurrent, therefore there exists a unique stationary measure, that will be reached when time goes to infinity, regardless of the initial configuration. 
We denote by $|\Psi\rangle$ the column vector that gives the stationary distribution (non-equilibrium steady state). This vector is the right eigenvector with vanishing eigenvalue of $H$, i.e. it solves 
\begin{equation}\label{definition-SteadyStateH}
	H|\Psi\rangle =0\dt
\end{equation}
The MPA states the following
\begin{equation}
	|\Psi\rangle=\frac{1}{Z_{L}} \langle \langle W|  \begin{pmatrix}
		X_{1}\\
		\vdots\\
		X_{N}
	\end{pmatrix}\otimes \ldots\otimes \begin{pmatrix}
		X_{1}\\
		\vdots\\
		X_{N}
	\end{pmatrix} |V\rangle\rangle 
\end{equation}
with the normalization 
\begin{equation}
	Z_{L}= \langle \langle W|  (X_{1}+\ldots +X_{N})^{L} |V\rangle\rangle \dt
\end{equation}
Here the operators $(X_{a})_{a\in \{1,\ldots,N\}}$ fulfil the commutators
\begin{equation}\label{bulk}
	\left[X_{a},X_{b}\right]=(\alpha_{a}-\beta_{a})X_{b}-(\alpha_{b}-\beta_{b})X_{a}
\end{equation}
and their action on the boundary vectors are
\begin{equation}\label{leftBoundary}
	 \langle \langle W|  \left(\alpha_{a}(X_{1}+\ldots+X_{N})-X_{a}\right)=(\alpha_{a}-\beta_{a}) \langle \langle W|  
\end{equation}
\begin{equation}\label{rightBoundary}
	\left(\beta_{a}(X_{1}+\ldots+X_{N})-X_{a}\right) |V\rangle\rangle =-(\alpha_{a}-\beta_{a}) |V\rangle\rangle 
\end{equation}
Without loss of generality, we further assume that $ \langle \langle W|  V\rangle\rangle =1$. 
\begin{remark} Only $N-1$ of the $N$ equations \eqref{leftBoundary}  are independent. This can be seen by summing them over the species $a\in \{1,\ldots,N\}$, i.e.
	\begin{equation}
		\sum_{a=1}^{N}	 \langle \langle W|  \left(\alpha_{a}(X_{1}+\ldots+X_{N})-X_{a}\right)=0,
	\end{equation}
	and using  \eqref{ratesConditions}.
 Similarly for the right boundary  \eqref{rightBoundary} there are only  $N-1$ independent equations. \end{remark} 
The MPA gives an abstract form of the non-equilibrium steady state in terms of an algebra of operators. However, the computation of  correlations are in general involved. 

\subsection{Strategy}
 In the next Section~\ref{subsection-exact} we will show that the DEHP algebra in \eqref{bulk}-\eqref{rightBoundary} can be simplified substantially using the ${gl}(N)$ invariance of the bulk. In fact, such simplification can always be achieved if 
  there exists an absorbing dual process as established in Section~\ref{sectionDuality}
 
 The idea is summarized in the following scheme. 
 We define a sequence of local similarity transformations such that 
 \begin{equation}\label{eq:scheme}
 	H\;\xleftrightarrow{S_{1}}\;H^{'}\;\xleftrightarrow{S_{2}}\;H^{''}
 \end{equation} 
 where $H^{'}= S_{1} H S_{1}^{-1} $ has both boundary terms in a triangular form, and $H^{''} = S_{2}H' S_{2}^{-1}$ has the left boundary in a triangular form and the right boundary is diagonal, we further refer the reader to \cite{alcaraz,melo2005bethe,SSEPReviewRagoucy,frassek2020eigenstates} where this idea was explored for the monospecies case. Using these transformations the commutators \eqref{bulk} and the action on the boundary vectors \eqref{leftBoundary}, \eqref{rightBoundary} simplify significantly. More precisely for the components of $\tilde X=S_2S_1 X$ we obtain the bulk relations
\begin{equation}\label{commutationsBula2}
	\left[\Xt_{a},\Xt_{N}\right]=(\alpha_{a}-\beta_{a})\Xt_{N}\dt 
\end{equation}
for $a=1,\ldots,N-1$. 
At the boundaries 
\begin{equation}\label{commLEFT2}
	 \langle \langle W|  \left((\alpha_{a}-\beta_{a})\Xt_{N}-\Xt_{a}\right)=(\alpha_{a}-\beta_{a}) \langle \langle W|  
	\co \qquad 
	\Xt_{a}  |V\rangle\rangle = (\alpha_{a}-\beta_{a}) |V\rangle\rangle
	\,.
\end{equation} 
where again we have  $a=1,\ldots,N-1$.   As a consequence, the ground state $|\Psi^{''}\rangle$ of $H^{''}$ can be written exactly in closed-form. Finally, reversing the transformations in \eqref{eq:scheme} we retrieve $|\Psi\rangle$, i.e. the vector whose components are the probabilities of a certain configuration of the process in the non-equilibrium steady state.
 The first transformation $S_{1}$ is closely related to the duality matrix \eqref{dualityMatrix} and we will see that $H^{'}$ is $\widetilde{H}^{T}$, up to the extra-site term described by the bosonic creation operator in \eqref{siteDualHamiltonian}. The Hamiltonian $H^{'}$ is not stochastic, however it turns out that the components of its eigenvector with zero eigenvalue $|\Psi^{'}\rangle$ are the correlations in the non-equilibrium steady state. 
\subsection{Correlations in the non-equilibrium steady state}\label{subsection-exact}
In this section we write a formula for the stationary non-equilibrium steady state correlations between $m$- points of the chain.
\begin{theorem}[Correlations in the non-equilibrium steady state]\label{thm-correlations}
Let $m\in \{1,\ldots,L\}$. Consider $m$ sites $1 \le x_{1} < x_2 < \ldots < x_m \le L $ and
{$m$ colours denoted by $a_{k}\in\{1,\ldots,N-1\}$  with $k=1,2,\ldots, m$, chosen among the $N-1$ available species.}  
Then the $m$-point correlations with respect to the non-equilibrium steady state measure are given by 
\begin{equation}\label{exactCorrelations}
\mathbb{E}\left[\prod_{k=1}^{m}\mathbbm{1}_{\{Y_{x_{k}}=a_{k}\}}\right]\\=\sum_{t_{1},\ldots,t_{m}=0}^{1}\left(\prod_{k=1}^{m}\alpha_{a_{k}}^{t_{k}}\beta_{a_{k}}^{1-t_{k}}\right)\mathcal{P}_{x_{1},\ldots,x_{m}}(t_{1},\ldots,t_{m})
\end{equation}
where
\begin{equation}\label{abs-probabilities}
	\mathcal{P}_{x_{1},\ldots,x_{m}}(t_{1},\ldots,t_{m})=\sum_{c_{1}=t_{1}}^{1}\ldots\sum_{c_{m}=t_{m}}^{1}f(c_{1},\ldots,c_{m})\prod_{j=1}^{m}(-1)^{c_{j}-t_{j}}g_{j}(x_{j},c_{j},\ldots,c_{m})
\end{equation}
with
\begin{equation}\label{powerCoeffNOspec}
	f(c_{1},\ldots,c_{m})=\frac{(L+1-\sum_{a=1}^{m}c_{a})!}{(L+1)!}
\end{equation}
and
\begin{equation}\label{powerCoeffSpecies}
	g_{j}(x_{j},c_{j},\ldots,c_{m})=\left(L+2-x_{j}-\sum_{k=j}^{m}c_{k}\right)^{c_{j}}\dt
\end{equation}
\end{theorem}

\subsubsection{Examples}
We give examples  of correlations for $m=1,2,3$ applying formula \eqref{exactCorrelations}.
\paragraph{One-point correlations}
We consider the average with respect to $\mu$ of the occupation variable of the species $a_{1}\in \{1,\ldots,N-1\}$ at coordinate $x_{1}\in \{1,\ldots,L\}$. 
Using \eqref{abs-probabilities} we obtain the absorption probabilities 
\begin{equation}
	\mathcal{P}_{x_1}(0)=1-\frac{(L+1-x_{1})}{(L+1)}\qquad \mathcal{P}_{x_1}(1)=\frac{(L+1-x_{1})}{(L+1)}
\end{equation} 
where $\mathcal{P}_{x_1}(1)$ is the probability that a single random walk started at $x_1$ is absorbed. Then, using \eqref{exactCorrelations}, we have
\begin{equation}\label{one-pts-corr}
		\langle\rho_{a_{1}}^{x_{1}}\rangle= \mathbb{E}\left[\mathbbm{1}_{\{Y_{x_{1}}=a_{1}\}}\right]=\frac{(L+1-x_{1})}{(L+1)}\alpha_{a_{1}}+\frac{x_{1}}{(L+1)}\beta_{a_{1}}\co
\end{equation} 
{where, for the sake of notation, we have introduced $\langle\rho_{a_{1}}^{x_{1}}\rangle$.}
\paragraph{Two-point correlations}
We consider the average with respect to $\mu$ of the occupation variable of the species  $a_{1},a_{2}\in \{1,\ldots,N-1\}$ at coordinates $x_{1},x_{2}\in \{1,\ldots,L\}$ such that $x_{1}< x_{2}$. 
Using \eqref{abs-probabilities} we obtain the absorption probabilities
\begin{equation}
	\begin{split}
	&\mathcal{P}_{x_{1},x_{2}}(0,0)=\frac{x_{1}(x_{2}-1)}{L(L+1)}\\
	&\mathcal{P}_{x_{1},x_{2}}(0,1)=\frac{x_{1}(L+1-x_{2})}{L(L+1)}
\end{split}\quad \begin{split}
	&\mathcal{P}_{x_{1},x_{2}}(1,0)=\frac{x_{2}(L+1-x_{1})}{L(L+1)}\\
	&\mathcal{P}_{x_{1},x_{2}}(1,1)=\frac{(L-x_{1})(L+1-x_{2})}{L(L+1)}
	\end{split}\dt
\end{equation}
Therefore, we compute the second cumulant, i.e. the two-point connected correlation
\begin{equation}\label{two-pts-corr}
	\begin{split}
		 \mathbb{E}\left[\left(\mathbbm{1}_{\{Y_{x_{1}}=a_{1}\}}-\langle\rho_{a_{1}}^{x_{1}}\rangle \right)\left(\mathbbm{1}_{\{Y_{x_{2}}=a_{2}\}}-\langle\rho_{a_{2}}^{x_{2}}\rangle \right)\right]=-\frac{x_{1}(L-x_{2}+1)}{L(L+1)^{2}}\left(\alpha_{a_{1}}-\beta_{a_{1}}\right)\left(\alpha_{a_{2}}-\beta_{a_{2}}\right)\dt
	\end{split}
\end{equation} 
\paragraph{Three-point correlations}
We consider the average with respect to $\mu$ of the occupation variable of the species  $a_{1},a_{2},a_{3}\in \{1,\ldots,N-1\}$ at coordinates $x_{1},x_{2},x_{3}\in \{1,\ldots,L\}$ such that $x_{1}<x_{2}<x_{3}$. Using \eqref{abs-probabilities} we compute the absorption probabilities 
\begin{equation}
	\begin{split}
	&\mathcal{P}_{x_{1},x_{2},x_{3}}(0,0,0)=\frac{x_{1}(x_{2}-1)(x_{3}-2)}{L(L^{2}-1)}\\
 &\mathcal{P}_{x_{1},x_{2},x_{3}}(0,1,1)=\frac{x_1 (L-x_2) (L-x_3+1)}{L \left(L^2-1\right)}\\
	&\mathcal{P}_{x_{1},x_{2},x_{3}}(0,1,0)=\frac{x_1 (L(1-x_3)+x_2 (x_3-2)+1)}{L(L^2-1)}\\
	&\mathcal{P}_{x_{1},x_{2},x_{3}}(0,0,1)=\frac{x_1 (x_2-1) (L-x_3+1)}{L \left(L^2-1\right)}
	\end{split}\quad \begin{split}
	&\mathcal{P}_{x_{1},x_{2},x_{3}}(1,1,0)=\frac{(L-1) x_2 (x_3-1)-x_1 (x_2-1)
		(x_3-2)}{L \left(L^2-1\right)}\\
	&\mathcal{P}_{x_{1},x_{2},x_{3}}(1,0,1)=\frac{(L-x_3+1) (x_2 (L-x_1-1)+x_1)}{L
		\left(L^2-1\right)}\\
	&\mathcal{P}_{x_{1},x_{2},x_{3}}(1,0,0)=\frac{ x_{2} (x_3-1)(L-1)-x_1 (x_2-1)
	(x_3-2)}{L \left(L^2-1\right)}\\
	&\mathcal{P}_{x_{1},x_{2},x_{3}}(1,1,1)=\frac{(L-x_{1}-1)(L-x_{2})(L-x_{3}+1)}{L(L^{2}-1)}
\end{split}
\end{equation}

Therefore, we compute the third cumulant, i.e. the three-point connected correlation as 
\begin{align}
	&\mathbb{E}\left[\left(\mathbbm{1}_{\{Y_{x_{1}}=a_{1}\}}-\langle\rho_{a_{1}}^{x_{1}}\rangle \right)\left(\mathbbm{1}_{\{Y_{x_{2}}=a_{2}\}}-\langle\rho_{a_{2}}^{x_{2}}\rangle \right)\left(\mathbbm{1}_{\{Y_{x_{3}}=a_{3}\}}-\langle\rho_{a_{3}}^{x_{3}}\rangle \right)\right]\nonumber\\
	=&-\frac{2x_{1}(L+1-2x_{2})(L+1-x_{3})}{(L+1)^{3}(L-1)L}(\alpha_{a_{1}}-\beta_{a_{1}})(\alpha_{a_{2}}-\beta_{a_{2}})(\alpha_{a_{3}}-\beta_{a_{3}})\dt
\end{align} 
\begin{remark} The first and second cumulants, computed in \eqref{one-pts-corr} and \eqref{two-pts-corr} respectively, match with those found in Section 4.3 of \cite{vanicat2017exact}.
	\end{remark}

\subsection{Proof of Theorem~\ref{thm-correlations}}
The proof of formula \eqref{exactCorrelations} is split into the following steps:
\begin{itemize}
	\item In Section~\ref{subsectionSTransf}, we introduce two similarity transformations $S_1$ and  $S_2$. 
	 The first turns the Hamiltonian $H$ into $H^{'}= S_{1} H S_{1}^{-1} $ with boundary terms in a triangular form;
	 the second  turns the Hamiltonian $H'$ into $H^{''} = S_{2}H' S_{2}^{-1}$ having the left boundary in a triangular form and the right boundary is diagonal.
	Associated to $H^{'}$ and $H^{''}$ there are two ground states denoted by $|\Psi^{'}\rangle$ and $|\Psi^{''}\rangle$ respectively. 
	\item In Section~\ref{subsectionSSHsec} we apply the MPA to $H^{''}$. Here the commutation relations defining the matrix algebra are simpler. The explicit expression for $|\Psi^{''}\rangle$ is determined (see \eqref{ResulsBasis}).
	\item In Section~\ref{subsectionSSdual}, we invert the similarity transformation $S_2$ to recover  the ground state $|\Psi^{'}\rangle$ from the explicit expression of $|\Psi^{''}\rangle$  (see \eqref{ABS}).
	\item In Section~\ref{correlation-section} we show that the correlations are in turn the components of the vector $|\Psi^{'}\rangle$. 
	By exploiting a binomial formula we rewrite the correlations in terms
of polynomials in the left and right boundary densities with coefficients given by the absorption probabilities, as claimed in Corollary~\ref{Corolollary-ABS_Corr-abstract}.
	For completeness, we also show that Corollary~\ref{Corollary-NESS-ABS} is in turn corresponding to the transformation $|\Psi\rangle = S_{1}^{-1} |\Psi'\rangle$ in Section~\ref{appB}. 
	\end{itemize}
\subsubsection{The similarity transformations}\label{subsectionSTransf}
Consider the matrix
\begin{align}
	\mathcal{S}_{1}:=\exp{\left(\sum_{a=1}^{N-1}e_{Na}\right)}\label{S1Gen}\co
\end{align}
and
\begin{equation}\label{transformationV}
 \mathcal{S}_{1}^{x}:=\exp{\left(\sum_{a=1}^{N-1}e_{Na}^{x}\right)}\co 
\end{equation}
to denote $\mathcal{S}_{1}$ (see \eqref{S1Gen}) when acting at site $x$. 
Observe that $\mathcal{S}_{1}^{x}$ is the transposed of the bulk part of the duality matrix $d_{x}$  at site $x\in \{1,\ldots,L\}$ (see \eqref{bulkElementDualityMatrix}). By taking the product over the chain we define
\begin{equation}\label{S1-Whole}
    S_{1}=\prod_{x=1}^{L}\mathcal{S}_{1}^{x}
\end{equation}
that is related with the bulk duality matrix by 
\begin{equation}
S_{1}=\prod_{x=1}^{L}d_{x}^{T}\dt
\end{equation}
 As the transformation $S_{1}$ is invertible,  we can introduce 
\begin{equation}\label{hatHamiltonian}
   H^{'}=H^{'}_{\text{left}}+H_{\text{bulk}}+H^{'}_{\text{right}}
\end{equation}
that is related to $H$ by
\begin{equation}\label{similarV}
   H^{'}=S_{1}HS_{1}^{-1}\dt
\end{equation}
The bulk part of the Hamiltonian is left unchanged (because it is written using the coproduct of the second Casimir, see equation \eqref{hamiltonianCasimir}) while the boundaries are given by 
\begin{equation}\label{eq:Hprimes}
	H^{'}_{\text{left}}=\sum_{a=1}^{N-1}\left(\alpha_{a}e_{aN}^{1}-e_{aa}^{1}\right)\qquad H^{'}_{\text{right}}=\sum_{a=1}^{N-1}\left(\beta_{a}e_{aN}^{L}-e_{aa}^{L}\right)\dt
\end{equation}
Indeed, using \eqref{HadTransfBoundary}, \eqref{ratesConditions} and \eqref{transpostionPropertyFund}, we obtain for the left boundary
\begin{equation}
	\begin{split}
		H_{\text{left}}^{'}&=S_{1}H_{\text{left}}S_{1}^{-1}=\exp{\left(\sum_{c=1}^{N-1}e_{Nc}^{1}\right)}\sum_{A,B=1}^{N}\alpha_{a}\left(e_{AB}^{1}-e_{BB}^{1}\right)\exp{\left(-\sum_{c=1}^{N-1}e_{Nc}^{1}\right)}
		\\&=
		\left\{\exp{\left(-\sum_{c=1}^{N-1}e_{cN}^{1}\right)}\sum_{A,B=1}^{N}\alpha_{A}\left(e_{BA}^{1}-e_{BB}^{1}\right)\exp{\left(\sum_{c=1}^{N-1}e_{cN}^{1}\right)}\right\}^{T}
		\\&=
		\sum_{a=1}^{N-1}\left(\alpha_{a}e_{Na}^{1}-e_{aa}^{1}\right)^{T}
		\\&=
		\sum_{a=1}^{N-1}\left(\alpha_{a}e_{aN}^{1}-e_{aa}^{1}\right)\dt
	\end{split}
\end{equation}
Similarly, for the right boundary we have 
\begin{equation}
H_{\text{right}}^{'}=S_{1}H_{\text{right}}S_{1}^{-1}=
\sum_{a=1}^{N-1}\left(\beta_{a}e_{aN}^{L}-e_{aa}^{L}\right)\dt
\end{equation}
\begin{remark}
	The boundary Hamiltonians $H_{\text{left}}^{'}$ and $H_{\text{right}}^{'}$ in \eqref{eq:Hprimes} resemble the transposed of  $\widetilde{H}_{\text{left}}$ and $\widetilde{H}_{\text{right}}$, i.e. to the transposed of the boundary part of the dual Hamiltonian  defined in \eqref{siteDualHamiltonian}. They can be identified  when replacing the  extra-site bosonic creation operators $(\mathbf{a}^{+})_{a}^{0}$  (resp. $(\mathbf{a}^{+})_{a}^{L+1}$) with the corresponding reservoir parameters $\alpha_a$ (resp. $\beta_a$).
\end{remark}
We have the following correspondence between eigenvectors with zero eigenvalues
\begin{equation}\label{S1-Inverse}
	|\Psi\rangle =S_{1}^{-1}|\Psi^{'}\rangle
\end{equation}
where $|\Psi^{'}\rangle$ satisfies $H^{'}|\Psi^{'}\rangle=0$.

We introduce
\begin{align} 
	\mathcal{S}_{2}:=\exp{\left(-\sum_{a=1}^{N-1}\beta_{a}e_{aN}\right)}\dt\label{S2Gen}
\end{align}
and
 \begin{equation}
 	\mathcal{S}_{2}^{x}:=\exp{\left(-\sum_{a=1}^{N-1}\beta_{a}e_{aN}^{x}\right)}\co
 \end{equation}
which denotes $\mathcal{S}_{2}$ (see \eqref{S2Gen}) acting at site $x$. By taking the product over the chain we obtain 
\begin{equation}\label{S2-Whole}
	S_{2}=\prod_{x=1}^{L}\mathcal{S}_{2}^{x}\dt
\end{equation}
We define 
\begin{equation}\label{HSecond}
	H^{''}=H_{\text{left}}^{''}+H_{\text{bulk}}+H_{\text{right}}^{''}\co
\end{equation}
that is related to $H'$ by
\begin{equation}
	H^{''}=S_{2}H^{'}S_{2}^{-1}\dt
\end{equation}
The boundary Hamiltonians of $H^{''}$ read
\begin{align}\label{tri-diag-boundaryHamiltonians}
	H_{\text{left}}^{''}=e_{NN}^{1}-\mathbbm{1}+\sum_{a=1}^{N-1}(\alpha_{a}-\beta_{a})e_{aN}^{1}\qquad H_{\text{right}}^{''}=e_{NN}^{L}-\mathbbm{1}\dt
\end{align}
The left one is lower triangular and depends on the differences between the boundary parameters, while the right one is diagonal. Equations \eqref{tri-diag-boundaryHamiltonians} are proved using the Hadamard formula \eqref{HadamardFormula} with $\mathcal{B}=\sum_{c=1}^{N-1}\beta_{c}e_{cN}$, $\mathcal{A}_{N}=e_{aN},$ and $\mathcal{A}_{a}=e_{aa}$.

Therefore, using \eqref{HT-BA} we obtain for the left boundary 
\begin{equation}
	\begin{split}
		H_{\text{left}}^{''}&=S_{2}H_{\text{left}}^{'}S_{2}^{-1}
		=
		\sum_{a=1}^{N-1}\left(\alpha_{a}e_{aN}^{1}-e_{aa}^{1}-\beta_{a}e_{aN}^{1}\right)
		=
		e_{NN}^{1}-\mathbbm{1}+\sum_{a=1}^{N-1}(\alpha_{a}-\beta_{a})e_{aN}^{1}\dt
	\end{split}
\end{equation}
Similarly, using again \eqref{HT-BA} we have for the right boundary 
\begin{equation}
	\begin{split}
			H_{\text{right}}^{''}=S_{2}H_{\text{right}}^{'}S_{2}^{-1}
			=
			\sum_{a=1}^{N-1}\left(\beta_{a}e_{aN}^{L}-e_{aa}^{L}-\beta_{a}e_{aN}^{L}\right)=
			e_{NN}^{L}-\mathbbm{1}\dt
	\end{split}
\end{equation}
The relation between eigenvectors with vanishing eigenvalue is 
\begin{equation}\label{S2-Inverse}
	|\Psi^{'}\rangle = S_{2}^{-1}|\Psi^{''}\rangle
\end{equation}
and $|\Psi^{''}\rangle$ solves
\begin{equation}\label{steadyS-SECOND-def}
	H^{''}|\Psi^{''}\rangle=0\dt
\end{equation}
It is convenient to introduce also the transformation
\begin{equation}\label{similarity}
	S=S_{2}S_{1}=\prod_{x=1}^{L}\exp{\left(-\sum_{a=1}^{N-1}\beta_{a}e_{aN}^{x}\right)}\exp{\left(\sum_{a=1}^{N-1}e_{aN}^{x}\right)}\dt
\end{equation}
This matrix $S$ connects $H$ with $H^{''}$ by 
\begin{equation}
H=S^{-1}H^{''}S
\end{equation}
and, the relation between eigenvectors with vanishing eigenvalue is
\begin{equation}
	|\Psi\rangle=S^{-1}|\Psi^{''}\rangle\dt
\end{equation}

\subsubsection{Closed formula for $|\Psi^{''}\rangle$}\label{subsectionSSHsec}

The ground state of the Hamiltonian $H^{''}$ is given by 
\begin{equation}\label{ResulsBasis}
	|\Psi^{''} \rangle= \sum_{\bm{\tau}\in \Omega^{'}}\frac{1}{\left(1+L\right)!}\prod_{x=1}^{L} (\alpha_{\tau_{x}}-\beta_{\tau_{x}})^{(1-\delta_{\tau_{x},N})}\left(1+\sum_{j=x}^{L}\delta_{\tau_{j},N}\right) |\bm{\tau}\rangle
\end{equation}

where 
\begin{equation}
    \delta_{\tau_{x},N}:=\begin{cases}
        1\quad \text{if}\quad \tau_{x}=N\\
        0\quad \text{otherwise}
    \end{cases}\dt
\end{equation}

\paragraph{Proof of formula \eqref{ResulsBasis}}  
We consider the vector with elements given by the operators of the MPA $(X_{1},\ldots,X_{N})$ and we act on it with the similarity transformation $S=S_{2}S_{1}$ obtaining new operators $(\Xt_{1},\ldots,\Xt_{N})$ that will satisfy simpler commutation relations. 
We define the transformed matrix product operators
via $\tilde X=S X$ such that
\begin{equation}\label{Xtildes2b}
 \tilde X_a=X_{a}-\beta_{a}(X_{a}+\ldots+X_{N})\,,\qquad \tilde X_N= 
				X_{1}+\ldots +X_{N}
\end{equation} 
where  $ a\in \{1,\ldots,N-1\}$.
We can also reverse the transformation by $S^{-1}$ and get: 
\begin{equation}\label{Xes}
 X_{a}=	\Xt_{a}+\beta_{a}\Xt_{N}\,,\qquad
	X_{N} = \beta_N\Xt_{N}-(\Xt_{1}+\ldots+\Xt_{N-1})\,.
\end{equation}

Summing over $b\in \{1,\ldots,N\}$ in commutation relations \eqref{bulk} we have
\begin{equation} 
	\left[X_{a},\Xt_{N}\right]=(\alpha_{a}-\beta_{a})\Xt_{N}
\end{equation}
with $\widetilde{X}_{N}= {X}_{1}+\ldots+ {X}_{N}$. 
Therefore, using \eqref{Xes}
we obtain the commutation relations for $\Xt_{a}$ 
\begin{equation}\label{commutationsBula}
	\left[\Xt_{a},\Xt_{N}\right]=(\alpha_{a}-\beta_{a})\Xt_{N}\co
\end{equation} 
cf.~\eqref{commutationsBula2}.
Moreover using \eqref{Xtildes2b} and \eqref{Xes}, the action of $\Xt_{a}$ on the boundary vectors are given by 
\begin{equation}\label{commLEFT}
	 \langle \langle W|  \left((\alpha_{a}-\beta_{a})\Xt_{N}-\Xt_{a}\right)=(\alpha_{a}-\beta_{a}) \langle \langle W|  
	\co
\end{equation}
\begin{equation}\label{commRIGHT}
	\Xt_{a}  |V\rangle\rangle = (\alpha_{a}-\beta_{a}) |V\rangle\rangle 
	\co
\end{equation}  
cf.~\eqref{commLEFT2}.
Using the transformed operators $(\widetilde{X}_{a})_{a\in\{1,\ldots,N\}}$, the vector $|\Psi^{''}\rangle$ of the matrix product ansatz is written as
\begin{equation}\label{toDetermineTilde}
	|\Psi^{''}\rangle = \frac{1}{Z_{L}}\sum_{\bm{\tau}\in \Omega^{'}} \langle \langle W| \widetilde{X}_{\tau_{1}}\cdots \widetilde{X}_{\tau_{L}}
	|V\rangle\rangle \,|\bm{\tau}\rangle
\end{equation} 
where
the normalization is given by 
\begin{align}
		Z_{L}&= \langle \langle W|  (X_{1}+\ldots+X_{N})^{L} |V\rangle\rangle = \langle \langle W|  \widetilde{X}_{N}^{L} |V\rangle\rangle \dt
\end{align}

To determine the eigenvector \eqref{toDetermineTilde} in closed form, we compute the coefficients and the normalization $Z_{L}$. Using \eqref{commutationsBula} we have
\begin{equation}\label{UsefulRelation}
	\widetilde{X}_{a}\widetilde{X}_{N}=\widetilde{X}_{N}\left((\alpha_{a}-\beta_{a})+\widetilde{X}_{a}\right)\dt
\end{equation}
Fix  $ \ell,n\in \mathbb{N}$. By applying the above formula \eqref{UsefulRelation} we have
\begin{align}\label{usefulRelaton-many}
	\widetilde{X}_{a}^{n}\widetilde{X}_{N}^{\ell}&=
	\widetilde{X}_{N}^{\ell}\left(\widetilde{X}_{a}+\ell(\alpha_{a}-\beta_{a})\right)^{n}\dt
\end{align}

Using \eqref{usefulRelaton-many} we obtain
\begin{equation}
	\sideset{}{'}\prod_{x=1}^L\widetilde{X}_{\tau_{x}}=\widetilde{X}_{N}^{\sum_{x=1}^{L}\delta_{\tau_{x},N}}\sideset{}{'}\prod_{x=1}^L\left(\widetilde{X}_{\tau_{x}}+(\alpha_{\tau_{x}}-\beta_{\tau_{x}})\sum_{j=x}^{L}\delta_{\tau_{j},N}\right)^{1-\delta_{\tau_{x},N}}\co
\end{equation}
where for convenience we introduce the ordered product
\begin{equation}
 \sideset{}{'}\prod_{x=1}^L \widetilde{X}_{\tau_{x}}= \widetilde{X}_{\tau_{1}}\cdots  \widetilde{X}_{\tau_{L}}\,.
\end{equation} 
Multiplying by the boundary vectors and by using \eqref{commRIGHT} we have 
\begin{align}
	 \langle \langle W| \widetilde{X}_{\tau_{1}}\cdots  \widetilde{X}_{\tau_{L}}
	|V \rangle\rangle=\langle \langle W|  \widetilde{X}_{N}^{\sum_{x=1}^{L}\delta_{\tau_{x},N}} |V\rangle\rangle
	\prod_{x=1}^{L}\left(\alpha_{\tau_{x}}-\beta_{\tau_{x}}\right)^{1-\delta_{\tau_{x},N}}\left(1+\sum_{j=x}^{L}\delta_{\tau_{j},N}\right)^{1-\delta_{\tau_{x},N}}   \dt
\end{align}
We now compute $ \langle \langle W|  \widetilde{X}_{N}^{\sum_{i=1}^{L}\delta_{\tau_{x},N}} |V\rangle\rangle $. For all $n\in \mathbb{N}$ we have that  
\begin{align}
	 \langle \langle W|  \widetilde{X}_{N}^{n} |V\rangle\rangle &= \langle \langle W|  \widetilde{X}_{N}\widetilde{X}_{N}^{n-1} |V\rangle\rangle = \langle \langle W|  \widetilde{X}_{N}^{n-1} |V\rangle\rangle  + \langle \langle W|  \frac{1}{(\alpha_{a}-\beta_{a})}\widetilde{X}_{a}\widetilde{X}_{N}^{n-1} |V\rangle\rangle \nonumber
	\\&=
	 \langle \langle W|  \widetilde{X}_{N}^{n-1} |V\rangle\rangle +\frac{1}{(\alpha_{a}-\beta_{a})} \langle \langle W|  \widetilde{X}_{N}^{n-1}\left(\widetilde{X}_{a}+(\alpha_{a}-\beta_{a})(n-1)\right) |V\rangle\rangle \nonumber
	\\&=
	 \langle \langle W|  \widetilde{X}_{N}^{n-1} |V\rangle\rangle +\left(n+1-1\right) \langle \langle W|  \widetilde{X}_{N}^{n-1} |V\rangle\rangle \nonumber
	\\&=
	\left(2+n-1\right) \langle \langle W|  \widetilde{X}_{N}^{n-1} |V\rangle\rangle \nonumber
	\\&=
	\frac{(1+n)!}{n!} \langle \langle W|  \widetilde{X}_{N}^{n-1} |V\rangle\rangle 
\end{align}
where in the second equality we used \eqref{usefulRelaton-many}. 
This leads to the recursion relation
\begin{align}
		 \langle \langle W|  \widetilde{X}_{N}^{n} |V\rangle\rangle =\frac{(1+n)!}{n!} \langle \langle W|  \widetilde{X}_{N}^{n-1} |V\rangle\rangle 
\end{align}
with $	 \langle \langle W|  \widetilde{X}_{N}^{0} |V\rangle\rangle=	 \langle \langle W |V\rangle\rangle  =1$. 
Therefore, we obtain 
\begin{align}
	 \langle \langle W|  \widetilde{X}_{N}^{n} |V\rangle\rangle &=
	\frac{(1+n)!}{n!}\frac{n!}{(n-1)!}\ldots \frac{3!}{2!}\frac{2!}{1!} \langle \langle W|  \widetilde{X}_{N}^{0} |V\rangle\rangle =
	(1+n)!\dt
\end{align}
By using this result we have that
\begin{equation}
	 \langle \langle W|  \widetilde{X}_{\tau_{1}}\cdots  \widetilde{X}_{\tau_{L}}
	|V\rangle\rangle=(1+{\textstyle \sum}_{x=1}^{N}\delta_{\tau_{x},N})!\prod_{x=1}^{L}\left(\alpha_{\tau_{x}}-\beta_{\tau_{x}}\right)^{1-\delta_{\tau_{x},N}}\left(1+\sum_{j=x}^{L}\delta_{\tau_{j},N}\right)^{1-\delta_{\tau_{x},N}}\dt
\end{equation}
The normalization constant is then given by
\begin{align}
	Z_{L}&= \langle \langle W|  (X_{1}+\ldots+X_{N})^{L} |V\rangle\rangle = \langle \langle W|  \widetilde{X}_{N}^{L}|V\rangle \rangle=
	(L+1)!\dt
\end{align}
Therefore, we write
\begin{equation}\label{resulEsteady}
	|\Psi^{''}\rangle= \sum_{\bm{\tau}\in \Omega^{'}}\frac{(1+\sum_{x=1}^{L}\delta_{\tau_{x},N})!}{(L+1)!}\prod_{x=1}^{L}\left(\alpha_{\tau_{x}}-\beta_{\tau_{x}}\right)^{1-\delta_{\tau_{x},N}}\left(1+\sum_{j=x}^{L}\delta_{\tau_{j},N}\right)^{1-\delta_{\tau_{x},N}}|\bm{\tau}\rangle\dt
\end{equation}
We observe that, for all $x$, we have
\begin{equation}
	\left(1+\textstyle\sum_{j=x}^{L}\delta_{\tau_{j},N}\right)^{1-\delta_{\tau_{x},N}}=\frac{(1+\sum_{j=x+1}^{L}\delta_{\tau_{j},N})!}{(\sum_{j=x}^{L}\delta_{\tau_{j},N})!}\dt
\end{equation}
It follows that 
\begin{align}\label{remove-exponent}
&\frac{(1+\sum_{x=1}^{L}\delta_{\tau_{x},N})!}{(L+1)!}\prod_{x=1}^{L}\left(\alpha_{\tau_{x}}-\beta_{\tau_{x}}\right)^{1-\delta_{\tau_{x},N}}\left(1+\sum_{j=x}^{L}\delta_{\tau_{j},N}\right)^{1-\delta_{\tau_{x},N}}\nonumber\\
=&\frac{(1+\sum_{x=1}^{L}\delta_{\tau_{x},N})!}{(L+1)!}\prod_{x=1}^{L}\frac{(1+\sum_{j=x+1}^{L}\delta_{\tau_{j},N})!}{(\sum_{j=x}^{L}\delta_{\tau_{j},N})!}\nonumber\\
=&\frac{1}{(L+1)!}\prod_{x=1}^{L}\frac{(1+\sum_{j=x}^{L}\delta_{\tau_{j},x})!}{(\sum_{j=x}^{L}\delta_{\tau_{j},x})!}\nonumber\\
=&\frac{1}{(L+1)!}\prod_{x=1}^{L}\left(1+\sum_{j=x}^{L}\delta_{\tau_{j},N}\right)\dt
\end{align}
Therefore, we obtain \eqref{ResulsBasis}.
\begin{flushright}
	$\square$
\end{flushright} 
\subsubsection{Closed formula for $|\Psi^{'}\rangle$}\label{subsectionSSdual}
By knowing the ground state  of the Hamiltonian $H^{''}$ we use \eqref{S2-Inverse} to retrieve the ground state of the Hamiltonian $H^{'}$. The result is the following 
\begin{equation}\label{ABS-vect}
    |\Psi^{'}\rangle =\sum_{\bm{\tau}\in \Omega^{'}}\Psi^{'}(\bm{\tau})|\bm{\tau}\rangle 
\end{equation}
where  
\begin{equation}\label{ABS}
		\begin{split}
			\Psi^{'}(\bm{\tau})=\frac{1}{(L+1)!}\sum_{c_{1}=\delta_{\tau_{1},N}}^{1}\ldots\sum_{c_{L}=\delta_{\tau_{L},N}}^{1}\prod_{x=1}^{L}\left(1+\sum_{j=x}^{L}c_{j}\right)(\alpha_{\tau_{x}}-\beta_{\tau_{x}})^{1-c_{x}}\beta_{\tau_{x}}^{c_{x}-\delta_{\tau_{x},N}}\dt
		\end{split}
	\end{equation} 
\paragraph{Proof of formula \eqref{ABS-vect}} We have that $|\Psi^{'}\rangle=S_{2}^{-1}|\Psi^{''}\rangle$, thus we  show how the transformation $S_{2}^{-1}$ acts on the vector $|\Psi^{''}\rangle$. Using the exponential series we have that  
\begin{equation}
   \mathcal{S}_{2}^{-1}=\mathbbm{1}+\sum_{a=1}^{N-1}\beta_{a}e_{aN}\co
\end{equation}
which action on the occupation variable $|\sigma\rangle$ of a site is 
\begin{equation}
  \mathcal{S}_{2}^{-1} |\sigma\rangle= |\sigma\rangle+\delta_{N,\sigma}\sum_{a=1}^{N-1}\beta_a|a\rangle\dt
\end{equation} 
Therefore, taking the tensor product over the chain we obtain 
\begin{equation}
 S_{2}^{-1} |\bm{\sigma}\rangle=\left( |\sigma_1\rangle+\delta_{N,\sigma_1}\sum_{a_1=1}^{N-1}\beta_{a_1}|a_1\rangle\right)\otimes\ldots\otimes \left( |\sigma_L\rangle+\delta_{N,\sigma_L}\sum_{a_L=1}^{N-1}\beta_{a_L}|a_L\rangle\right)\dt
\end{equation} 
By projecting over a vector $\langle \bm{\tau}|$ we have 
\begin{equation}\label{goodTP-S2}
  \langle \bm{\tau}|S_{2}^{-1} |\bm{\sigma}\rangle=\prod_{x=1}^L\left[ \delta_{\tau_x,\sigma_x}+\delta_{N,\sigma_x}\beta_{\tau_x} (1-\delta_{\tau_x,N})\right]\dt
\end{equation} 
From this it follows that
\begin{align}\label{ABS_intermediate}
		\Psi^{'}(\bm{\tau})=&\langle\bm{\tau}|\Psi^{'}\rangle=\langle \bm{\tau}|S_{2}^{-1}|\Psi^{''}\rangle = \sum_{\bm{\sigma}\in \Omega'}  \langle \bm{\tau}|S_{2}^{-1} |\bm{\sigma}\rangle\langle\bm{\sigma}|\Psi^{''}\rangle \nonumber\\=&\sum_{\bm{\sigma}\in \Omega^{'}}\frac{1}{(L+1)!}\prod_{x=1}^L(\alpha_{\sigma_{x}}-\beta_{\sigma_{x}})^{1-\delta_{\sigma_{x},N}}\left(1+\sum_{j=x}^{L}\delta_{\sigma_{j},N}\right)\left[ \delta_{\tau_x,\sigma_x}+\beta_{\tau_x} \delta_{N,\sigma_x} (1-\delta_{\tau_x,N})\right]\dt
	\end{align}
	where we used the expression \eqref{ResulsBasis} for $\Psi^{''}(\bm{\sigma})$.
We observe that, for any fixed $x\in \{1,\ldots,L\}$ and for any fixed $\bm{\tau},\bm{\sigma}\in\Omega^{'}$, we have 
\begin{align}
	\left[\delta_{\tau_{x},\sigma_{x}}+\delta_{\sigma_{x},N}(1-\delta_{\tau_{x},N})\beta_{\tau_{x}}\right]\left(\alpha_{\sigma_{x}}-\beta_{\sigma_{x}}\right)^{1-\delta_{\sigma_{x},N}}=	\left[\delta_{\tau_{x},\sigma_{x}}+\delta_{\sigma_{x},N}(1-\delta_{\tau_{x},N})\beta_{\tau_{x}}\right]\left(\alpha_{\tau_{x}}-\beta_{\tau_{x}}\right)^{1-\delta_{\sigma_{x},N}}\dt
\end{align}
Indeed, on one hand if $\sigma_{x}\neq N$ only the term $\delta_{\tau_{x},\sigma_{x}}$ does survive in the square brackets of the above equation, that is non-vanishing only if $\sigma_{x}=\tau_{x}$. This implies that both the left-hand-side and the right-hand-side are either $(\alpha_{\tau_{x}}-\beta_{\tau_{x}})$ or $0$. On the other hand, if $\sigma_{x}=N$ we have $(\alpha_{\sigma_{x}}-\beta_{\sigma_{x}})^{1-\delta_{\sigma_{x},N}}=(\alpha_{\tau_{x}}-\beta_{\tau_{x}})^{1-\delta_{\sigma_{x},N}}=1$, thus both sides of the equality are either $\beta_{\tau_{x}}$ or $0$.  \\
Moreover, we have that 
\begin{align}
	\delta_{\sigma_{x},\tau_{x}}+\delta_{\sigma_{x},N}(1-\delta_{\tau_{x},N})\beta_{\tau_{x}}=&\beta_{\tau_{x}}^{\delta_{\sigma_{x},N}(1-\delta_{\tau_{x},N})}\left(\delta_{\tau_{x},\sigma_{x}}+\delta_{\sigma_{x},N}(1-\delta_{\tau_{x},N})\right)\nonumber\\
	=&\beta_{\tau_{x}}^{\delta_{\sigma_{x},N}(1-\delta_{\tau_{x},N})}\left(\delta_{\tau_{x},N}\delta_{\sigma_{x},N}+(1-\delta_{\tau_{x},N})(1-\delta_{\sigma_{x},N})+(1-\delta_{\tau_{x},N})\delta_{\sigma_{x},N}\right)\nonumber\\
	=&\beta_{\tau_{x}}^{\delta_{\sigma_{x},N}(1-\delta_{\tau_{x},N})}\left(\delta_{\sigma_{x},N}+(1-\delta_{\sigma_{x},N})(1-\delta_{\tau_{x},N})\right)
\end{align}
Indeed, considering the first and last equalities we have that: if $\tau_{x}=\sigma_{x}$, both sides are $1$; if $\tau_{x}\neq N$ and $\sigma_{x}=N$ both sides are $\beta_{\tau_{x}}$; while, in all other cases, both sides are $0$. 

As a consequence we can write \eqref{ABS_intermediate} as 
\begin{align}
\Psi^{'}(\bm{\tau})
=&\frac{1}{(L+1)!}\sum_{\bm{\sigma}\in \Omega^{'}}^{N}\prod_{x=1}^{L}\left(\alpha_{\tau_{x}}-\beta_{\tau_{x}}\right)^{1-\delta_{\sigma_{x},N}}\left(1+\sum_{j=x}^{L}\delta_{\sigma_{j},N}\right)\nonumber\\&\times\left[\beta_{\tau_{x}}^{\delta_{\sigma_{x},N}(1-\delta_{\tau_{x},N})}\left(\delta_{\sigma_{x},N}+(1-\delta_{\sigma_{x},N})(1-\delta_{\tau_{x},N})\right)\right]\dt
\end{align}
We observe that the argument of the summation above does not distinguish the colours of the $\sigma$'s but only whether, at each site $x$, $\sigma_{x}$ is occupied or empty. Therefore, we can replace the summation over $\sigma_{1},\ldots,\sigma_{L}$ with a summation over the indices  $c_{1},\ldots,c_{L}\in \{0,1\}$. Thus, we obtain 
\begin{align}
\Psi^{'}(\bm{\tau})	=&\frac{1}{(L+1)!}\sum_{c_{1}=0}^{1}\ldots\sum_{c_{L}=0}^{1}\prod_{x=1}^{L}\left(\alpha_{\tau_{x}}-\beta_{\tau_{x}}\right)^{1-c_{x}}\left(1+\sum_{j=x}^{L}c_{j}\right)\left[\beta_{\tau_{x}}^{c_{x}}\left(c_{x}+(1-c_{x})(1-\delta_{\tau_{x},N})\right)\right]\dt
\end{align}
Moreover, because of the term $(1-\delta_{\tau_{x},N})(1-c_{x})$ we can make the indices $c_{x}$ vary in the set $\{\delta_{\tau_{x},N},1\}$. Indeed, when $c_{x}=0$, this term is non-vanishing only if $\tau_{x}\neq N$. Therefore, we obtain 
\begin{equation}
	\Psi^{'}(\bm{\tau})=\frac{1}{(L+1)!}\sum_{c_{1}=\delta_{\tau_{1},N}}^{1}\ldots\sum_{c_{L}=\delta_{\tau_{L},N}}^{1}\prod_{x=1}^{L}\left(\alpha_{\tau_{x}}-\beta_{\tau_{x}}\right)^{1-c_{x}}\left(1+\sum_{j=x}^{L}c_{j}\right)\beta_{\tau_{x}}^{c_{x}(1-\delta_{\tau_{x},N})}\dt
\end{equation}

Observing now that $c_{x}(1-\delta_{\tau_{x},N})$ can be replaced by  $c_{x}-\delta_{\tau_{x},N}$, we finally have
\begin{equation}\label{elementsABS}
	\begin{split}
		\Psi^{'}(\bm{\tau})=\frac{1}{(L+1)!}\sum_{c_{1}=\delta_{\tau_{1},N}}^{1}\ldots\sum_{c_{L}=\delta_{\tau_{L},N}}^{1}\prod_{x=1}^{L}(\alpha_{\tau_{x}}-\beta_{\tau_{x}})^{1-c_{x}}\left(1+\sum_{j=x}^{L}c_{j}\right)\beta_{\tau_{x}}^{c_{x}-\delta_{\tau_{x},N}}\dt
	\end{split}
\end{equation} 
\begin{flushright}
    $\square$
\end{flushright}
\begin{remark} By changing the summation indices from $c_x\in \{\delta_{\tau_{x},N},1\}$ to $c_{x}^{'}\in \{0,1-\delta_{\tau_{x},N}\}$ and by  the fact that occupation of each site is bounded by $1$ we rewrite \eqref{ABS} as 
\begin{equation}\label{ABS_1}
	\Psi^{'}(\bm{\tau})=\sum_{c^{'}_{1}=0}^{1-\delta_{\tau_{1},N}}\ldots\sum_{c^{'}_{L}=0}^{1-\delta_{\tau_{L},N}}\frac{(L+1-\sum_{x=1}^{L}c^{'}_{x})!}{(L+1)!}\prod_{x=1}^{L}\left((\alpha_{\tau_{x}}-\beta_{\tau_{x}})\left(2+L-x-\sum_{j=x}^{L}c^{'}_{j}\right)\right)^{c^{'}_{x}}\beta_{\tau_{x}}^{(1-c^{'}_{x}-\delta_{\tau_{x},N})}
\end{equation}
where we used the fact that 
\begin{equation}
	\left(1+\sum_{j=x}^{L}c_{j}\right) =\left(2+L-x-\sum_{j=x}^{L}c^{'}_{j}\right)\dt
\end{equation}
Equation \eqref{ABS_1} will be useful in the following. 
\end{remark}
\subsubsection{Closed formula for correlations}\label{correlation-section}
We consider $m\in \{1,\ldots,L\}$ sites with coordinates $x_{1},\ldots,x_{m}\in \{1,\ldots,L\}$ such that $x_{k}<x_{k+1}$ for all $k\in \{1,\ldots,m-1\}$. We fix $a_{k}\in\{1,\ldots,N-1\}$ and we would like to compute
\begin{equation}
	\mathbb{E}\left[\prod_{k=1}^{m}\mathbbm{1}_{\{Y_{x_{k}}=a_{k}\}}\right]=\sum_{\bm{\tau}\in \Omega^{'}}\left(\prod_{k=1}^{m}\mathbbm{1}_{\{\tau_{x_{k}}=a_{k}\}}\right)\langle\bm{\tau}|\Psi\rangle\dt
\end{equation}
We fix the dual configuration $\bm{\xi}=\sum_{k=1}^{m}\delta_{a_{k}}^{x_{k}}$, corresponding to a vector $|\bm{\xi}\rangle$, and we have 
\begin{equation}
	\prod_{k=1}^{m}\mathbbm{1}_{\{\tau_{x_{k}}=a_{k}\}}=D(\bm{\tau},\bm{\xi})\dt
\end{equation}
Thus it follows that 
\begin{align}
	\mathbb{E}\left[\prod_{k=1}^{m}\mathbbm{1}_{\{Y_{x_{k}}=a_{k}\}}\right]=&\sum_{\bm{\tau}\in \Omega^{'}}D(\bm{\tau},\bm{\xi})\langle\bm{\tau}|\Psi\rangle
	=\sum_{\bm{\tau}\in \Omega^{'}}\langle \bm{\tau}|D|\bm{\xi}\rangle\langle\bm{\tau}|\Psi\rangle\dt
\end{align}
We denote by $|\hat{\bm{\xi}}\rangle=|\xi^{1},\ldots,\xi^{L}\rangle$, i.e. the vector constructed from $|\bm{\xi}\rangle$ by removing the components at sites $0$ and $L+1$. Recalling that {in the case $\nu=1$ the matrix $R_{x}$ defined in \eqref{Revmatrix} reduces to the identity, we have that the matrix $d_{x}$, defined in \eqref{bulkElementDualityMatrix}, becomes
\begin{equation}
	d_{x}=\exp{(\sum_{a=1}^{N-1}e_{aN}^{x})}\dt
	\end{equation} Then, considering equation \eqref{transformationV}, the following holds}
\begin{equation}
	S_{1}^{T}=\prod_{x=1}^{L}d_{x}\dt
\end{equation}  
{As a consequence we have that }
\begin{equation}
	\langle \bm{\tau}|S_{1}^{T}|\hat{\bm{\xi}}\rangle=\langle \bm{\tau}|D|\bm{\xi}\rangle \dt
\end{equation}
Therefore, it follows that, using the resolution of the identity
\begin{align}
\label{corr-psi1}
	\mathbb{E}\left[\prod_{k=1}^{m}\mathbbm{1}_{\{Y_{x_{k}}=a_{k}\}}\right]=& \sum_{\bm{\tau}\in \Omega^{'}}\langle \bm{\tau}|D|\bm{\xi}\rangle\langle\bm{\tau}|\Psi\rangle
	=\sum_{\bm{\tau}\in \Omega^{'}}\langle \bm{\tau}|S_{1}^{T}|\hat{\bm{\xi}}\rangle\langle\bm{\tau}|\Psi\rangle\nonumber\\
	=&\sum_{\bm{\tau}\in \Omega^{'}}\langle \hat{\bm{\xi}}|S_{1}|\bm{\tau}\rangle\langle\bm{\tau}|\Psi\rangle
	=\langle \hat{\bm{\xi}}|S_{1}|\Psi\rangle
	=\langle \hat{\bm{\xi}}|\Psi^{'}\rangle\dt
\end{align}

Using $\Psi^{'}(\bm{\tau})$ in \eqref{ABS_1} we have 
\begin{align}
	\mathbb{E}\left[\prod_{k=1}^{m}\mathbbm{1}_{\{Y_{x_{k}}=a_{k}\}}\right]=&\sum_{c_{1}=0}^{1}\ldots\sum_{c_{m}=0}^{1}\frac{(L+1-\sum_{a=1}^{m}c_{a})!}{(L+1)!}\prod_{k=1}^{m}(\alpha_{a_{k}}-\beta_{a_{k}})^{c_{k}}\beta_{a_{k}}^{1-c_{k}}\left(2+L-x_{k}-\sum_{j=k}^{m}c_{j}\right)^{c_{k}}\nonumber\\
	=&\sum_{c_{1},\ldots,c_{m}=0}^{1}
	f(c_{1},\ldots,c_{m})\prod_{k=1}^{m}(\alpha_{a_{k}}-\beta_{a_{k}})^{c_{k}}\beta_{a_{k}}^{1-c_{k}}g_{k}(x_{k},c_{k},\ldots,c_{m})\co
\end{align}
where in the last equality we have used the definition of \eqref{powerCoeffNOspec} and \eqref{powerCoeffSpecies}.
As a last step, using the binomial theorem, we rewrite the previous expression as a polynomial in the boundary densities: 
\begin{align}
	&\mathbb{E}\left[\prod_{k=1}^{m}\mathbbm{1}_{\{Y_{x_{k}}=a_{k}\}}\right]\nonumber\\=&
	\sum_{c_{1},\ldots,c_{m}=0}^{1}
	f(c_{1},\ldots,c_{m})\prod_{k=1}^{m}\left(\sum_{t_{k}=0}^{c_{k}}(-1)^{c_{k}-t_{k}}\alpha_{a_{k}}^{t_{k}}\beta_{a_{k}}^{1-t_{k}}g_{k}(x_{k},c_{k},\ldots,c_{m})\right)\nonumber\\
	=&\sum_{c_{1},\ldots,c_{m}=0}^{1}
	f(c_{1},\ldots,c_{m})\sum_{t_{1}=0}^{c_{1}}\ldots\sum_{t_{m}=0}^{c_{m}}\left(\prod_{k=1}^{m}\alpha_{a_{k}}^{t_{k}}\beta_{a_{k}}^{1-t_{k}}\right)\left(\prod_{k=1}^{m}(-1)^{c_{k}-t_{k}}g_{k}(x_{k},c_{k},\ldots,c_{m})\right)\nonumber\\
	=&\sum_{t_{1},\ldots,t_{m}=0}^{1}\left(\prod_{k=1}^{m}\alpha_{a_{k}}^{t_{k}}\beta_{a_{k}}^{1-t_{k}}\right)\sum_{c_{1}=t_{1}}^{1}\ldots\sum_{c_{m}=t_{m}}^{1}f(c_{1},\ldots,c_{m})\left(\prod_{k=1}^{m}(-1)^{c_{k}-t_{k}}g_{k}(x_{k},c_{k},\ldots,c_{m})\right)\dt
\end{align}
From third to fourth line we have changed the summation order. Therefore, comparing the above formula and \eqref{absProbabilitiesIntegrable}, we can read off the absorption probabilities given in \eqref{abs-probabilities}.
\begin{flushright}
	$\square$
\end{flushright}
\begin{remark}
From equation \eqref{corr-psi1}, we observe that, for any configuration $\bm{\tau}$ that has at least a particle, we interpret the coefficient $\Psi^{'}(\tau)$ of the vector $|\Psi^{'}\rangle$ as a correlation. However, this interpretation does not cover the case when $\bm{\tau}=(N,\ldots,N)$, i.e. when no particles are present in the configuration. In Appendix \ref{subsection-interpretationPSIp} we show that, for any configuration $\bm{\tau}\in \Omega^{'}$ we have 
\begin{equation}\label{interpretation-PsiP}
	\Psi^{'}(\bm{\tau})=\mathbb{E}\left[\prod_{x=1}^{L}\left(\mathbbm{1}_{\{Y_{x}=\tau_{x}\}} \right)^{(1-\delta_{\tau_{x},N})}\right]\dt
\end{equation} 
The above formula \eqref{interpretation-PsiP} allows to interpret these coefficients as marginals of the non-equilibrium steady distribution $\mu$. As a by-product, we obtain that $\Psi^{'}(N,\ldots,N)=1$, i.e. it is the normalization of the measure $\mu$. In particular, these marginal are the non-equilibrium steady state correlations as soon as the configuration $\bm{\tau}$ does contain at least a particle. 
\end{remark}

\begin{remark} 
Using  the definition $|\Psi^{'}\rangle=S_1|\Psi\rangle$ and that
\begin{equation}
 \langle \tau |S_1=\begin{cases}
                    \sum_{\sigma=1}^N\langle \sigma|  \quad\text{if}\quad \tau=N\\
                    \langle \tau| \qquad\qquad\text{else} 
                   \end{cases}\,,
\end{equation} 
  we immediately verify 
\begin{equation}\label{eq:EE}
\begin{split}
\mathbb{E}\left[\prod_{k=1}^{m}\mathbbm{1}_{\{Y_{x_{k}}=a_{k}\}}\right]=\sum_{\bm{\tau}\in\Omega^{'}}\prod_{k=1}^{m}\delta_{{{\tau}_{x_k}},a_{k}}\langle \bm{\tau}|\Psi\rangle=
\langle N,\ldots,N,\underset{\uparrow}{a_1}_{x_{1}},N,\ldots,N,\underset{\uparrow}{a_m}_{x_{m}},N,\ldots,N|\Psi^{'}\rangle\dt
 \end{split}
\end{equation} 
{Here the notation means that the entries $a_k\neq N$ in \eqref{eq:EE} are at position $x_k$ for $k\in\{1,\ldots,m\}$.} 
\end{remark}

\subsubsection{Steady state}\label{appB}
By the knowledge of the closed formula for correlations \eqref{exactCorrelations}, we find the probability mass-function in the non-equilibrium steady state by using Corollary~\ref{Corollary-NESS-ABS}. Indeed we show here that 
$|\Psi\rangle=S_{1}^{-1}|\Psi^{'}\rangle$ is nothing but Corollary~\ref{Corollary-NESS-ABS}.
On a single site we have that 
\begin{equation}\label{S1-L1}
	\mathcal{S}_{1}^{-1}=\mathbbm{1}+(-1)\sum_{a=1}^{N-1}e_{Na}
\end{equation}
thus,
\begin{equation}
	\mathcal{S}_{1}^{-1}|\sigma\rangle=|\sigma\rangle+(-1)(1-\delta_{\sigma,N})|N\rangle
\end{equation}
By taking the tensor product over the chain we obtain 
\begin{equation}
	S_{1}^{-1}|\bm{\sigma}\rangle=\left(|\sigma_{1}\rangle+(-1)(1-\delta_{\sigma_{1},N})|N\rangle\right)\otimes \ldots \otimes \left(|\sigma_{L}\rangle+(-1)(1-\delta_{\sigma_{L},N})|N\rangle\right)
\end{equation}
therefore, projecting on $\langle \bm{\tau}|$ we have
\begin{equation}\label{Tensor-S1}
	\langle \bm{\tau}|	S_{1}^{-1}|\bm{\sigma}\rangle=\prod_{x=1}^{L}\left(\delta_{\tau_{x},\sigma_{x}}+(-1)\delta_{\tau_{x},N}\sum_{a=1}^{N-1}\delta_{\sigma_{x},a}\right)\dt
\end{equation}
	By inserting the resolution of identity $\sum_{\bm{\sigma}\in \Omega^{'}} |\bm{\sigma}\rangle\langle\bm{\sigma}|$ we have that 
{
	\begin{align}
		\Psi(\bm{\tau})=\langle \bm{\tau}|\Psi\rangle=\langle \bm{\tau}|S^{-1}_{1}|\Psi^{'}\rangle=\sum_{\bm{\sigma}\in \Omega^{'}}\langle \bm{\tau}|S_{1}^{-1}|\bm{\sigma}\rangle \langle \bm{\sigma}|\Psi^{'}\rangle\dt
	\end{align} 
	Using \eqref{Tensor-S1} we obtain 
			 \begin{align}\label{implicit-PSI-PSIp}
			\Psi(\bm{\tau})=&\sum_{m=0}^{L}\sum_{1\leq x_{1}<x_{2}<\ldots<x_{m}\leq L}\left(\prod_{x\notin\{x_{1},\ldots,x_{m}\}}\delta_{\tau_{x},N}\right)\sum_{b_{1},\ldots,b_{m}=1}^{N-1}\left(\prod_{k=1}^{m}(\delta_{\tau_{x_{k}},b_{k}}-\delta_{\tau_{x_{k}},N})\right)\nonumber\\
			\times &			
			\Psi^{'}(N,\ldots,N,\underset{\uparrow}{b_{1}}_{x_{1}},N,\ldots,N,\underset{\uparrow}{b_{m}}_{x_{m}},N,\ldots,N)
		\end{align} 
			In the second equality we have rewritten the product by introducing the summation over $m$, the coordinates $x_{1},\ldots,x_{m}$ and the occupation variables $b_{1},\ldots,b_{m}$. Here, the notation
		\begin{equation}\label{notation-PSI}
			\Psi^{'}(N,\ldots,N,\underset{\uparrow}{b_{1}}_{x_{1}},N,\ldots,N,\underset{\uparrow}{b_{m}}_{x_{m}},N,\ldots,N)
		\end{equation}
		indicates that in positions $x_{1},\ldots,x_{m}$ there are particles of species $b_{1},\ldots,b_{m}$, while there are holes $N$ in all other positions.} 
\begin{remark}[Equilibrium case]
	Using equation \eqref{implicit-mass-function} with correlation computed in \eqref{exactCorrelations} in the equilibrium case ($\beta_{a}=\alpha_{a}$), we obtain 
	\begin{equation}
		\Psi(\bm{\tau})=\prod_{a=1}^{N}\alpha_{a}^{\sum_{x=1}^{L}\delta_{\tau_{x},a}}
	\end{equation}
As it has to be, this is the probability of the configuration $\bm{\tau}$ under the reversible measure \eqref{reversibleMeasure}, with site marginals distributed as $\text{Multinomial }(1,\alpha_{1},\ldots,\alpha_{N})$. 
\end{remark}
\subsection{The single species case (SSEP)}
The computations made for the multispecies stirring process can be specialized to the single species case $N=2$, {i.e. when only one type of particles and the holes are present.} In this situation we retrieve the SSEP, that was studied using the same approach as presented here in \cite{frassek2020eigenstates}. We report the Hamiltonians and the similarity transformations of SEEP: 
\begin{equation}
	H=H_{\text{left}}+\sum_{x=1}^{L-1}H_{x,x+1}+H_{\text{right}}
\end{equation}
where the bulk Hamiltonian densities are 
\begin{equation}
	H_{x,x+1}=\sum_{a,b=1}^{2}\left(e_{ab}\otimes e_{ba}-e_{bb}\otimes e_{aa}\right)\dt
\end{equation}
Here $e_{ab}$ are the basis elements of the fundamental representation of $gl(2)$. The boundary Hamiltonians are
\begin{equation}
	H_{\text{left}}=\begin{pmatrix}
		\alpha_{1}-1&\alpha_{1}\\
		\alpha_{2}&\alpha_{2}-1
	\end{pmatrix},\qquad H_{\text{right}}=\begin{pmatrix}
	\beta_{1}-1&\beta_{1}\\
	\beta_{2}&\beta_{2}-1
\end{pmatrix}\dt
\end{equation} 
where we assume that $(\alpha_{1}+\alpha_{2})=(\beta_{1}+\beta_{2})=1$. In this case, the similarity transformations read
\begin{equation}
	\mathcal{S}_{1}=\begin{pmatrix}
		1&0\\
		1&1
	\end{pmatrix},\qquad \mathcal{S}_{2}=\begin{pmatrix}
	1&-\beta_{1}\\
	0&1
\end{pmatrix},\qquad \mathcal{S}=\mathcal{S}_{2}\mathcal{S}_{1}=\begin{pmatrix}
1-\beta_{1}&-\beta_{1}\\
1&1
\end{pmatrix}\dt
\end{equation}
Therefore, we have that $H^{'}$ has the following boundaries
\begin{equation}
	H^{'}_{\text{left}}=\begin{pmatrix}
		-1&\alpha_{1}\\
		0&0
	\end{pmatrix},\qquad H^{'}_{\text{right}}=\begin{pmatrix}
	-1&\beta_{1}\\
	0&0
\end{pmatrix}\co
\end{equation}
while the boundary Hamiltonians of $H^{''}$ are
\begin{equation}
	H^{''}_{\text{left}}=\begin{pmatrix}
		-1&\alpha_{1}-\beta_{1}\\
		0&0
	\end{pmatrix},\qquad H^{''}_{\text{right}}=\begin{pmatrix}
	-1&0\\
	0&0
\end{pmatrix}\dt
\end{equation}
{As a check of our results we report the steady state correlations and the vectors $|\Psi^{''}\rangle$, $|\Psi^{'}\rangle$ and $|\Psi\rangle$ for the SSEP, i.e. the situation where the process has only one species of particles and holes, with maximal occupancy $\nu=1$ and compare the results to \cite{frassek2020eigenstates}. }
\paragraph{Correlations of SSEP.}
We retrieve the closed formula for correlations in the non-equilibrium steady state found in \cite[(4.26)]{frassek2020eigenstates}:  
\begin{equation}\label{correlations-SSEP}
	\mathbb{E}\left[\prod_{k=1}^{m}\mathbbm{1}_{\{Y_{x_{k}}=1\}}\right]=\sum_{q=0}^{m}(\alpha-\beta)^{q}\beta^{m-q}\sum_{1\leq \ell_{1}<\ldots<\ell_{q}\leq m}\prod_{k=1}^{q}\frac{\left(L+1-x_{\ell_{k}}-q+k\right)}{(2+L-k)}\dt
\end{equation} 
\textbf{Proof of formula \eqref{correlations-SSEP}:} Starting from \eqref{exactCorrelations} we exploit {the fact that for $N=2$  each site can be either occupied by a particle or empty.} 
We introduce $q=\sum_{k=1}^{m}c_{x_{k}}$. There are many values of $c$'s that give the same $q$, therefore we define $q$ variables denoted by  
$1\leq \ell_{1}<\ell_{2}<\ldots<\ell_{q}\leq m$.  
To each of these variables we associate coordinates $x_{\ell_{1}},\ldots,x_{\ell_{q}}\in\{x_{1},\dots,x_{m}\}$ that are such that $c_{x_{\ell_{k}}}=1$ for all $k\in \{1,\ldots,q\}$ and $c_{x}=0$ for $x\notin \{x_{\ell_{1}},\ldots,x_{\ell_{q}}\}$. Therefore, for any fixed $c_{x_{1}},\ldots,c_{x_{m}}$, we find $q$ { (by the summation $\textstyle{\sum_{k=1}^{m}}c_{x_{k}}=q$)} 
 and $\ell_{1},\ldots,\ell_{q}$ from which we fix $x_{\ell_{1}},\ldots,x_{\ell_{q}}$ { as explained.} Then, we obtain  
	\begin{equation}
		\prod_{k=1}^{m}\left(2+L-x_{k}-\sum_{j=k}^{m}c_{j}\right)^{c_{k}}=\prod_{k=1}^{q}\left(1+L-x_{\ell_{k}}-q+k\right)\dt
	\end{equation}
Indeed, in the product on the left hand side only $q$ terms at coordinates $x_{\ell_{1}},\ldots x_{\ell_{q}}$ do survive. Moreover, any fixed $x_{\ell_{k}}$ has $q-k$ occupied sites at its right.
Furthermore, using the properties of the factorials, we have that 
\begin{equation}\label{Gamma-product}
	\frac{(L+1-\sum_{k=1}^{m}c_{k})!}{(L+2)!}=\prod_{k=1}^{q}\frac{1}{(2+L-k)}\dt
\end{equation}
Varying all possible $c$'s in \eqref{exactCorrelations} is equivalent of varying all possible $q$ and all possible $\ell_{1},\ldots,\ell_{q}$, therefore we obtain 
\begin{equation}
	\sum_{c_{1}\ldots,c_{m}=0}^{m}\frac{(L+1-\sum_{k=1}^{m}c_{k})!}{(L+1)!}\prod_{k=1}^{m}\left(2+L-x_{k}-\sum_{j=k}^{m}c_{j}\right)=\sum_{q=0}^{m}\sum_{1\leq \ell_{1}<\ldots<\ell_{q}\leq L}\prod_{k=1}^{q}\frac{(1+L-x_{\ell_{k}}-q+k)}{(2+L-k)}
\end{equation} 
and \eqref{correlations-SSEP} follows. 
\begin{flushright}
	$\square$
\end{flushright}
\paragraph{The vector $|\Psi^{''}\rangle$ for SSEP.} The ground state of $H^{''}$ is 
\begin{equation}\label{psiII-SSEP}
	|\Psi^{''}\rangle=\sum_{q=0}^{L}\sum_{1\leq \ell_{1}<\ldots<\ell_{q}\leq L}(\alpha_{1}-\beta_{1})^{q}\prod_{k=1}^{q}\frac{\left(1+L-\ell_{k}-q+k\right)}{(2+L-k)}|\bm{\ell},q\rangle
\end{equation}
where $\bm{\ell}=(\ell_{1},\ldots,\ell_{q})$. \\ \\
\textbf{Proof of formula \eqref{psiII-SSEP}:} We specialize equation \eqref{resulEsteady} to the {$N=2$ case} and introduce $q=\sum_{x=1}^{L}\delta_{\tau_{x},1}$. For each $q\in \{1,\ldots,L\}$ we call $1\leq\ell_{1}<\ldots,\ell_{q}\leq L$ the coordinates where particles of type $1$ are present. Thus, arguing as in the proof of \eqref{correlations-SSEP} and using \eqref{Gamma-product} we have that 
\begin{align}
	&\frac{\left(1+\sum_{x=1}^{L}\delta_{\tau_{x},2}\right)!}{\left(1+L\right)!}\prod_{x=1}^{L}\left[(\alpha_{\tau_{x}}-\beta_{\tau_{x}})\left(1+\sum_{j=x}^{L}\delta_{\tau_{j},2}\right)\right]^{1-\delta_{\tau_{x},2}}\nonumber\\
	=&\frac{\left(1+L-\sum_{x=1}^{L}(1-\delta_{\tau_{x},2})\right)!}{\left(1+L\right)!}\prod_{x=1}^{L}\left[(\alpha_{\tau_{x}}-\beta_{\tau_{x}})\left(2+L-x-\sum_{j=x}^{L}(1-\delta_{\tau_{j},2})\right)\right]^{1-\delta_{\tau_{x},2}}\nonumber\\
	=&(\alpha_{1}-\beta_{1})^{q}\prod_{k=1}^{q}\frac{\left(1+L-\ell_{k}-q+k\right)}{(2+L-k)}\dt
\end{align}
By varying $q$ and $\ell_{1},\ldots,\ell_{q}$, we have a one-to-one mapping with $\bm{\tau}$, therefore we obtain \eqref{psiII-SSEP}.
\begin{flushright}
	$\square$
\end{flushright}
\paragraph{The vector $|\Psi^{'}\rangle$ for SSEP.} The ground state of $H^{'}$ reads
\begin{equation}
	|\Psi^{'}\rangle=\sum_{\bm{\tau}\in\Omega^{'}}\Psi^{'}(\bm{\tau})|\bm{\tau}\rangle
\end{equation}
where 
\begin{equation}\label{PsiI-N2}
	\Psi^{'}(\bm{\tau})=\sum_{q=0}^{|\bm{\tau}|}(\alpha-\beta)^{q}\beta^{L-q}\sum_{1\leq \ell_{1}<\ldots<\ell_{q}\leq |\bm{\tau}|}\prod_{k=1}^{q}\frac{\left(L+1-x_{\ell_{k}}-q+k\right)}{(2+L-k)}\dt
\end{equation}
and where $|\bm{\tau}|=\sum_{x=1}^{L}\delta_{\tau_{x},1}$ . \\ \\
 \textbf{Proof of formula \eqref{PsiI-N2}:}  For any configuration $\bm{\tau}\in \Omega$ we denote by  $|\bm{\tau}|=\sum_{x=1}^{L}\delta_{\tau_{x},2}$ the number of occupied sites. Then, considering equation \eqref{ABS_1} we have the result by adapting the proof of \eqref{correlations-SSEP} in the following way. We select $q\in\{0,\ldots,|\bm{\tau}|\}$ particles among the occupied ones. We call $\ell_{1},\ldots\ell_{q}\in \{1,\ldots,|\tau|\}$ such that $\ell_{1}<\ell_{2}<\ldots<\ell_{q}$. These $(\ell_{k})_{k\in {1,\ldots,q}}$ are labels for the $q$ particles previously selected. For each $\ell_{k}$ we introduce $x_{\ell_{k}}$, i.e. the coordinate where the particle with label $\ell_{k}$ is placed. Thus, we have that  
\begin{equation}
	\prod_{x=1}^{L}\left(L+2-x-\sum_{j=x}^{L}c_{j}\right)^{c_{x}}=\prod_{k=1}^{q}\left(L+2-x_{\ell_{k}}-\sum_{j=x_{\ell_{k}}}^{L}c_{j}\right)=\prod_{k=1}^{q}\left(1+L-x_{\ell_{k}}-q+k\right)\dt
\end{equation}
It follows that
\begin{equation}
	\sum_{c_{1}=0}^{1}\ldots\sum_{c_{L}=0}^{1}\mathbbm{1}_{\{c_{1}+\cdots+c_{L}=q\}}	\prod_{x=1}^{L}\left(L+2-x-\sum_{j=x}^{L}c_{j}\right)^{c_{x}}=\sum_{1\leq \ell_{1}<\ldots<\ell_{q}\leq |\bm{\tau}|}\prod_{k=1}^{q}\left(1+L-x_{\ell_{k}}-q+k\right)\dt
\end{equation} 
Therefore, using \eqref{Gamma-product}  one finds $|\Psi^{'}\rangle$ for the SSEP.
\begin{flushright}
	$\square$
\end{flushright}
\paragraph{The vector $|\Psi\rangle$ for SSEP.}
In the single species case {$N=2$} we have that  
\begin{equation}\label{NESS-SSEP}
	|\Psi\rangle=\sum_{\bm{\tau}\in \Omega^{'}}\Psi(\bm{\tau})|\bm{\tau}\rangle
\end{equation}
where
\begin{equation}
	\Psi(\bm{\tau})=\sum_{b=0}^{L}(-1)^{b}\sum_{1\leq q_{1}<\ldots<q_{b}\leq L}\left(\prod_{k=1}^{q}\delta_{\tau_{q_{k}},2}\right)\sum_{s=0}^{|\bm{\tau}|+b}(\alpha-\beta)^{s}\beta^{L-s}\sum_{1\leq \ell_{1}<\ldots<\ell_{s}\leq |\bm{\tau}|}\prod_{k=1}^{s}\frac{\left(L+1-x_{\ell_{k}}-s+k\right)}{(2+L-k)}\dt
\end{equation}

\textbf{Proof of \eqref{NESS-SSEP}:} the proof follows from \eqref{implicit-PSI-PSIp} and from the coefficients $\Psi^{'}(\bm{\tau})$ computed in \eqref{PsiI-N2}. 
\begin{flushright}
	$\square$
\end{flushright}

\section{Extensions beyond the integrable chain}
\label{sec5}
In Section~\ref{sectionIntegrabiliy} we have shown that, combining duality and integrability of the multispcies stirring process with $\nu=1$ and on the geometry of chain with two boundaries, the non-equilibrium steady state can be found in an explicit form. As soon as $\nu>1$ the integrability property is lost, but duality still holds. Therefore, in this section we first address the  question whether we can use duality to characterize the non-equilibrium steady state of the boundary driven multispecies stirring process on a chain with $\nu>1$ via absorption probabilities. For the sake of simplicity and  for the sake of clarity, we keep the geometry of a chain with two reservoirs attached to the end sites $1$ and $L$ of the chain. However, this question can be extended to the geometry of the graph $G$ introduced in Section~\ref{sec1}.  \\
Inspired by the results for the thermalized version of the symmetric exclusion process reported in \cite{carinci2013duality}, in this section we also investigate its multispecies counterpart. More precisely, we ask whether a thermalized version of the boundary driven multispecies stirring process can be defined and if it keeps the absorbing duality properties as in the single species case. Again, we restrict our analysis on the chain with two reservoirs at the end sites. Finally we add a reaction mechanism of transition to the boundary driven multispecies stirring process. The aim is to answer the following question: Can we generalize to the case with $\nu>1$ with arbitrary number of species the boundary driven model introduced in \cite{casini2022uphill}? Moreover, in \cite{casini2022uphill} the duality for this reaction-diffusion process was proved only for the equilibrium case and the result was achieved by direct computation (without using a $gl(3)$ symmetry). Thus, we also investigate if this last duality result can be extended to a boundary driven process by using an approach similar to the one of Section~\ref{sectionDuality} and based on Lie algebras. The section is organized as follows: 
\begin{enumerate}
	\item \textit{Correlations} (Section~\ref{Subsection-ss-nonI}): exploiting the duality, the steady state correlations {\color{black} of the boundary driven multispecies stirring process on a chain with two reservoirs} is expressed in terms of absorption probabilities of the dual particles. These probabilities are hard to find when $\nu>1$. As an example we give, 
	in Appendix~\ref{appendix-1pts-general}, the expression for the one-point correlation.  
	\item \textit{Instantaneous thermalization} (Section~\ref{Subsection-Thermalized}): starting from the multispecies stirring dynamics, we define the thermalized process, where particles are redistributed on each bond with a measure that is the reversible one conditioned to the conservation of the number of particles on this bond. Also for this thermalized process we prove duality. 	
	\item \textit{Reaction-diffusion process} (Section~\ref{Subsection-RD}): {\color{black} starting from the usual stirring dynamics (see Section~\ref{sec1}) where particles and holes are exchanged in two nearest neighbour sites (now at a constant rate $\sigma_{11}$),} {\color{black}two additional reaction} dynamics are added. Particle can now change type {\color{black} with two different mechanism: a \textit{pure mutation} (with rate $\Upsilon$) and a \textit{stirring-mutation} (at rate $\sigma_{12}$), i.e. a transition where particles perform an exchange of their occupancies at nearest neighbour sites but, at the same time, they also mutate their own species.} {\color{black} This reaction-diffusion} model   is  constructed with the goal of having the average densities evolving as a closed system of $N-1$ coupled difference-differential equation, providing a generalization of the model studied in \cite{casini2022uphill}. {\color{black} Exploiting the $gl(N)$ symmetry of this reaction-diffusion process, absorbing duality} for the boundary driven process is also proved.
\end{enumerate}
{\color{black}In this section we always assume that the geometry is a chain of length $L$ where two reservoirs are connected to the end sites $1$ and $L$. In all the three investigated situations, we denote the original processes by the variables $\left(\bm{n}(t)\right)_{t\geq0}$, while their dual processes is denoted by the variables  $(\bm{\xi}(t))_{t\geq 0}$. Moreover, we always denote the configuration spaces of the original processes by
\begin{equation}\label{stateSpace-Chain}
	\Omega_{L}=\bigotimes_{x=1}^{L} \Omega_{x}\,.
\end{equation}
and their dual configuration spaces 
\begin{equation}\label{dualSpace-chain}
	\widetilde{\Omega}_{L}=\mathbb{N}_{0}^{N-1}\otimes \Omega_{L}\otimes \mathbb{N}_{0}^{N-1}\dt
\end{equation}
We recall that the dual processes are defined on an enlarged chin with two extra-site $0$ and $L+1$ attached to $1$ and $L$ respectively.}
\subsection{Duality for a boundary driven chain}\label{Subsection-ss-nonI}
 As already pointed out, the dual process voids the chain by piling up particles of species $\{1,\ldots,N-1\}$ in the extra sites $0$ and $L+1$ and replacing them by the holes $N$. This property allows to characterize the steady state non-equilibrium distribution (that is unique) by the probability that dual particles are absorbed in the extra-sites, called \textit{absorption probabilities}. This is a generalization for the non-integrable chain of the Corollary~\ref{Corolollary-ABS_Corr-abstract}. The process $\left(\bm{n}(t)\right)_{t\geq0}$ has generator
\begin{equation}
	\mathcal{L}=\mathcal{L}_{\text{left}}+\sum_{x=1}^{L-1}\mathcal{L}_{x,x+1}+\mathcal{L}_{\text{right}}\dt
\end{equation} 
where $\mathcal{L}_{x,x+1}$ is the edge generator defined in \eqref{edgeGenerator} acting on the bond $(x,x+1)$ and $\mathcal{L}_{\text{left}}$ and $\mathcal{L}_{\text{right}}$ are the generators \eqref{siteGenerator} acting on sites $1$ and $L$ with parameters $(\alpha_{A})_{A\in \{1,\ldots,N\}}$ and $(\beta_{A})_{A\in \{1,\ldots,N\}}$ respectively. 
The process is dual to $(\bm{\xi}(t))_{t\geq0}$, defined on the dual state space $\widetilde{\Omega}_{L}$, and has generator
\begin{equation}
	\widetilde{\mathcal{L}}=\widetilde{\mathcal{L}}_{\text{left}}+\sum_{x=1}^{L-1}\mathcal{L}_{x,x+1}+\widetilde{\mathcal{L}}_{\text{right}}
\end{equation} 
where $\mathcal{L}_{x,x+1}$ is given by \eqref{Generator} and where the boundary generators $\widetilde{\mathcal{L}}_{\text{left}}$ and $\widetilde{\mathcal{L}}_{\text{right}}$ are given in \eqref{siteDualGenerator}: they absorb particles from sites $1$ and $L$ and put them at extra-sites $0$ and $L+1$ respectively. The duality function is \eqref{dualityElements} specified on the chain. 
We call $\mu_{\text{NESS}}$ the non-equilibrium stationary steady state of the the process $(\bm{n}(t))_{t\geq 0}$, we denote by $\mathbb{E}_{\mu_{\text{NESS}}}[\cdot]$ the expectation with respect to this measure and and we call $\bm{\mathcal{Y}} = (\mathcal{Y}^{1},\ldots,\mathcal{Y}^{L})$ the random vector with law $\mu_{\text{NESS}}$. For every $m\in\{1,\ldots,L\}$ we consider {\color{black}the coordinates $x_{1},\ldots,x_{m}\in\{1,\ldots,L\}$ and we select $m$ colours $a_{1},\ldots,a_{m}\in \{1,\ldots,N-1\}$ among the $N-1$ available ones}. We introduce the dual configuration  $\bm{\xi}=\sum_{k=1}^{m}\delta_{a_{k}}^{x_{k}}$, meaning that 
\begin{equation}
	\xi_{A}^{x}=	\begin{cases}
	\sum_{y=1}^{L}\sum_{b=1}^{N-1}\mathbbm{1}_{\{y=x_{k}\}}\mathbbm{1}_{\{b=a_{k}\}}\qquad &\text{if}\quad x=x_{k}\;\text{and} \; A=a_{k}\\
		 0\qquad &\text{otherwise}
	\end{cases}\dt
\end{equation}
The duality function is \eqref{dualityElements}, adapted to the chain. Therefore, we have that
\begin{equation}
	D(\bm{n},\bm{\xi})= \frac{\prod_{k=1}^{m}n_{a_{k}}^{x_{k}}}{C(\nu,m)}\dt
\end{equation}
where $C(\nu,m)$ is a constant that depends on $m$ and {\color{black} on the maximal occupancy} per site $\nu$. 
Therefore, up to this constant, $\mathbb{E}_{\mu_{\text{NESS}}}\left[D(\bm{n},\bm{\xi})\right]$ gives the $m-$point correlations in the non-equilibrium steady state. The following equalities hold using ergodicity and duality:
\begin{equation}\label{ExptationSS}
	\begin{split}
		\frac{\mathbb{E}_{\mu_{\text{NESS}}}\left[\prod_{k=1}^{m}\mathbbm{1}_{\{\mathcal{Y}_{a_{k}}^{x_{k}}=n_{a_{k}}^{x_{k}}\}}\right]}{C(\nu,m)}&=\lim_{t\to\infty}\mathbb{E}_{\bm{n}}\left[D(\bm{n}(t),\bm{\xi})\right]=\lim_{t\to\infty}\mathbb{E}_{\bm{\xi}}\left[D(\bm{n},\bm{\xi}(t))\right]
		\\&=
		\sum_{t_{1}=0}^{|\xi_{1}|}\ldots\sum_{t_{N}=0}^{|\xi_{N}|}\prod_{a=1}^{N-1}\left(\rho_{a}^{\text{left}}\right)^{t_{a}}\left(\rho_{a}^{\text{right}}\right)^{|\xi_{a}|-t_{a}}\mathcal{P}_{x_{1},\ldots,x_{m}}(t_{1},\ldots,t_{N})
	\end{split}
\end{equation}
here we recall that, generalizing Corollary~\ref{Corolollary-ABS_Corr-abstract},
\begin{equation}\label{Pass} 
	\mathcal{P}_{x_{1},\ldots,x_{m}}(t_{1},\ldots,t_{N})	=\mathbb{P}\left(\bm{\xi}(\infty)=\sum_{a=1}^{N-1}\left(t_{a}\delta_{a}^{0}+(|\xi_{a}|-t_{a})\delta_{a}^{L+1}\right)\Big|  \bm{\xi}(0)=\bm{\xi}\right)
\end{equation}
are the absorption probabilities for the non-integrable chain, i.e. the probabilities that $t_{a}$ particles of species $a$ are absorbed at $0$. Here we have denoted $|\xi_{a}|=\sum_{x=1}^{L}\xi_{a}^{x}$.
 These correlations can be computed once the absorption probabilities are known. For example, we report in Appendix~\ref{appendix-1pts-general} the solution for the simplest case of one-point correlations that is exactly solvable, since the single dual particle behaves as an independent random walker. 
 The absorption probabilities for higher point correlations fulfil  difference equations.

\subsection{Instantaneous thermalization}\label{Subsection-Thermalized}
In this subsection we aim to extend to the multispecies stirring process the concept of \textit{instantaneous thermalization limit} (see \cite{carinci2013duality,KMP}). The thermalized process gives rise, for any pair nearest neighbour sites of the chain, to a redistribution of the total number of particles. The thermalized model that we present in this paper is obtained from the boundary driven model with generator \eqref{Generator} as follows: for each bond $(x,x+1)$ the total number of particles $\epsilon=n^{x}+n^{x+1}$ is redistributed according to the reversible measure, conditioned on the conservation of $\epsilon$. 
\subsubsection{The thermalization measure}
We introduce the reversible measure \eqref{reversibleMeasure} conditioned on the conservation of the particles on the bond $(x,x+1)$
\begin{align}
	\mu_{TH}(r|\epsilon):&=\mathbb{P}\left(n^{x}=r|n^{x}+n^{x+1}=\epsilon\right)\nonumber
	\\&
	=\mathbb{P}\left((n_{1}^{x},\ldots,n_{N}^{x})=
	(r_{1},\ldots,r_{N})\;|\;(n_{1}^{x}+n_{1}^{x+1},\ldots,n_{N}^{x}+n_{N}^{x+1})=(\epsilon_{1},\ldots,\epsilon_{N})\right)\dt
\end{align}
Here $r=(r_{1},\ldots,r_{N})$ and  $\epsilon=(\epsilon_{1},\ldots,\epsilon_{N})$ are such that $\forall A\in\{1,\ldots,N\}$ we have $r_{A}\in \{0,\ldots,\nu\}$ and $\epsilon_{A}\in \{0,\ldots,2\nu\}$, under the constraints 
\begin{equation}
	\epsilon_{N}=2\nu-\sum_{a=1}^{N-1}\epsilon_{a},\qquad r_{N}=\nu-\sum_{a=1}^{N-1}r_{a}\dt
\end{equation}
 Using the conditional probability and the fact that the reversible measure \eqref{reversibleMeasure} is the product over the sites of multinomial distributions we have
\begin{equation}
	\begin{split}
\mu_{TH}(r|\epsilon)&=\frac{\mathbb{P}\left(\left\{n^{x}=r\right\}\,\cap\,\left\{n^{x}+n^{x+1}=\epsilon\right\}\right)}{\mathbb{P}\left(n^{x}+n^{x+1}=\epsilon\right)}=\frac{\mathbb{P}\left(n^{x}=r\right)\,\mathbb{P}\left(n^{x+1}=\epsilon-r\right)}{\mathbb{P}\left(n^{x}+n^{x+1}=\epsilon\right)}\dt
	\end{split}
\end{equation}
Therefore the thermalization measure reads
\begin{equation}
	\begin{split}
	\mu_{TH}(r|\epsilon)=\frac{\binom{\nu}{r_{1},\ldots,r_{N}}\binom{\nu}{(\epsilon_{1}-r_{1}),\ldots,(\epsilon_{N}-r_{N})}}{\binom{2\nu}{\epsilon_{1}\ldots,\epsilon_{N}}}\mathbbm{1}_{\{r_{1}+\ldots+r_{N}=\nu\}}
	\end{split}
\end{equation}
where we recall that the multinomial coefficient reads
\begin{equation}
	\binom{\nu}{n_{1},\ldots,n_{N}}=\frac{\nu!}{n_{1}!\cdots n_{N}!}\dt
\end{equation}

\subsubsection{Thermalized multispecies stirring process}
The thermalized multispecies stirring process is defined on the configuration space $\Omega_{L}$ introduced in \eqref{stateSpace-Chain} and its generator reads
\begin{equation}\label{thermalizedGenerator}
	\mathcal{L}^{TH}=\mathcal{L}_{\text{left}}^{TH}+\sum_{x=1}^{L-1}\mathcal{L}_{x,x+1}^{TH}+\mathcal{L}_{\text{right}}^{TH}
\end{equation}
where 
\begin{equation}\label{bulkThermalized}
	\begin{split}
	\mathcal{L}_{x,x+1}^{TH}f(\bm{n})=&\sum_{r_{1}=0}^{n_{1}^{x}+n_{1}^{x+1}}\ldots\sum_{r_{N}=0}^{n_{N}^{x}+n_{N}^{x+1}}\mathbbm{1}_{\{r_{1}+\ldots+r_{N}=\nu\}}
	\\&\times
	\left\{f(n^{1},\ldots,n^{x-1},r,n^{x}+n^{x+1}-r,n^{x+2},\ldots,n^{L})-f(\bm{n})\right\}\mu_{TH}(\bm{r}|\bm{\epsilon})
	\end{split}
\end{equation}
while
\begin{equation}\label{boundaryThermalized}
	\begin{split}
		\mathcal{L}_{\text{left}}^{TH}f(\bm{n})=&\sum_{r_{1}=0}^{\nu}\ldots\sum_{r_{N}=0}^{\nu}\mathbbm{1}_{\{r_{1}+\ldots+r_{N}=\nu\}}\left\{f(r,n^{2},\ldots,n^{L})-f(\bm{n})\right\}\\&\times
		\binom{\nu}{r_{1},\ldots, r_{N}}\prod_{a=1}^{N-1}\left(\frac{\alpha_{a}}{\alpha_{N}}\right)^{r_{a}}\left(\frac{\alpha_{N}}{|\alpha|}\right)^{\nu}
		\end{split}
	\end{equation} 
and $\mathcal{L}_{\text{right}}^{TH}$ acting similarly on the site $L$ and with parameters $\beta$'s. 
In the thermalized bulk generator particles on a bond $x$ and $x+1$ are redistribute according to the measure $\mu_{TH}(r|\epsilon)$, while in the thermalized boundary generators the configuration at site $1$ ($L$) is replaced by $r$, according to a rate depending on $r$ itself and on the reservoirs parameters. 
\subsubsection{The dual thermalized process}\label{subsection-dualTHM}
Here we state the duality result for the thermalized multispecies stirring process with boundary driving.
\begin{proposition}[Duality for the thermalized multispecies stirring process]\label{proposition-dualityTHM}
The thermalized multispecies stirring process $(\bm{n}(t))_{t\geq 0}$ on the state space $\Omega_{L}$, with generator $\mathcal{L}^{TH}$ defined in \eqref{thermalizedGenerator} is dual to the process $(\bm{\xi}(t))_{t\geq 0}$ on the enlarged configuration space $\widetilde{\Omega}_{L}$, with dual generator
\begin{equation}
	\widetilde{\mathcal{L}}^{TH}=\widetilde{\mathcal{L}}_{\text{left}}^{TH}+\sum_{x=1}^{L-1}\mathcal{L}_{x,x+1}^{TH}+\widetilde{\mathcal{L}}_{\text{right}}^{TH}
\end{equation}
where $\mathcal{L}^{TH}_{x,x+1}$ is \eqref{bulkThermalized} and $\widetilde{\mathcal{L}}_{\text{left}}^{TH}$ and $\widetilde{\mathcal{L}}_{\text{right}}^{TH}$ are absorbing with rate $1$:
\begin{align}
	\widetilde{\mathcal{L}}_{\text{left}}^{TH}f(\bm{\xi})&=\left\{f(\xi^{0}+\xi^{1},0,\xi^{2},\ldots,\xi^{L})-f(\bm{\xi})\right\}\nonumber\\ \widetilde{\mathcal{L}}_{\text{right}}^{TH}f(\bm{\xi})&=\left\{f(\xi^{0},\ldots,\xi^{L-1},0,\xi^{L}+\xi^{L+1})-f(\bm{\xi})\right\}\dt
\end{align}
The duality function is the same of the multispecies stirring process, i.e. 
\begin{equation}
	D(\bm{n},\bm{\xi})=\prod_{a=1}^{N-1}\left(\rho_{a}^{\text{left}}\right)^{\xi_{a}^{0}}\prod_{x=1}^{L}\frac{(\nu-\sum_{a=1}^{N-1}\xi_{a}^{x})!}{\nu!}\prod_{a=1}^{N-1}\frac{n_{a}^{x}!}{(n_{a}^{x}-\xi_{a}^{x})!}\prod_{a=1}^{N-1}\left(\rho_{a}^{\text{right}}\right)^{\xi_{a}^{L+1}}\co
\end{equation}
cf.~\eqref{dualityElements}.
\end{proposition}
This dual process is absorbing since the dual boundary generator $\mathcal{L}_{\text{left}}^{TH}$ ($\mathcal{L}_{\text{left}}^{TH}$) remove all dual particles at site $1$ ($L$) and put these particles in the extra-site $0$ ($L+1$) with rate 1.
The proof of Proposition~\ref{proposition-dualityTHM} is standard. For the sake of completeness we report it in Appendix~\ref{appendix-dualityThermalized}.
\subsection{Multispecies stirring process with reaction} \label{Subsection-RD}
{\color{black}  Considering a generic $N-1$ species Markov process $(\bm{n}(t))_{t\geq 0}$ (where in the spirit of Section~\ref{sec1}, $n_{A}^{x}$ denotes the number of particles or the number of holes at site $x$), we define the average density occupation of species $a\in\{1,\ldots,N-1\}$ at site $x$ and at time $t$ as
\begin{equation}
	\rho_{a}(x,t)=\mathbb{E}_{\mu_{0}}\left[n_{a}^{x}(t)\right]
\end{equation}
where $\mathbb{E}_{\mu_{0}}$ denotes the expectation with respect to the law of the Markov process $(\bm{n}(t))_{t\geq 0}$ initialized with arbitrary initial measure $\mu_{0}$. Calling
$\mathcal{L}$ the generator of this process, these average occupations evolves according to the evolution equation
\begin{equation}\label{EE-density}
	\frac{d \rho_{a}(x,t)}{dt}=\mathbb{E}_{\mu_{0}}\left[\mathcal{L}n_{a}^{x}(t)\right]\dt
\end{equation} }
In \cite{casini2022uphill} {\color{black}it has been defined a process with two species $a=1,2$ and the holes (denoted by $0$ in that context)} whose evolution equation for the average occupation $\rho_{a}(x,t)$ of species $a\in \{1,2\}$ is given by a difference-differential equation with two discrete Laplacians (one for each species, multiplied by a proper diffusivity constant) and a linear reaction term (mutation), i.e.
\begin{equation}\label{2SpeciesRD}
	\frac{d}{dt} \rho_{a}(x,t)=\sigma_{11}\Delta_{L}\rho_{a}(x,t)+\sigma_{12}\Delta_{L}\rho_{\overline{a}}(x,t)+\Upsilon\left(\rho_{\overline{a}}(x,t)-\rho_{a}(x,t)\right)
\end{equation} 
for $ x\in\{1,\ldots,L\}\quad a\in\{1,2\}$  and where $\overline{a}=1$ if $a=2$ and $\overline{a}=1$ if $a=2$. Here $\Delta_{L}\rho(x,t)=\rho_{a}(x+1,t)+\rho_{a}(x-1,t)-2\rho_{a}(x,t)$.
This equation is endowed with boundary conditions $\rho_{a}^{\text{left}},\rho_{\overline{a}}^{\text{left}}$ on the left and $\rho_{a}^{\text{right}},\rho_{\overline{a}}^{\text{right}}$ on the right. 
The generator of this model consists in a sum of three Markov generators: a stirring, a stirring-mutation and a pure mutation one (see section 4 of \cite{casini2022uphill}). For the lattice $\mathbb{Z}$, self-duality has been showed and the duality function is exactly \eqref{elementsABS} in case of hard-core exclusion (see section 5 of \cite{casini2022uphill}). In view of what have been proved in the previous sections of this work, some questions arise naturally:
\begin{enumerate}
		\item Is it possible to extend the results of \cite{casini2022uphill} to a case with arbitrary number of species $(N-1)\in \mathbb{N}$ and arbitrary maximal occupation $\nu\in \mathbb{N}$?
			\item Is it possible to extend the duality result to the boundary-driven case in order to have an absorbing dual process, using the same ${gl}(N)$ Lie algebraic structure?
\end{enumerate}
Therefore, we aim to find a process whose average occupations $\rho_{a}(x,t)$ for any species $a\in\{1,\ldots,N-1\}$ evolves as a system of difference-differential equation given by 
\begin{equation}\label{NSpeciesRD}
	\frac{d}{dt} \rho_{a}(x,t)=\nu\sigma_{11}\Delta_{L}\rho_{a}(x,t)+\nu\sigma_{12}\sum_{b=1,b\neq a}^{N-1}\Delta_{L}\rho_{b}(x,t)+\Upsilon\sum_{b=1,b\neq a}^{N-1}\left(\rho_{b}(x,t)-\rho_{a}(x,t)\right)\dt
\end{equation} 
 We assume that the boundary condition are given and denoted by $\rho_{a}^{\text{left}}$ and $\rho_{a}^{\text{right}}$ for all $a\in\{1,\ldots,N-1\}$ at left and right respectively, further assuming that $\sum_{A=1}^{N}\rho_{A}^{\text{left}}=\sum_{A=1}^{N}\rho_{A}^{\text{right}}=\nu$. In the following of this section, we first construct the process on a chain in order to obtain \eqref{NSpeciesRD} for the evolution of the average occupation variable of species $a$. Once we have found such a process, we construct the Lie algebraic description and we use it to prove absorbing duality.
\subsubsection{The reaction-diffusion process on a chain}
We consider the geometry of a chain of length $L$, where we assume that a reservoir with parameters $(\alpha_{A})_{A\in\{1,\ldots,N\}}$ is attached at site $1$ and a reservoir with parameters $(\beta_{A})_{A\in\{1,\ldots,N\}}$ is attached at site $L$. The state space is $\Omega_{L}$ given in \eqref{stateSpace-Chain} and the generator reads
\begin{equation}\label{RDGenerator}
	\mathcal{L}^{rd}=\mathcal{L}_{\text{right}}^{rd}+\sum_{x=1}^{L-1}\mathcal{L}_{x,x+1}^{rd}+\mathcal{L}_{\text{right}}^{rd}
\end{equation}
where
\begin{equation}\label{edgeGeneratorRD}
	\mathcal{L}_{x,x+1}^{rd}=\nu \sigma_{11}\mathcal{L}_{x,x+1}+\nu \sigma_{12}\sum_{c=1}^{N-2}\mathcal{L}_{x,x+1}^{c}+(\Upsilon-2\nu\sigma_{12})\mathcal{L}_{x,x+1}^{m}
\end{equation}
where $\mathcal{L}_{x,x+1}$ is the edge-generator \eqref{edgeGenerator} of the multispecies stirring process and, for any function $f:\Omega\to \mathbb{R}$ and for any $c\in \{1,\ldots,N-2\}$ we introduce the \textit{stirring-mutation} generator
\begin{equation}
	\mathcal{L}_{x,x+1}^{c}f(\bm{n})=\sum_{A,B=1}^{N}n_{A}^{x}n_{B}^{x+1}\left(f(\bm{n}-\bm{\delta}_{A}^{x}+\bm{\delta}_{h_{c}(B)}^{x}+\bm{\delta}_{h_{c}(A)}^{x+1}-\bm{\delta}_{B}^{x+1})-f(\bm{n})\right)
\end{equation}
with the mapping 
\begin{equation}
	\begin{split}
		h_{c}(\cdot):\{1,\ldots,N\}&\to\{1,\ldots,N\}\\
		A&\to h_{c}(A)
	\end{split}
\end{equation}
defined as
\begin{equation}
	h_{c}(A)=\begin{cases}
		A+c \quad &\text{if}\quad A+c< N \\
		A+c-N+1\quad &\text{if}\quad A+c\geq N\quad \text{and}\quad A\neq N\\
		N\quad &\text{if}\quad A=N
	\end{cases}\dt
\end{equation}
Observe that $h_{c}(\cdot)$ is surjective and injective, and thus invertible. 

Moreover, we define the \textit{pure-mutation} generator
\begin{equation}
	\mathcal{L}_{x,x+1}^{m}f(\bm{n})=\sum_{a,b=1}^{N-1}n_{a}^{x}\left(f(\bm{n}-\bm{\delta}_{a}^{x}+\bm{\delta}_{b}^{x})-f(\bm{n})\right)\dt
\end{equation}
The left boundary generator $\mathcal{L}_{\text{left}}^{rd}$ is given by \eqref{siteGenerator} at site $1$,where the $\alpha$'s parameters are defined in function of the given boundary condition of the difference-differential system as
\begin{align}\label{boundaryParamRD}
		\alpha_{a}&=\sigma_{11}\rho_{a}^{\text{left}}+\sigma_{12}\sum_{b=1\,:\,b\neq a}^{N-1}\rho_{b}^{\text{left}}\qquad \text{if} \quad a\in \{1,\ldots,N-1\}\nonumber\\
		\alpha_{N}&=\nu\left(\sigma_{11}+(N-2)\sigma_{12}\right)-\sum_{a=1}^{N-1}\alpha_{a}\dt
\end{align}
As a consequence, from the above conditions we write the boundary values in terms of the reservoir parameters as
\begin{equation}
	\rho_{a}^{\text{left}}=\frac{\alpha_{a}\sigma_{11}-\sigma_{12}\sum_{b=1\,:\,b\neq a}^{N-1}\alpha_{b}}{(\sigma_{11}-(N-2)\sigma_{12})(\sigma_{11}+(N-2)\sigma_{12})}\qquad \rho_{N}^{\text{left}}=\frac{\alpha_{N}}{(\sigma_{11}+(N-2)\sigma_{12})}
\end{equation}
for every $a\in\{1,\ldots,N-1\}$. Moreover, with this choice we have that $\sum_{A=1}^{N}\alpha_{A}=\nu(\sigma_{11}+(N-1)\sigma_{12})$ and $\sum_{A=1}^{N}\rho_{A}^{\text{left}}=\nu$. 
The boundary generator $\mathcal{L}_{\text{right}}^{rd}$ is defined analogously, but the $\alpha$'s are replaced by $\beta$'s and the boundary values $\rho_{A}^{\text{right}}$.
\paragraph{Action on the occupation variable}
We only report the final result. The details are written in Appendix~\ref{appendix-RD}. For arbitrary site $x\in \{2,\ldots,L-1\}$, the action on the occupation variable $f(\bm{n})=n_{a}^{x}$ of particle of type $a\in \{1,\ldots,N-1\}$ at site $x$ reads
\begin{equation}\label{DifferenceEquation}
	\mathcal{L}^{rd}n_{a}^{x}=\nu\sigma_{11}\Delta_{L}n_{a}^{x}+\nu\sigma_{12}\sum_{b=1\,:\,b\neq a}^{N-1}\Delta_{L}n_{b}^{x}+\Upsilon\sum_{b=1\,:\,b\neq a}^{N-1}(n_{b}^{x}-n_{a}^{x})
\end{equation}
where $\Delta_{L}n_{a}^{x}= n_{a}^{x+1}+n_{a}^{x-1}-2n_{a}^{x}$. For $x=1$ and for all $a\in \{1,\ldots,N-1\}$ we have 
\begin{equation}\label{actionBDLine}
	\mathcal{L}^{rd}n_{a}^{1}=\sigma_{11}\nu \left(\rho_{a}^{\text{left}}-2n_{a}^{1}+n_{a}^{2}\right)+\sigma_{12}\nu\sum_{b=1\,:\,b\neq a}^{N-1}\left(\rho_{b}^{\text{left}}-2n_{b}^{1}+n_{b}^{2}\right)+\Upsilon\sum_{b=1\,:\,b\neq a}^{N-1}\left(n_{b}^{1}-n_{a}^{1}\right)\dt
\end{equation}
On the occupation variable $n_{a}^{L}$ the action is similar. Therefore, using the evolution equation \eqref{EE-density} for the average densities and  the action of $\mathcal{L}^{rd}$ expressed by \eqref{DifferenceEquation} and \eqref{actionBDLine} we obtain the desired system of $N-1$ difference-differential equations \eqref{NSpeciesRD}. For these reasons, this model is a generalization to $N-1$ species and to maximal occupation $\nu$ of the one studied in \cite{casini2022uphill}. 
\subsubsection{Reversible measure for the reaction-diffusion process (equilibrium)} 
The process described by the generator $\mathcal{L}^{rd}$ introduced in \eqref{RDGenerator} is reversible with respect to the homogeneous product measure \begin{equation}\label{reversibleMeasureRD}
	\Lambda_{\text{rev}}=\bigotimes_{x=1}^{L}\Lambda_{\text{rev}}^{x}
\end{equation}
when
\begin{equation}\label{reversibilityConditionRD}
	\alpha_{A}=\beta_{A}\quad\forall A\in\{1,\ldots,N\}\qquad \text{and}\qquad \alpha_{a}=\alpha_{b}=\alpha \quad \forall a,b\in \{1,\ldots,N-1\}\dt
\end{equation}
This measure has marginals $\Lambda_{\text{rev}}^{x}$ given by 
\begin{equation}
	\Lambda^{x}_{\text{rev}}\sim \text{Multinomial}\left(\twoj,p,\ldots,p,p_{N}\right)
\end{equation}
where 
\begin{equation}
	p=\frac{\alpha}{\alpha_{N}+(N-1)\alpha}\qquad \text{and}\qquad p_{N}=\frac{\alpha_{N}}{\alpha_{N}+(N-1)\alpha}
\end{equation}
Namely,
\begin{equation}
	\Lambda_{\text{rev}}^{x}(n^{x})=\frac{\nu!}{\prod_{A=1}^{N}n_{A}^{x}!}p^{\sum_{a=1}^{N-1}n_{a}^{x}}p_{N}^{n_{N}^{x}}
\end{equation} 
This can be proved by imposing the detailed balance conditions for the bond $(x,x+1)$ and for the boundaries $\{1,L\}$.\\

\subsubsection{Duality for the reaction-diffusion process}
In this section we formulate duality for the multispecies reaction diffusion process. This answer to the second question addressed at the beginning of Section~\ref{Subsection-RD}.
\begin{proposition}[Duality for the reaction-diffusion process]\label{propositin-duality-RD}
The reaction-diffusion multispecies stirring process $(\bm{n}(t))_{t\geq 0}$, on the state space $\Omega_{L}$, with generator $\mathcal{L}^{rd}$ defined in \eqref{RDGenerator} is dual to the process $(\bm{\xi}(t))_{t\geq 0}$ on the state space $\widetilde{\Omega}_{L}$ with dual generator
\begin{equation}\label{DualGeneratorRD}
	\widetilde{\mathcal{L}}_{\text{left}}^{rd}=\widetilde{\mathcal{L}}^{rd}+\sum_{x=1}^{L-1}\mathcal{L}_{x,x+1}^{rd}+\widetilde{\mathcal{L}}_{\text{right}}^{rd}
\end{equation}
where 
$\mathcal{L}_{x,x+1}^{rd}$ is defined in \eqref{edgeGeneratorRD} and, $\widetilde{\mathcal{L}}_{\text{left}}^{rd}$ and $\widetilde{\mathcal{L}}_{\text{right}}^{rd}$ are the absorbing dual generators $\widetilde{\mathcal{L}}_{x}$ defined in \eqref{siteDualGenerator} and acting at sites $1$ and $L$, with the specific choice of the reservoirs parameters \eqref{boundaryParamRD}. The duality function
is the same of the stirring process, i.e.
\begin{equation}
	D(\bm{n},\bm{\xi})=\left(\prod_{a=1}^{N-1}\left(\rho_{a}^{\text{left}}\right)^{\xi_{a}^{0}}\right)\left(\prod_{x=1}^{L}\frac{(\nu -\sum_{a=1}^{N-1}\xi_{a}^{x})!}{\nu!}\prod_{a=1}^{N-1}\frac{n_{a}^{x}!}{(n_{a}^{x}-\xi_{a}^{x})!}\right)\left(\prod_{a=1}^{N-1}\left(\rho_{a}^{\text{right}}\right)^{\xi_{a}^{L+1}}\right)\co
\end{equation}
cf. \eqref{dualityElements}.
\end{proposition} 

\paragraph{Proof of Proposition~\ref{propositin-duality-RD}:}
it is enough to prove duality for the bulk generator, since the boundary generators are the same of the multispecies stirring process. In the spirit of Section~\ref{sectionDuality}, it is convenient to write the Hamiltonian of the process through the basis elements \eqref{actionE} of the ${gl}(N)$ Lie algebra. We have that
\begin{equation}
	H^{rd}=H_{\text{left}}+\sum_{x=1}^{L-1}\left(\sigma_{11}\mathcal{H}_{x,x+1}+\sigma_{12}\sum_{c=1}^{N-2}\mathcal{H}_{x,x+1}^{c}+(\Upsilon-2\sigma_{12})\mathcal{H}_{x,x+1}^{m}\right)+H_{\text{right}}\dt
\end{equation}
Here, $\mathcal{H}_{x,x+1}$ is  \eqref{edgeHamiltonian}  and $H_{\text{left}},H_{\text{right}}$ is \eqref{siteHamiltonian} (with the choice of the boundary parameters done in \eqref{boundaryParamRD}). We, then have that 
\begin{equation}
	\mathcal{H}_{x,x+1}^{c}=\sum_{A,B=1}^{N}\left(E_{h_{c}(A)B}\otimes E_{h_{c}(B)A}-E_{BB}\otimes E_{AA}\right)
\end{equation}
and 
\begin{equation}
	\mathcal{H}_{x,x+1}^{m}=\sum_{A,B=1}^{N}\left(E_{BA}\otimes \mathbbm{1}-E_{AA}\otimes \mathbbm{1}\right)\dt
\end{equation}
We prove that, for arbitrary $x\in\{1,\ldots,L\}$,
\begin{align}
	&(\mathcal{H}_{x,x+1}^{c})^{T}D=D\mathcal{H}_{x,x+1}^{c}\qquad \forall c\in \{1,\ldots,N-2\}\label{cDualityRelation}\\
	&(\mathcal{H}_{x,x+1}^{m})^{T}D=D\mathcal{H}_{x,x+1}^{m}\label{mDualityRelation}\dt
\end{align}
Since the matrix $R_{x}$ is the same as in Section~\ref{sectionDuality}, we shall show that $\mathcal{H}_{x,x+1}^{c}$ and $\mathcal{H}_{x,x+1}^{m}$ commute with $\exp{(E^{x})}\exp{(E^{x+1})}$, where $E$ is defined in \eqref{EquationEx}. Using \eqref{eq:comgl}, the bilinearity and associativity of the Kronecker product and the bilinearity of the brackets we obtain 
\begin{align*}
	&\left[\left(\sum_{a=1}^{N-1}E_{aN}\otimes\sum_{a=1}^{N-1}E_{aN}\right), \sum_{A,B=1}^{N}\left(E_{h_{c}(A)B}\otimes E_{h_{c}(B)A} -E_{BB}\otimes E_{AA}\right)\right]
	\\=&
	\sum_{a,b=1}^{N-1}\sum_{A,B=1}^{N}\left\{\left(E_{AN}-E_{h_{c}(B)N}\right)\otimes \left(E_{AN}-E_{h_{c}(B)N}\right)\right\}=0\dt
\end{align*}
In the last equality we used the fact that map $h_{c}(\cdot)$ is surjective. 
Concerning the commutator of $\mathcal{H}_{x,x+1}^{m}$ the proof is similar and gives
\begin{equation}
	\left[\sum_{a=1}^{N-1}E_{aN}\otimes \sum_{b=1}^{N-1}E_{bN},\;\sum_{A,B=1}^{N}\left(E_{B a}\otimes \mathbbm{1}-E_{AA}\otimes \mathbbm{1}\right)\right]=0\dt
\end{equation} 
\begin{flushright}
	$\square$
\end{flushright}

\appendix 
\section*{Appendix}
\section{Interpretation of $\Psi^{'}$ as marginals of the steady state distribution}\label{subsection-interpretationPSIp} 
We still miss the interpretation of $\Psi^{'}(\bm{\tau})$ when  $\bm{\tau}= (N,\ldots,N)$. We show here that for this particular configuration
$\Psi^{'}(N,\ldots,N) =1$. Indeed, for any $\bm{\tau}\in \Omega^{'}$ we have that 
\begin{align}
	\Psi^{'}(\bm{\tau})=&\langle\bm{\tau}|\Psi^{'}\rangle=\sum_{\bm{\sigma}\in \Omega^{'}}\langle \bm{\tau}|S_{1}|\bm{\sigma}\rangle\langle \bm{\sigma}|\Psi\rangle=\sum_{\bm{\sigma}\in \Omega^{'}}\langle \bm{\sigma}|S_{1}^{T}|\bm{\tau}\rangle\langle \bm{\sigma}|\Psi\rangle\dt
\end{align} 
{\color{black}We introduce the dual configuration constructed by attaching two extra-sites (without any particle) to the vector $\bm{\tau}$, transforming its dimension from $L$ to $L+2$. We denote it by $\bm{\xi}(\bm{\tau})=(0,\tau_{1},\ldots,\tau_{L},0)$, where we wrote $0$ in the two extra-sites to indicate that no dual particles are present there.}  More explicitly, this dual configuration is given by
\begin{equation}
	\begin{cases}
		\xi_{A}^{x}(\bm{\tau})= \delta_{\tau_{x},A}\qquad \forall A\in \{1,\ldots N\}\quad \forall x\in \{1,\ldots,L\}\\
		\xi_{1}^{0}(\bm{\tau})=\ldots=\xi_{N-1}^{0}(\bm{\tau})=0\\
		\xi_{1}^{L+1}(\bm{\tau})=\ldots=\xi_{N-1}^{L+1}(\bm{\tau})=0
	\end{cases}\dt
\end{equation} Now we consider the elements $\langle\bm{\sigma}|D|\xi(\bm{\tau})\rangle$ of the duality matrix $D$ and, {\color{black} since no dual particles are present in the extra-sites $0$ and $L+1$}, we have
\begin{align}\label{useful-InInterpretation}
	\Psi^{'}(\bm{\tau})=&\sum_{\bm{\sigma}\in \Omega^{'}}\langle \bm{\sigma}|D|\bm{\xi}(\bm{\tau})\rangle\langle \bm{\sigma}|\Psi\rangle\dt
\end{align}
We observe that for any $x\in \{1,\ldots,L\}$
\begin{equation}
	\prod_{a=1}^{N-1}\mathbbm{1}_{\{\delta_{\sigma_{x},a}\geq\xi_{a}^{x}(\bm{\tau})\}}=\left(\mathbbm{1}_{\{\sigma_{x}=\tau_{x}\}}\right)^{(1-\delta_{\tau_{x},N})}\dt
\end{equation}
Indeed, we have that 
\begin{equation}
	\prod_{a=1}^{N-1}\mathbbm{1}_{\{\delta_{\sigma_{x},a}\geq\xi_{a}^{x}(\bm{\tau})\}}=\begin{cases}
		\mathbbm{1}_{\{\sigma_{x}=\tau_{x}\}}\qquad &\text{if}\quad \tau_{x}\neq N\\
		1\qquad  &\text{if}\quad \tau_{x}= N
	\end{cases}\dt
\end{equation}
Therefore using the duality function \eqref{ElementsDualityMatrixChain_1} and equation~\ref{useful-InInterpretation}, we obtain 
\begin{equation}
	\Psi^{'}(\bm{\tau})=\mathbb{E}\left[\prod_{x=1}^{L}\left(\mathbbm{1}_{\{Y_{x}=\tau_{x}\}} \right)^{(1-\delta_{\tau_{x},N})}\right]\dt
\end{equation} 
This last equation completes and generalizes \eqref{corr-psi1} by showing that $\Psi^{'}(N,\ldots,N) =1$.

Equivalently, we can interpret the components of $\Psi'(\bm{\tau})$ as follows.
Let $\ell$ be the number of sites {\color{black}occupied by a particle} in the configuration $\bm{\tau}$, i.e.  $\ell=\sum_{x=1}^{L}(1-\delta_{\tau_{x},N})$. We denote by $x_{1},\ldots,x_{\ell}$, such that $x_{k}<x_{k+1}$ for all $k\in \{1,\ldots,\ell-1\}$, the coordinates where a particle of any species is present. Finally, we call 
\begin{equation}
	\Omega_{\ell}^{'}(\bm{\tau}):=\left\{\bm{\sigma}\in \Omega^{'}\;:\; \sigma_{x_{1}}=\tau_{x_{1}},\ldots,\sigma_{x_{\ell}}=\tau_{x_{\ell}}\right\}\dt
\end{equation} 
Observe that, if $\ell=0$ then $\Omega_{\ell}^{'}(\bm{\tau})$ does coincide with the whole state space $\Omega^{'}$, while, if $\ell=L$ then $\Omega_{\ell}^{'}(\bm{\tau})$ reduces to the configuration $\bm{\tau}$. Denoting the marginal on $\ell$ sites of the steady state distribution by 
\begin{equation}
	\mu_{x_{1},\ldots,x_{\ell}}(\bm{\tau})=\sum_{\bm{\sigma}\in \Omega_{\ell}^{'}(\bm{\tau})}\mu(\bm{\sigma})
\end{equation}
we have that 
\begin{align}
	\Psi^{'}(\bm{\tau})=&\mathbb{E}\left[\prod_{x=1}^{L}\left(\mathbbm{1}_{\{Y_{x}=\tau_{x}\}} \right)^{(1-\delta_{\tau_{x},N})}\right]=\mathbb{E}\left[\prod_{k=1}^{\ell}\mathbbm{1}_{\{Y_{x_{k}}=\tau_{x_{k}}\}}\right]=\sum_{\bm{\sigma}\in \Omega_{\ell}^{'}(\bm{\tau})}\mu(\bm{\sigma})=\mu_{x_{1},\ldots,x_{\ell}}(\bm{\tau})\dt
\end{align} 
It follows that $\Psi^{'}(\bm{\tau})$ is the marginal on $\ell$ sites $x_{1},\ldots,x_{\ell}$ of the non-equilibrium steady state distribution. In particular, when the configuration $\bm{\tau}$ has $\ell$ particles, $\Psi^{'}(\bm{\tau})$ does coincide with the $\ell$-point correlation function.

\section{One point correlations for $\nu\geq 1$}\label{appendix-1pts-general}
We take $m=1$ and we fix a site $x\in\{1,\ldots,L\}$ and a species $a\in \{1,\ldots,N-1\}$. We consider the dual configuration $\bm{\xi}=\delta_{a}^{x}$, then 
\begin{equation}
	\nu D(\bm{n},\delta_{a}^{x})=n_{a}^{x}\dt
\end{equation}
The average occupation reads
\begin{equation}\label{OnePtsCORR}
	\begin{split}
			\mathbb{E}_{\mu_{\text{NESS}}}\left[\mathbbm{1}_{\{\mathcal{Y}_{a}^{x}=n_{a}^{x}\}}\right]&=\nu\lim_{t\to\infty}\mathbb{E}_{\bm{n}}\left[D(\bm{n}(t),\delta_{a}^{x})\right]\\&=\nu\sum_{t_{a}=0}^{1}\left(\rho_{a}^{\text{left}}\right)^{t_{s}}\left(\rho_{s}^{\text{right}}\right)^{1-t_{s}}\mathbb{P}_{\delta_{a}^{x}}\left(\xi_{\infty}=t_{a}\delta_{0}+(1-t_{a})\delta_{L+1}\right)\dt
	\end{split}
\end{equation}
To explicitly find the result, we aim to determine $\mathcal{P}(1,\delta_{a}^{x})=\mathbb{P}_{\delta_{a}^{x}}\left(\xi_{\infty}=t_{a}\delta_{a}^{0}+(1-t_{a})\delta_{a}^{L+1}\right)$. This is the probability that a random walker 
(started at $x$) on the chain $\{1,\ldots,L\}$ with absorbing boundaries {\color{black}is absorbed at extra-site $0$. It fulfils the following equations discrete Laplace equation}
\begin{equation}\label{gamblers-ruin}
	\begin{cases}
		\Delta_{L}\mathcal{P}(1,\delta_{a}^{x})=0\\
		\mathcal{P}(1,\delta_{a}^{1})=\frac{|\alpha|}{1+|\alpha|}+\mathcal{P}(1,\delta_{a}^{2})\frac{1}{1+|\alpha|}\\
		\mathcal{P}(1,\delta_{a}^{L})=\mathcal{P}(1,\delta_{a}^{L-1})\frac{1}{1+|\beta|}
	\end{cases}
\end{equation}
where $\Delta_{L}$ is the discrete Laplace operator on the chain of length $L$. The equations \eqref{gamblers-ruin} can be solved and using $\mathcal{P}(1,\delta_{a}^{x})=1-\mathcal{P}(0,\delta_{a}^{x})$ we obtain 
\begin{equation}
	\mathbb{E}_{\mu_{\text{NESS}}}\left[\mathbbm{1}_{\{\mathcal{Y}_{a}^{x}=n_{a}^{x}\}}\right]=\nu\frac{\alpha_{a}(L|\beta|-|\beta|x+1)+\beta_{a}(|\alpha|x+1-|\alpha|)}{|\alpha||\beta|L-|\alpha||\beta|+|\alpha|+|\beta|}\dt
\end{equation}
\begin{remark}: In case $|\alpha|=|\beta|=\nu=1$ (when the chain is integrable) we have 
\begin{equation}
		\mathbb{E}\left[\mathbbm{1}_{\{Y_{x}=a\}}\right]=\frac{\alpha_{a}(L-x+1)+\beta_{a}x}{L+1}
\end{equation}
that is the same result obtained in \eqref{one-pts-corr}.  
\end{remark}

\section{Proof of duality of thermalized multispecies stirring}\label{appendix-dualityThermalized}
The proof of the duality consists in showing
\begin{align}
	\left(\mathcal{L}_{x,x+1}^{TH}D(\cdot,\bm{\xi})\right)(\bm{n})&=\left(\widetilde{\mathcal{L}}_{x,x+1}^{TH}D(\bm{n},\cdot)\right)(\bm{\xi})\qquad x\in \{1,\ldots,L-1\}\label{bulkDualityTH}\\
	\left(\mathcal{L}_{\text{left}}^{TH}D(\cdot,\bm{\xi})\right)(\bm{n})&=\left(\widetilde{\mathcal{L}}_{\text{left}}^{TH}D(\bm{n},\cdot)\right)(\bm{\xi})\dt\label{boundaryDualityTH}
\end{align}
Duality for the right boundary is similar. \\
Equation \eqref{bulkDualityTH} follows from the proof of the edge duality for the multispecies stirring process. To prove \eqref{boundaryDualityTH} we directly apply the generators on the duality function. For the sake of notation, we introduce 
\begin{equation}
	d(n^{x},\xi^{x}):=\frac{(\nu-\sum_{a=1}^{N-1}\xi_{a}^{x})!}{\nu!}\prod_{a=1}^{N-1}\frac{n_{a}^{x}!}{(n_{a}^{x}-\xi_{a}^{x})!}\dt
\end{equation}
First, we act with $\mathcal{L}_{\text{left}}^{TH}$ on $D(\bm{n},\bm{\xi})$. By writing explicitly $\rho_{a}^{L}=\frac{\alpha_{a}}{|\alpha|}$, this action gives
\begin{align}\label{ProvaBordoTH}
	\left(\mathcal{L}_{\text{left}}^{TH}D(\cdot,\bm{\xi})\right)(\bm{n})=\nonumber&\sum_{r_{1}=0}^{\nu}\ldots\sum_{r_{N}=0}^{\nu}\mathbbm{1}_{\{r_{1}+\ldots+r_{N}=\nu\}}\left\{\left(\prod_{a=1}^{N-1}\left(\frac{\alpha_{a}}{|\alpha|}\right)^{\xi_{a}^{0}}\right)\frac{\left(\nu-\sum_{c=1}^{N-1}\xi_{c}^{1}\right)!}{\nu!}\right.
	\\
	\times &\nonumber
	\left.\left(\prod_{b=1}^{N-1}\frac{r_{b}!}{(r_{b}-\xi^{1}_{b})!}\right)\left(\prod_{x=2}^{L}d(n^{x},\xi^{x})\right)\left(\prod_{d=1}^{N-1}(\rho_{d}^{R})^{\xi_{d}^{L+1}}\right)-D(\bm{n},\bm{\xi})\right\}
	\\\times &
	\frac{\nu!}{r_{1}!\cdots r_{N}!}\left(\prod_{a=1}^{N-1}\left(\frac{\alpha_{a}}{\alpha_{N}}\right)^{r_{a}}\left(\frac{\alpha_{N}}{|\alpha|}\right)^{\nu}\right)\dt
\end{align}
We consider the first addend in the curly bracket of the \eqref{ProvaBordoTH}. We add and remove $\xi_{a}^{1}$ at the exponent of $\left(\frac{\alpha_{a}}{|\alpha|}\right)$ and then we multiplying this first addend by $\left(\frac{\alpha_{N}}{|\alpha|}\right)^{-(r_{1}^{1}+\ldots+r_{N-1})+(r_{1}^{1}+\ldots+r_{N-1})}$ to get
\begin{align}\label{pezzo1}
	&\sum_{r_{1}=0}^{\nu}\ldots\sum_{r_{N}=0}^{\nu}\mathbbm{1}_{\{r_{1}+\ldots+r_{N}=\nu\}}\nonumber
	\\
	\times&
	\left\{\left(\prod_{a=1}^{N-1}\left(\frac{\alpha_{a}}{|\alpha|}\right)^{\xi_{a}^{0}-\xi_{a}^{1}+\xi_{a}^{1}}\right)\frac{\left(\nu-\sum_{c=1}^{N-1}\xi_{c}^{1}\right)!}{\nu!}\left(\prod_{b=1}^{N-1}\frac{r_{b}!}{(r_{b}-\xi^{1}_{b})!}\right)\left(\prod_{x=2}^{L}d(n^{x},\xi^{x})\right)\left(\prod_{d=1}^{N-1}(\rho_{d}^{R})^{\xi_{d}^{L+1}}\right)\right\}\nonumber\\
	\times	&\frac{\nu!}{r_{1}!\cdots r_{N}!}\left(\prod_{a=1}^{N-1}\left(\frac{\alpha_{a}}{\alpha_{N}}\right)^{r_{a}}\right)\left(\frac{\alpha_{N}}{|\alpha|}\right)^{-(r_{1}^{1}+\ldots+r_{N-1})+(r_{1}^{1}+\ldots+r_{N-1})}\nonumber
	\\=&
	\left(\prod_{a=1}^{N-1}\left(\frac{\alpha_{a}}{|\alpha|}\right)^{\xi_{a}^{0}+\xi_{a}^{1}}\right)\sum_{r_{1}=0}^{\nu}\ldots\sum_{r_{N}=0}^{\nu}\mathbbm{1}_{\{r_{1}+\ldots+r_{N}=\nu\}}	\frac{\left(\nu-\xi_{1}^{1}-\ldots-\xi_{N-1}^{1}\right)!}{(r_{1}-\xi_{1}^{1})!\cdots(r_{N-1}-\xi_{N-1}^{1})!(\nu-r_{1}-\ldots-r_{N-1})!}\nonumber\\
	\times& \left(\prod_{b=1}^{N-1}\left(\frac{\alpha_{b}}{|\alpha|}\right)^{r_{b}-\xi_{b}^{1}}\right)\left(\frac{\alpha_{N}}{|\alpha|}\right)^{\nu-r_{1}-\ldots-r_{N-1}}\left(\prod_{x=2}^{N}d_{x}(n^{x},\xi^{x})\right)\left(\prod_{d=1}^{N-1}(\rho_{d}^{R})^{\xi_{d}^{L+1}}\right)\nonumber
	\\=&
	\left(\prod_{a=1}^{N-1}\left(\rho_{a}^{\text{left}}\right)^{\xi_{a}^{0}+\xi_{a}^{1}}\right)\left(\prod_{x=2}^{L}d(n^{x},\xi^{x})\right)\left(\prod_{d=1}^{N-1}(\rho_{d}^{\text{right}})^{\xi_{d}^{L+1}}\right)\dt
\end{align}
Where we used 
\begin{equation*}
\frac{\left(\nu-\sum_{c=1}^{N-1}\xi_{c}^{1}\right)!}{\nu!}\left(\prod_{b=1}^{N-1}\frac{r_{b}!}{(r_{b}-\xi^{1}_{b})!}\right)\frac{\nu!}{r_{1}!\cdots r_{N}!}=\frac{\left(\nu-\xi_{1}^{1}-\ldots-\xi_{N-1}^{1}\right)!}{(r_{1}-\xi_{1}^{1})!\cdots(r_{N-1}-\xi_{N-1}^{1})!(\nu-r_{1}-\ldots-r_{N-1})!}
\end{equation*}
and 
\begin{align*}
&\left(\prod_{a=1}^{N-1}\left(\frac{\alpha_{a}}{|\alpha|}\right)^{\xi_{a}^{0}-\xi_{a}^{1}+\xi_{a}^{1}}\right)\left(\prod_{b=1}^{N-1}\left(\frac{\alpha_{b}}{\alpha_{N}}\right)^{r_{b}}\right)\left(\prod_{c=1}^{N-1}\left(\frac{\alpha_{N}}{|\alpha|}\right)^{\nu-r_{c}+r_{c}}\right)\\=&	\left(\prod_{a=1}^{N-1}\left(\frac{\alpha_{a}}{|\alpha|}\right)^{\xi_{a}^{0}+\xi_{a}^{1}}\right)\left(\prod_{b=1}^{N-1}\left(\frac{\alpha_{b}}{|\alpha|}\right)^{r_{b}-\xi_{b}^{1}}\right)\left(\frac{\alpha_{N}}{|\alpha|}\right)^{\nu-r_{1}-\ldots-r_{N-1}}
\end{align*}
and the multinomial theorem in the last equality. \\ We now consider the second addend in the curly brackets of \eqref{ProvaBordoTH}.
We multiplying it by $\prod_{a=1}^{N-1}\left(\frac{\alpha_{N}}{|\alpha|}\right)^{-r_{a}+r_{a}}$ to get
\begin{equation}\label{pezzo2}
	\begin{split}
		&D(\bm{n},\bm{\xi})\sum_{r_{1}=0}^{\nu}\ldots\sum_{r_{N}=0}^{\nu}\mathbbm{1}_{\{r_{1}+\ldots+r_{N}=\nu\}}\frac{\nu!}{r_{1}!\cdots r_{N}!}\left(\prod_{a=1}^{N-1}\left(\frac{\alpha_{a}}{\alpha_{N}}\right)^{r_{a}}\right)\left(\frac{\alpha_{N}}{|\alpha|}\right)^{\nu}\left(\frac{\alpha_{N}}{|\alpha|}\right)^{-(r_{1}+\ldots+r_{N-1})+(r_{1}+\ldots+r_{N-1})}\\=&D(\bm{n},\bm{\xi})
	\end{split}
\end{equation} 
where we used the multinomial theorem. Therefore, replacing \eqref{pezzo1} and \eqref{pezzo2} in \eqref{ProvaBordoTH} we have that
\begin{align}
	\left(\mathcal{L}_{\text{left}}^{TH}D(\cdot,\bm{\xi})\right)(\bm{n})&=\left\{\prod_{a=1}^{N-1}\left(\frac{\alpha_{k}}{|\alpha|}\right)^{\xi_{a}^{0}+\xi_{a}^{1}}\prod_{x=2}^{N}d_{x}(n^{x},\xi^{x})\prod_{d=1}^{N-1}(\rho_{a}^{R})^{\xi_{d}^{L+1}}-D(\bm{n},\bm{\xi})\right\}
	\\&=
	\left(\widetilde{\mathcal{L}}_{\text{left}}^{TH}D(\bm{n},\cdot)\right)(\bm{\xi})\dt
\end{align}

\begin{flushright}
	$\square$
\end{flushright}
\section{Action on the occupation variable of $\mathcal{L}^{rd}$}\label{appendix-RD}
\paragraph{Action of $\mathcal{L}_{x,x+1}^{rd}$ on the occupation variable}
For all $x\in \{1,\ldots,L-1\}$ and for all $d\in \{1,\ldots,N-1\}$ we have
\begin{equation}\label{actionGraphRD}
	\begin{split}
		\mathcal{L}_{x,x+1}^{rd}n_{d}^{x}&=\nu \sigma_{11}(n_{d}^{x+1}-n_{d}^{x})+\nu\sigma_{12}\sum_{b=1\,:	,b\neq d}^{N-1}(n_{b}^{x+1}-n_{d}^{x})+(\Upsilon-2\nu\sigma_{12})\sum_{b=1}^{N-1}(n_{b}^{x}-n_{d}^{x})\dt
	\end{split}
\end{equation}
Indeed, 
\begin{enumerate}
	\item \textit{\underline{Stirring generator}}: {using edge generator $\mathcal{L}_{x,x+1}$ of the multispecies stirring process introduced in Section~\ref{sec1} (with $y=x+1$) we have that} 
\begin{equation}
	\mathcal{L}_{x,x+1}n_{d}^{x}=\sum_{A,B=1}^{N}n_{A}^{x}n_{B}^{x+1}\left((n_{d}^{x}-\bm{\delta}_{A}^{x}+\bm{\delta}_{B}^{x}+\bm{\delta}_{A}^{x+1}-\bm{\delta}_{B}^{x+1})-n_{d}^{x}\right)\dt
\end{equation}
In the brackets of the right-hand-side of the above equation we have: a contribution $-n_{d}^{x}n_{B}^{x+1}$, when $A=d$; a contribution $n_{A}^{x}n_{d}^{x+1}$, when $B=d$. Thus we obtain 
\begin{align*}
	\mathcal{L}_{x,x+1}n_{d}^{x}=&\sum_{A=1}^{N}n_{A}^{x}n_{d}^{x+1}-\sum_{B=1}^{N}n_{d}^{x}n_{B}^{x+1}\\
	=&\sum_{a=1}^{N-1}n_{a}^{x}n_{d}^{x+1}+(\nu-\sum_{c=1}^{N-1}n_{c}^{x})n_{d}^{x+1}-\sum_{b=1}^{N-1}n_{d}^{x}n_{b}^{x+1}-n_{d}^{x}(\nu-\sum_{c=1}^{N-1}n_{c}^{x+1})\\
	=&\nu\left(n_{d}^{x+1}-n_{d}^{x}\right)
\end{align*} 
where we have used the fact that $n_{N}^{x}=\nu-\sum_{c=1}^{N-1}n_{c}^{x}$.
\item \textit{\underline{Stirring-mutation generator}}: for every $c\in \{1,\ldots,N-2\}$ we have 
\begin{equation}
	\begin{split}
		\mathcal{L}_{x,x+1}^{c}n_{d}^{x}&=\sum_{A,B=1}^{N}n_{A}^{x}n_{B}^{x+1}\left((n_{d}^{x}-\bm{\delta}_{A}^{x}+\bm{\delta}_{h_{c}(B)}^{x}+\bm{\delta}_{h_{c}(A)}^{x+1}-\bm{\delta}_{B}^{x+1})-n_{d}^{x}\right)\dt
	\end{split}
\end{equation}
In the brackets of the right-hand-side of the above equation we have: a contribution $-n_{d}^{x}n_{D}^{x+1}$, when $A=d$ (here we call $D=h_{c}(B)$); a contribution $n_{A}^{x}n_{h_{c}(B^{'})}^{x+1}$, where we denoted by $B^{'}\in \{1,\ldots,N\}$ the species of particle such that $h_{c}(B^{'})=d$. Let us observe that, by the definition of the map $h_{c}(\cdot)$ and for the fact that $1\leq c\leq N-2$, we have that $h_{c}(B^{'})\neq d$ and $h_{c}(B^{'})\neq N$. Thus we obtain 
\begin{align}
		\mathcal{L}_{x,x+1}^{c}n_{d}^{x}&=\sum_{A=1}^{N}n_{A}^{x}n_{h_{c}(B^{'})}^{x+1}-\sum_{D=1}^{N}n_{d}^{x}n_{D}^{x+1}
		\\&=
		\sum_{a=1}^{N-1}n_{a}^{x}n_{h_{c}(B^{'})}^{x+1}+\left(\nu-\sum_{b=1}^{N-1}n_{h_{c}(B^{'})b}^{x}\right)n_{b}^{x+1}-\sum_{a=1}^{N-1}n_{d}^{x}n_{a}^{x+1}-n_{d}^{x}\left(\nu-\sum_{b=1}^{N-1}n_{b}^{x+1}\right)
		\\&=
		\nu(n_{h_{c}(B^{'})}^{x+1}-n_{d}^{x})
\end{align}
where  we used the fact that $n_{N}^{x}=\nu-\sum_{b=1}^{N-1}n_{b}^{x}$. 
\item \textit{\underline{Pure-mutation generator}}: here we have that 
\begin{equation}
	\mathcal{L}_{x,y}^{m}n_{d}^{x}=\sum_{A,B=1}^{N-1}n_{A}^{x}\left((n_{d}^{x}-\bm{\delta}_{A}^{x}+\bm{\delta}_{B}^{x})-n_{d}^{x}\right)=\sum_{b=1}^{N-1}(n_{b}^{x}-n_{d}^{x})
\end{equation}
\end{enumerate}
Then, summing over all indices $c\in\{1,\ldots,N-2\}$, we obtain \eqref{actionGraphRD} and \eqref{DifferenceEquation} follows.
\paragraph{Action of $\mathcal{L}_{\text{left}}^{rd}$ on the occupation variable}
Acting on $n_{d}^{1}$  with $\mathcal{L}_{\text{left}}^{rd}$ we obtain
\begin{equation}\label{actionBoudaryRD}
	\begin{split}
		\mathcal{L}_{\text{left}}^{rd}n_{d}^{1}&=\sum_{A,B=1}^{N}\alpha_{A}n_{B}^{1}\left((n_{d}^{1}-\bm{\delta}_{B}^{1}+\bm{\delta}_{A}^{1})-n_{d}^{1}\right)=\alpha_{d}\sum_{B=1\,:\,B\neq d}^{N}n_{B}^{1}-n_{d}^{1}\sum_{A=1\,:A\neq d}^{N}\alpha_{A}
	\end{split}
\end{equation}
Using equation \eqref{boundaryParamRD}, we obtain 
\begin{align}
			\mathcal{L}_{\text{left}}^{rd}n_{d}^{1}&=
		\alpha_{d}\left(\nu-n_{d}^{1}\right)-n_{d}^{1}\left(\nu(\sigma_{11}+(N-2)\sigma_{12})-\alpha_{d}\right)\nonumber
		\\&=\alpha_{d}\nu-n_{d}^{1}\nu(\sigma_{11}+(N-2)\sigma_{12})\nonumber
		\\&=
		\nu\sigma_{11}\rho_{d}^{\text{left}}+\nu\sigma_{12}\left(\sum_{b=1\,:\,b\neq d}^{N-1}\rho_{b}^{\text{left}}\right)-n_{d}^{1}\nu(\sigma_{11}+(N-2)\sigma_{12})\nonumber
		\\&=
		\nu \sigma_{11} (\rho_{d}^{\text{left}}-n_{d}^{1})+\nu\sigma_{12}\sum_{b=1\,:\,b\neq d}^{N-1}(\rho_{b}^{\text{left}}-n_{d}^{1})
	\end{align}
Therefore, \eqref{actionBDLine} follows.
\bibliography{ref}

\begin{thebibliography}{10}

\bibitem{schutzSandow}
Gunter~M. Schütz and Sven Sandow.
\newblock Non-abelian symmetries of stochastic processes: Derivation of
  correlation functions for random-vertex models and
  disordered-interacting-particle systems.
\newblock {\em Phys.Rev. E}, 49:2726--2741, 1994.

\bibitem{KMP}
Claude Kipnis, Carlo Marchioro, and Errico Presutti.
\newblock Heat flow in an exactly solvable model.
\newblock {\em Journal of Statistical Physics}, 27:65--74, 1982.

\bibitem{giardina2009duality}
Cristian Giardin{\`a}, Jorge Kurchan, Frank Redig, and Kiamars Vafayi.
\newblock Duality and hidden symmetries in interacting particle systems.
\newblock {\em Journal of Statistical Physics}, 135(1):25--55, 2009.

\bibitem{carinci2013duality}
Gioia Carinci, Cristian Giardin{\`a}, Claudio Giberti, and Frank Redig.
\newblock Duality for stochastic models of transport.
\newblock {\em Journal of Statistical Physics}, 152(4):657--697, 2013.

\bibitem{zhou2021orthogonal}
Zhengye Zhou.
\newblock Orthogonal polynomial stochastic duality functions for multi-species
  {SEP}(2j) and multi-species {IRW}.
\newblock {\em SIGMA. Symmetry, Integrability and Geometry: Methods and
  Applications}, 17:113, 2021.

\bibitem{vanicat2017exact}
Matthieu Vanicat.
\newblock Exact solution to integrable open multi-species {SSEP} and
  macroscopic fluctuation theory.
\newblock {\em Journal of Statistical Physics}, 166(5):1129--1150, 2017.

\bibitem{casini2023density}
Francesco Casini, Cristian Giardin{\`a}, and Frank Redig.
\newblock Density fluctuations for the multi-species stirring process.
\newblock {\em preprint arXiv:2307.05111}, 2023.

\bibitem{derrida1993exact}
Bernard Derrida, Martin~R. Evans, Vincent Hakim, and Vincent Pasquier.
\newblock Exact solution of a {1D} asymmetric exclusion model using a matrix
  formulation.
\newblock {\em Journal of Physics A: Mathematical and General}, 26(7):1493,
  1993.

\bibitem{1993JSP....72..277S}
Gunter~M. {Sch{\"u}tz} and E.~{Domany}.
\newblock {Phase transitions in an exactly soluble one-dimensional exclusion
  process}.
\newblock {\em Journal of Statistical Physics}, 72(1-2):277--296, July 1993.

\bibitem{spohn1983long}
Herbert Spohn.
\newblock Long range correlations for stochastic lattice gases in a
  non-equilibrium steady state.
\newblock {\em Journal of Physics A: Mathematical and General}, 16(18):4275,
  1983.

\bibitem{derrida2007non}
Bernard Derrida.
\newblock Non-equilibrium steady states: fluctuations and large deviations of
  the density and of the current.
\newblock {\em Journal of Statistical Mechanics: Theory and Experiment},
  2007(07):P07023, 2007.

\bibitem{derrida1998exact}
Bernard Derrida and Joel~L. Lebowitz.
\newblock Exact large deviation function in the asymmetric exclusion process.
\newblock {\em Physical review letters}, 80(2):209, 1998.

\bibitem{mallick2022exact}
Kirone Mallick, Hiroki Moriya, and Tomohiro Sasamoto.
\newblock Exact solution of the macroscopic fluctuation theory for the
  symmetric exclusion process.
\newblock {\em Physical Review Letters}, 129(4):040601, 2022.

\bibitem{bodineau2005current}
Thierry Bodineau and Bernard Derrida.
\newblock Current large deviations for asymmetric exclusion processes with open
  boundaries.
\newblock {\em Journal of Statistical Physics}, 123:277--300, 2006.

\bibitem{schutzManyBody}
Gunter~M. Sch{\"u}tz.
\newblock Exactly solvable models for many-body systems far from equilibrium.
\newblock In {\em Phase transitions and critical phenomena}, volume~19, pages
  1--251. Elsevier, 2001.

\bibitem{SSEPReviewRagoucy}
Nicolas Cramp{\'e}, Eric Ragoucy, and Matthieu Vanicat.
\newblock Integrable approach to simple exclusion processes with boundaries.
  review and progress.
\newblock {\em Journal of Statistical Mechanics: Theory and Experiment},
  2014(11):P11032, 2014.

\bibitem{frassek2020eigenstates}
Rouven Frassek.
\newblock Eigenstates of triangularisable open {XXX} spin chains and
  closed-form solutions for the steady state of the open {SSEP}.
\newblock {\em Journal of Statistical Mechanics: Theory and Experiment},
  2020(5):053104, 2020.

\bibitem{frassek2020duality}
Rouven Frassek, Cristian Giardin{\`a}, and Jorge Kurchan.
\newblock Duality and hidden equilibrium in transport models.
\newblock {\em SciPost Physics}, 9(4):054, 2020.

\bibitem{schuetz2022reverse}
Gunter~M. Sch{\"u}tz.
\newblock A reverse duality for the asep with open boundaries.
\newblock {\em Journal of Physics A: Mathematical and Theoretical}, 2022.

\bibitem{barraquand2022markov}
Guillaume Barraquand and Ivan Corwin.
\newblock Markov duality and {Bethe} ansatz formula for half-line open {ASEP}.
\newblock {\em preprint arXiv:2212.07349}, 2022.

\bibitem{casini2022uphill}
Francesco Casini, Cristian Giardin{\`a}, and Cecilia Vernia.
\newblock Uphill in reaction-diffusion multi-species interacting particles
  systems.
\newblock {\em Journal of Statistical Physics}, 190(8):132, 2023.

\bibitem{schutzReaction}
Gunter~M. Sch{\"u}tz.
\newblock Reaction-diffusion processes of hard-core particles.
\newblock {\em Journal of statistical physics}, 79:243--264, 1995.

\bibitem{belitsky2015self}
Vladimir Belitsky and Gunter~M. Sch{\"u}tz.
\newblock Self-duality for the two-component asymmetric simple exclusion
  process.
\newblock {\em Journal of mathematical physics}, 56(8):083302, 2015.

\bibitem{jansen2014notion}
Sabine Jansen and Noemi Kurt.
\newblock On the notion(s) of duality for {M}arkov processes.
\newblock {\em Probability Surveys}, 11:59 -- 120, 2014.

\bibitem{alcaraz}
Francisco~C. Alcaraz, Michel Droz, Malte Henkel, and Vladimir Rittenberg.
\newblock Reaction-diffusion processes, critical dynamics, and quantum chains.
\newblock {\em Annals of Physics}, 230(2):250--302, 1994.

\bibitem{melo2005bethe}
C.S. Melo, Giuliano~A.P. Ribeiro, and Marcio~Josè Martins.
\newblock Bethe ansatz for the {XXX-S} chain with non-diagonal open boundaries.
\newblock {\em Nuclear Physics B}, 711(3):565--603, 2005.

\end{thebibliography}
\bibliographystyle{unsrt}

\end{document}